\documentclass[12pt]{article}
\usepackage{epsfig,amsfonts,amssymb}
\usepackage[english]{babel}
\usepackage{amsmath}
\usepackage{comment}
\usepackage{cite}
\usepackage[a4paper, top=2.5cm, bottom=2.5cm, left=2.2cm, right=2.2cm]
{geometry}
\usepackage{graphicx}

\usepackage{caption, subcaption}
\usepackage{float}
\usepackage{appendix}

\usepackage{tcolorbox}

\usepackage{tikz}
\usetikzlibrary{shapes.misc}

\usepackage{tikz-3dplot}
\definecolor{mycolor1}{rgb}{0.9, 0.1, 0.4}

\def\ZZZ{{\hbox{ Z\kern-1.6mm Z}}}
\def\RRR{{\hbox{ R\kern-2.4mm R}}}
\def\CCC{{\hbox{ C\kern-2.0mm C}}}
\def\zzz{{\hbox{z\kern-1mm z}}}

\newcommand{\qeq}{{\hbox{=\kern-2.3mm ? \kern.5mm }}}
\renewcommand{\qeq}{=}

\newcommand{\tp}{\tilde}

\newcommand{\be}{\begin{equation}}
\newcommand{\ee}{\end{equation}}
\newcommand{\ben}{\begin{eqnarray}\displaystyle}
\newcommand{\een}{\end{eqnarray}}

\def\one{{\hbox{ 1\kern-.8mm l}}}
\def\zero{{\hbox{ 0\kern-1.5mm 0}}}

\newcommand{\bea}[1]{\begin{eqnarray}\label{#1} }
\newcommand{\eea}{\end{eqnarray}}

\usepackage{bm}

\usepackage{enumitem}
\setlist[itemize]{noitemsep, topsep=0pt}

\def\figone{

\def\JPicScale{0.4}
\ifx\JPicScale\undefined\def\JPicScale{1}\fi
\unitlength \JPicScale mm
\begin{picture}(140,90)(0,0)
\linethickness{0.1mm}
\put(30,60){\line(1,0){100}}
\linethickness{0.1mm}
\put(30,20){\line(1,0){100}}
\linethickness{0.3mm}
\multiput(30,90)(0.12,-0.18){167}{\line(0,-1){0.18}}
\linethickness{0.3mm}
\multiput(100,60)(0.12,0.12){250}{\line(1,0){0.12}}
\linethickness{0.3mm}
\multiput(100,20)(0.18,-0.12){167}{\line(1,0){0.18}}
\linethickness{0.3mm}
\multiput(30,0)(0.18,0.12){167}{\line(1,0){0.18}}
\put(150,62){\makebox(0,0)[cc]{$x^{\prime0}=T$}}

\put(155,22){\makebox(0,0)[cc]{$x^{\prime0}=-T$}}

\put(95,65){\makebox(0,0)[cc]{$c_{(1)}$}}

\put(140,90){\makebox(0,0)[cc]{$r_{(1)}$}}

\put(20,90){\makebox(0,0)[cc]{$r_{(2)}$}}

\put(140,0){\makebox(0,0)[cc]{$r_{(3)}$}}

\put(20,0){\makebox(0,0)[cc]{$r_{(4)}$}}

\put(55,65){\makebox(0,0)[cc]{$c_{(2)}$}}

\put(95,15){\makebox(0,0)[cc]{$c_{(3)}$}}

\put(65,15){\makebox(0,0)[cc]{$c_{(4)}$}}

\end{picture}

}

\newenvironment{claim}[1]{\par\noindent\underline{Claim:}\space#1}{}
\newenvironment{claimproof}[1]{\par\noindent\underline{Proof:}\space#1}{\hfill $\blacksquare$}

\begin{document}

\baselineskip 24pt

\begin{center}

{\Large \bf On the Positive Geometry of Quartic Interactions III : One Loop Integrands from Polytopes}\\
{\bf    }

\end{center}

\vskip .6cm
\medskip

\vspace*{4.0ex}

\baselineskip=18pt

\centerline{\large \rm Mrunmay Jagadale and Alok Laddha}

\vspace*{4.0ex}
\centerline{\large \it ~Chennai Mathematical Institute, Siruseri, Chennai, India}


\vspace*{1.0ex}
\centerline{\small E-mail: mrunmay@cmi.ac.in, aladdha@cmi.ac.in}

\vspace*{5.0ex}





\centerline{\bf Abstract} \bigskip

Building on the seminal work of Arkani-Hamed, He, Salvatori and Thomas (AHST) \cite{Arkani-Hamed:2019vag} we explore the positive geometry encoding one loop scattering amplitude for quartic scalar interactions. We define a new class of combinatorial polytopes that we call pseudo-accordiohedra whose poset structures are associated to singularities of the one loop integrand associated to scalar quartic interactions. Pseudo-accordiohedra parametrize a family of projective forms on the abstract kinematic space defined by AHST and restriction of these forms to the type-D associahedra can be associated to one-loop integrands for quartic interactions. The restriction (of the projective form) can also be thought of as a canonical top form on certain geometric realisations of pseudo-accordiohedra. Our work explores a large class of geometric realisations of the type-D associahedra which include all the AHST realisations. These realisations are  based on the pseudo-triangulation model for type-D cluster algebras discovered by Ceballos and Pilaud\cite{ceballos-pilaud}.

\vfill \eject

\baselineskip 18pt


\section{Introduction}
In \cite{Arkani-Hamed:2017mur}, Arkani-Hamed, Bai, He and Yun introduced the amplituhedron program to the world of non-supersymmetric quantum field theories. In a striking development they proved that the tree-level scattering amplitude of cubic bi-adjoint scalar theory is a canonical form on a polytope located inside the positive region of kinematic space of Mandelstam variables. The polytope is known as associahedron \cite{Stasheff} and is the amplituhedron for the bi-adjoint scalar theories. Some of the fundamental  postulates in the S-matrix program are simply consequences of combinatorial and geometric properties of the associahedron. The singularities of scattering form ensured that the amplitude was local and the factorisation properties of associahedron implied that the amplitude was unitary. These ideas have been generalised to include polynomial scalar interactions \cite{Banerjee:2018tun, Raman:2019utu, Aneesh:2019ddi} as well as (a class of) non-planar amplitudes whose poles are associated to boundary of a polytope called permutahedron\cite{Early:2017lku}.\\
In \cite{Arkani-Hamed:2019vag},  Arkani-Hamed, He, Salvatori and Thomas (AHST) brought the 1-loop  scalar scattering amplitudes within the paradigm of the amplituhedron program. (For earlier work on exploring the relationship between 1-loop $\phi^{3}$ integrand and positive geometry, we refer the reader to \cite{Salvatori:2018aha}.)
The catalyst for this development was the fact that associahedra are combinatorial models for type-A cluster algebras. All the finite dimensional cluster algebras (classified by the Dynkin classification \cite{2003InMat.15463F}) are generated using clusters which can be associated to (dual of) Feynman diagrams of $\phi^{3}$ theory. In particular, clusters in the type-D cluster algebras can be labelled by triangulations of a ``marked" polygon \cite{2006math8367F}. These ideas were made precise in \cite{Arkani-Hamed:2019vag} where it was shown that the combinatorial polytope associated to  type-D cluster algebra is the ``amplituhedron: for 1-loop $\phi^{3}$ integrand. The geometric realisation of these polytopes in an abstract kinematic space were obtained by the intriguing structures called Causal diamonds which mapped the problem of finding a geometric realisation into a finding solution (satisfying certain positivity property) to the wave equation on a two dimensional null lattice. The boundary structure of type-D associahedra ensures the expected factorisation properties of the 1-loop $\phi^{3}$ integrand and shows a clear relationship of the loop integrand with the forward scattering limit of tree-level amplitudes.\\
In \cite{Aneesh:2019cvt} along with other authors, we explored the ``upgrade" of amplitude as differential forms for $\phi^{4}$ interactions \cite{Banerjee:2018tun, Raman:2019utu, Aneesh:2019ddi}. By using the results of Padrol, Palu, Pilaud and Plamondon \cite{1906ppp}, it was shown in \cite{Aneesh:2019cvt} that tree-level planar amplitudes in $\phi^{4}$ theory could be understood as scattering forms in one of the two ways.\vspace*{-0.2in}
\begin{enumerate}
\item There is a set of unique lower forms on the kinenatic space associahedra which generate the quartic amplitudes
\item There is a set of convex polytopes (called accordiohedra or Stokes polytope) with canonical top forms which generate the amplitude. 
\end{enumerate}
Although the analysis in \cite{Aneesh:2019cvt} was explicitly for quartic interactions, it can be easily generalised to $\phi^{p}$ interactions.\\ 
In this paper, we investigate these connections between 1-loop scattering amplitudes in $\phi^{4}$ theory on one hand and   lower forms on associahedra, and polytopes in kinematic space on the other hand to loop level integrands.\\ In a nutshell, we show the following. 
\begin{itemize}
\item There exists a combinatorial polytope that we call a pseudo-accordiohedron whose facets are in bijection with various singularities of the 1-loop integrand.We suspect that these polytopes have been known to mathematicians in other guises \cite{1807ppp}, but the definition we propose here appears new. 
\item The one-loop integrand for quartic interactions are obtained from certain lower forms on type-D associahedra such that the poles of this lower form correspond to poles of $\phi^{4}$ integrand. 
\item There exists geometric realisation of the pseudo-accordiohedron in the abstract kinematic space defined in \cite{Arkani-Hamed:2019vag} obtained via certain projection of the type-D associahedra.
\item The lower form mentioned in point 2 is the canonical top form on these realisations
\end{itemize}
Hence both the (planar) tree-level amplitudes and one-loop integrands in $\phi^{4}$ theory show a ``universal" relationship with the amplituhedron of the bi-adjoint $\phi^{3}$ theory. In both the cases, the desired object (amplitudes at tree-level and integrands at one loop level)  are generated by certain unique lower forms on the (type-A or type-D) associahedron. These lower forms are simply restrictions of projective forms in kinematic space. Projectivity is a consequence of the fact that these forms are uniquely determined by their singularity structure and a class of combinatorial polytopes, (accordiohedron for tree-level amplitudes and pseudo-accordiohedron for loop integrands). 
For planar tree level amplitudes it was shown in \cite{Aneesh:2019cvt} that there is an entire family of lower $\frac{n-4}{2}$ forms which generate an $n$ point amplitude, such that each of these forms is a lower form on certain geometric realisations of the associahedron. These realisations were obtained in \cite{1906ppp} using an algebra called gentle algebra that is determined uniquely using triangulation of a planar polygon.\\
We show that precisely in the same way, there is an entire class of geometric realisation of type-D associahedron. Just as in the case of associahedron, different realisations of a type-D associahedron are obtained using an algebra that we refer to as colored gentle algebra. The colored gentle algebra is determined uniquely by using pseudo-triangulation of a polygon with an annulus inside \cite{ceballos-pilaud}. These realisations include all the AHST type-D associahedra obtained in \cite{Arkani-Hamed:2019vag}.\\ 
We thus show that there is a beautiful synthesis between combinatorial structures related to dissections of polygons, gentle algebras, geometric realisation of polytopes such as associahedron and scattering amplitudes at tree and loop level.\\
This paper is organised as follows. In section \ref{rev}, we give a rather quick review of some of the key tenets of the theory of dissection quiver and gentle algebra in the context of $\phi^{4}$ theory. In section \ref{Pseudo-Triangulation Model}, we review a combinatorial model for type-D associahedron called pseudo-triangulation model proposed by Ceballos and Pilaud. We use this model to generate a class of geometric realisations of the type-D associahedra in an abstract kinematic space ${\cal KL}_{n}$ introduced in \cite{Arkani-Hamed:2019vag}. These realisations are obtained using an algebra which is a generalisation of the gentle algebra defined in \cite{1906ppp}. In section \ref{fi41l} we introduce a combinatorial polytope that we call pseudo-accordiohedron.We show that the combinatorial and geometric realisation of the pseudo-accordiohedron generates loop integrands in $\phi^{4}$ theory. We end with some remarks and a sampling of open questions.
\section{A lightening review of Associahedra and Accordiohedra in Kinematic Space} \label{rev}
In this section, we briefly review the essential ideas from \cite{Arkani-Hamed:2017mur, Banerjee:2018tun, Raman:2019utu, 1906ppp, Aneesh:2019cvt}. We only review the results which are central to our analysis in this paper. This review is by no means self-contained and the interested reader should consult the cited papers (and references therein) for more details.

We consider ``planar" amplitudes in massless scalar field theories with $\phi^{3}$ or $\phi^{4}$ interactions.\footnote{For generic scalar interactions, we refer the reader to \cite{Raman:2019utu, Aneesh:2019ddi}.} Planar amplitudes should be thought of as scattering amplitudes where the order of scattering particles placed on the boundary of a disc is fixed once and for all. These amplitudes are basis of scattering amplitudes in theories with color (such as bi-adjoint $\phi^{3}$ theory) but for the purpose of our paper, we suppress color indices and just refer to them as planar amplitudes. 

The basic arena in which $n$-point planar scattering amplitudes live is known as planar kinematic space ${\cal K}_{n}$. This space is $\frac{n(n-3)}{2}$ dimensional.\footnote{We assume here that we are in arbitrary number of dimensions larger then $n$, so that complicated constraints imposed by Gram determinant conditions do not complicate matters.} A nice choice of basis in this space is provided by planar kinematic variables defined as
\begin{flalign}
X_{ij}\, =\, (p_{i}\, +\, \dots,\, +\, p_{j-1})^{2}
\end{flalign}
$\forall\, 1\, \leq\, i\, \leq\, j-1\, \leq\, n$. All the Mandelstam invariants can be expressed in terms of $X_{ij}$ via
\begin{flalign}
s_{ij}\, =\, X_{ij}\, +\, X_{i+1,j+1}\, -\, X_{i j+1}\, -\, X_{i+1 j}
\end{flalign}
We note that if we consider a planar polygon with vertices labelled as $1,\, \dots,\, n$ in clockwise fashion, then the planar scattering variables are labelled by chords $ij$ that dissect  the polygon. 

Arkani-Hamed, Bai, He and Yun (ABHY) discovered a large class of geometric realisations of the associahedron in  the positive region of planar kinematic space ${\cal K}_{n}$. It was shown in \cite{HughThomas} that the ABHY realisations are polytopal realisations corresponding to certain g-vector fans of type-A cluster algebra.  In \cite{1906ppp}, it was shown that  there is a larger class of geometric realisations of an associahedron which include ABHY realisations. In practice, it was shown that given any triangulation $T$ of an $n$-gon, the following system of linear constraints generate a class of realisations which include ABHY realisations. 
\begin{flalign}\label{eq1con}
s_{ij}\, =\, -\, c_{ij}\, \forall\, (ij)\, \notin\, T^{c}
\end{flalign}
where $T^{c}$ is the triangulation of the n-gon obtained by rotating each diagonal in $T$ counter-clockwise by $\frac{2\pi}{n}$\footnote{In the $n=5$ case, if $T\, =\, (13,35)$ then $T^{c}\, =\, (25,24)$. Here $c_{ij}$ are positive constants and as the number of $(ij)\, \in\, T^{c}$ are $n-3$, the number of constraints we have in \eqref{eq1con} are $\frac{n(n-3)}{2}\, -\, (n - 3)\, =\, \frac{(n-3)(n-2)}{2}$.}.\\
On the other hand, as was shown in \cite{Arkani-Hamed:2017mur}, there is a unique canonical form of rank $n-3$ on the kinematic space ${\cal K}_{n}$ which is projective.
\begin{flalign}
\Omega_{n}\, =\, \sum_{T}\, (-1)^{\sigma(T, T_{0})}\, \bigwedge{(ij)\, \in\, T}\, d \ln X_{ij}
\end{flalign}
where $\sigma(T, T_{0})\, =\, \pm 1$ depending on the number of mutations (modulo two) required to reach $T$ from $T^{0}$. The choice of $T^{0}$ is arbitrary and the form $\Omega_{n}$ is independent of such a reference.\footnote{The overall sign is unimportant,and the final amplitude is independent of this sign.} It was shown in \cite{Arkani-Hamed:2017mur} that this form projects to the canonical top form on the (any) realisation of the associahedron and this canonical form is precisely the planar amplitude in massless $\phi^{3}$ theory. 
\begin{flalign}
\Omega_{n}\vert_{{\cal A}_{n}^{T}}\, =\, m_{n}(p_{1},\, \dots,\, p_{n})\, \bigwedge_{(ij)\, \in\, T}\, dX_{ij}
\end{flalign}
Using the results of \cite{Banerjee:2018tun, Raman:2019utu , Aneesh:2019ddi} and especially \cite{1906ppp}, in \cite{Aneesh:2019cvt} we showed the following two results.  
\vspace*{-0.2in}
\begin{enumerate}
\item Pull-back of the unique $(n-3)$ planar scattering form $\Omega_{n}$ on ${\cal A}_{n}^{T}$ for any reference triangulation $T$ is the canonical form on the kinematic space associahedron, ${\cal A}_{n}^{T}$.
\item The restriction of the $\frac{n-4}{2}$ planar scattering forms $\Omega_{n}^{Q}$ (parametrized by $Q$) on ${\cal A}_{n}^{T}$ is given by $m_{n}^{Q}\, \bigwedge_{i_{k} j_{k} \in Q } d X_{i_{k}j_{k}}$
\end{enumerate}
\vspace*{-0.2in}
Both of these results were proved in \cite{Aneesh:2019cvt}, but involved a rather involved set of arguments from \cite{1906ppp}. In lieu of reviewing earlier results, we give simpler proofs for both of these statements. 
\begin{claim}\label{cl1}
On any geometric realisation of the associahedron ${\cal A}_{n}^{T}$ in ${\cal K}_{n}^{+}$, the pull back of the scattering form equals the canonical form on ${\cal A}_{n}$.  That is, if $T\, =\, \{i_{1}j_{1},\, \dots,\, i_{n}j_{n}\, \}$ then 
\begin{flalign}
\Omega_{n}\vert_{{\cal A}_{T}}\, =\, m_{n}(p_{1},\, \dots,\, p_{n})\, \bigwedge_{i_{m}j_{m}\in T} d X_{i_{m}j_{m}}
\end{flalign}
where $m_{n}(p_{1},\, \dots,\, p_{n})$  is the (planar) $\phi^{3}$ amplitude.\\
\end{claim}
The proof is simply a generalisation of the proof given in \cite{Arkani-Hamed:2017mur} which was for the specific case when $T\, =\, \{13,\, \dots,\, 1(n-1)\, \}$ or cyclic permutations thereof. 
\begin{claimproof}
Let $T$ be a reference triangulation labelled by the diagonals $\{\, i_{1}j_{1},\, \dots,\, i_{n}j_{n}\, \}$. As every triangulation is related to $T$ by an even or an odd number of flips, it suffices for us to show that any two triangulations, say $T^{\prime}$ and $T^{\prime\prime}$ which are related by one flip between say $(ij)$ and $(k\ell)$
with  $ i< k< j< \ell<i$, then
\begin{flalign}
dX_{ij}\, \wedge d\Omega_{T^{\prime}\setminus(ij)}\, =\, -\, d X_{kl}\, \wedge\, d\Omega_{T^{\prime\prime}\setminus (kl)}
\end{flalign}
This can be shown as follows. 
As $(ij)$ and $(k\ell)$ are intersecting diagonals of a polygon, where the cyclic order of the vertices is given by $ i< k< j< \ell<i$ we have, 
\begin{align}\label{xcoitos}
X_{ij} + X_{k \ell} -X_{i \ell} -X_{k j} &= -\sum_{p = i}^{k-1} \sum_{q= j}^{\ell -1} s_{p q}\\
X_{i j} +X_{k\ell} - X_{\ell j} - X_{i k} &=  -\sum_{p = \ell }^{i-1} \sum_{q= k}^{j -1} s_{p q}.
\end{align}
The vertices $i,k,j,\ell$ divide the vertices of the polygon in four sets, viz. $V_{i,k}=\{ i+1,i+2,\ldots, k \}$, $V_{k,j}=\{ k+1,k+2,\ldots, j \}$, $V_{j,\ell}=\{ j+1,j+2,\ldots, \ell  \}$ and $V_{\ell,i}=\{ \ell +1 ,\ell+2,\ldots, i \}$.  Any diagonal with one end point in $V_{i,k}$ and other end point in $V_{j ,\ell}$ intersects all the diagonals with one end point in $V_{k,j}$ and other end point in $V_{\ell, i}$. Therefore, any triangulation either doesn't have diagonals with one end point in $V_{i,k}$ and other end point in $V_{j ,\ell}$ or it  doesn't have diagonals with one end point in $V_{k,j}$ and other end point in $V_{\ell, i}$. Suppose, the reference triangulation $T$ doesn't have any diagonals with one end point in $V_{i,k}$ and other end point in $V_{j ,\ell}$ then using 
\begin{flalign}
s_{ij}\, =\, -\, c_{ij}\, \forall\, (ij)\, \notin\, T^{c}
\end{flalign}
we can verify that,  
\begin{equation}
X_{ij} + X_{k \ell} -X_{i \ell} -X_{k j} = \sum_{p = i}^{k-1} \sum_{q= j}^{\ell -1} c_{p q},
\end{equation}
Which implies 
\begin{equation}\label{iftoos}
d X_{ij} + d X_{k \ell} = d X_{i \ell} + d X_{k j}.
\end{equation}
On the other hand if the reference triangulation $T$ does not have any diagonal with one end point in $V_{k,j}$ and other end point in $V_{\ell, i}$, then 
\begin{equation}\label{oftmc}
X_{i j} +X_{k\ell} - X_{\ell j} - X_{i k} =  \sum_{p = \ell }^{i-1} \sum_{q= k}^{j -1} c_{p q}.
\end{equation}
Which implies
\begin{equation}\label{iftotos}
d X_{ij} + d X_{k \ell} = d X_{\ell j }  + d X_{i k }.
\end{equation}
As $T^{\prime}$ and $T^{\prime \prime}$ only differ by one flip, $(i \ell),\, (k j)\, (\ell j),\, (i k)$  are either in $T^{\prime}\, \setminus (ij)\, =\, T^{\prime\prime}\, \setminus (kl)$ or are in the edge set of the polygon. Using either eqn. \eqref{iftoos} or eqn.\eqref{iftotos}, the proof follows. 
\end{claimproof}
We now prove the second claim which equates planar tree-level amplitudes in $\phi^{4}$ theory with  a linear combination of lower forms on the associahedra ${\cal A}_{n}^{T}$.

\begin{claim}\label{cl2}
Let $Q$ be a quadrangulation of an $n$-gon (where $n$ is even) and let $T$ be any triangulation such that $Q\, \subset\, T$. Then,
\begin{flalign}
\Omega^{Q}_{n}\vert_{{\cal A}_{n}^{T}}\, =\, m_{n}^{Q}(p_{1},\, \dots,\, p_{n})\, \bigwedge_{(ij) \in Q}\, d X_{ij}
\end{flalign}
\end{claim}
\begin{claimproof}
As the Q-compatible scattering form is defined as, 
\begin{flalign}
\Omega^{Q}_{n}\, =\, \sum_{Q^{\prime}\leftarrow Q}\, (-1)^{\sigma(Q, Q^{\prime})}\, \bigwedge_{(mn) \in Q^{\prime}}\, d\ln X_{mn}
\end{flalign}
This amounts to showing that for any $T\, \supset\, Q$ and $Q$-compatible quadrangulation $Q\prime $
\begin{flalign}\label{qcomflof}
\bigwedge_{(mn) \in Q^{\prime}}\, d X_{mn}\vert_{{\cal A}_{n}^{T}}\, =\, (-1)^{\sigma(Q, Q^{\prime})}\, \bigwedge_{(ij) \in Q}\, d X_{ij}\vert_{{\cal A}_{n}^{T}}
\end{flalign}
Now as $Q^{\prime}$ can be reached from $Q$ by a finite number of (Q-compatible) mutations, eqn.\eqref{qcomflof} is equivalent to showing that for any $Q$-compatible quadrangulation $Q_{1}$ that differs from another $Q$-compatible quadrangulation $Q_{2}$ by a single mutation, $(ij)\, \rightarrow\, (k\ell)$ 
\begin{flalign}
d X_{ij}\, \bigwedge_{(mn) \in Q_{1} \setminus (ij)}\, d X_{mn}\, = - d X_{kl}\, \bigwedge_{(mn) \in Q_{2} \setminus (k\ell)}\, d X_{mn}
\end{flalign}
That this is true follows from application of  Eqn.\eqref{oftmc} and by meditating over the proof of the lemma (1.2) in \cite{Chapoton}.  That is, consider a hexagon cell $H$ in the polygon made out of diagonals in $Q\, -\, (ij)$ and the external edges such that $(ij)$ dissects $H$ into two quadrilaterals.  Let the (clockwise and cyclic) labelling of the vertices is $\{\, a_{0},\, b_{0},\, a_{1},\, b_{1},\, a_{2},\, b_{2}\, \}$ where four of these vertices are $i\, <\, k\, <\, j\, <\, \ell$ with $(ij)$ and $k\ell$ being two compatible diagonals. Without loss of generality we can assume that $i=b_{2}, k = a_{0}, j= a_{1},  \ell = b_{1}$. As Chapoton argues in \cite{Chapoton},  if $(i,j)$ and $(k,l)$ are compatible with the quadrangulation $Q$ then no diagonal of $Q$ have one vertex in $V_{k,j}=\{ k+1,k+2,\ldots, j \}$ and  other vertex in $V_{\ell,i}=\{ \ell +1 ,\ell+2,\ldots, i \}$. The second of the two equations in eqn.\eqref{xcoitos} is satisfied. That is,
\begin{flalign}
X_{ij} +X_{k\ell} - X_{\ell j} - X_{i k} &=  -\sum_{p = \ell }^{i-1} \sum_{q= k}^{j -1} s_{p q}\nonumber
\end{flalign}
where $X_{\ell j}$ and $X_{i k}$ either belong to $Q$ or are external edges of the polygon. 
However if $T$ is any triangulation that contains $Q$, then 
\begin{flalign}\label{secfsr}
s_{mn}\, =\, -\, c_{mn}\, \forall\, (m,n)\, \in\, ( [l, i-1],\, [k, j-1] )
\end{flalign}
Hence using eqns. \eqref{secfsr}, \eqref{xcoitos}, the proof follows. 
\end{claimproof}

\section{Pseudo-triangulation Model for Type-D associahedron}\label{Pseudo-Triangulation Model}
As AHST have shown, the positive geometry for one loop integrand in bi-adjoint $\phi^{3}$ theory is a geometric realisation of type-D associahedra which are related to type-D cluster algebras\cite{2003InMat.15463F}. There exists many combinatorial models for type-D associahedra including the one proposed by Fomin and Zelevinsky which was based on so-called tagged triangulations of a punctured polygon. In this paper, we use a different combinatorial model for the type-D associahedra which was proposed by Ceballos and Pilaud in \cite{ceballos-pilaud}.\footnote{We believe that all our results are independent of the choice of the combinatorial model but our definition of pseudo-accordiohedra in section \ref{fi41l} is perhaps simpler to state in this model.}

In this section we briefly review the Ceballos-Pilaud pseudo-triangulation model for type-$D$ polytopes from \cite{ceballos-pilaud}. We consider a regular convex $2n$-gon together with a disc at the center of the $2n$-gon, whose radius is small enough such that only the long diagonals (diagonals between antipodal points) of the $2n$-gon intersect the disc. For concreteness, let's consider the $2n$-gon given by the set of points $\{  i e^{-i \pi \frac{j-1}{n}}  | 1\leq j \leq 2n      \}$. We label the points  $ i e^{-i \pi \frac{j-1}{n}}  $ with $1 \leq j\leq n$ by $j$ and the points $ - i e^{-i \pi \frac{j-1}{n}}  $  with $1 \leq j\leq n$ by $ \bar j$.  At the center of this $2n$-gon we consider a small disc whose diameter is one-third of the side of the $2n$-gon (that is $\frac{2}{3} \sin(\frac{\pi}{2n})$). 
A Pseudo-triangulation of the $2n$-gon with a disc is it's dissection by chords of two type such that each cell is a pseudo triangle.\footnote{A Pseudo-triangle is a polygon with precisely three convex corners}
\begin{itemize}[topsep=0pt]
\item all the diagonals of the $2n$-gon, except the diagonals between antipodal points, we call these chords linear chords,
\item segments tangent to the disc with one end point on the boundary of the disc and one end-point among the vertices of the $2n$-gon, we call these chords central chords.
\end{itemize}
For each vertex of the $2n$-gon there are two segments tangent to the disc with one end point at that vertex. We denote by $j_{L}$ (resp. $j_{R}$) the chord emanating from the vertex $j$ and tangent on the left (resp. right) of the disc. 
\begin{figure}[h]
\centering
\begin{subfigure}{0.45\textwidth}
  \centering
\includegraphics[scale=1.65]{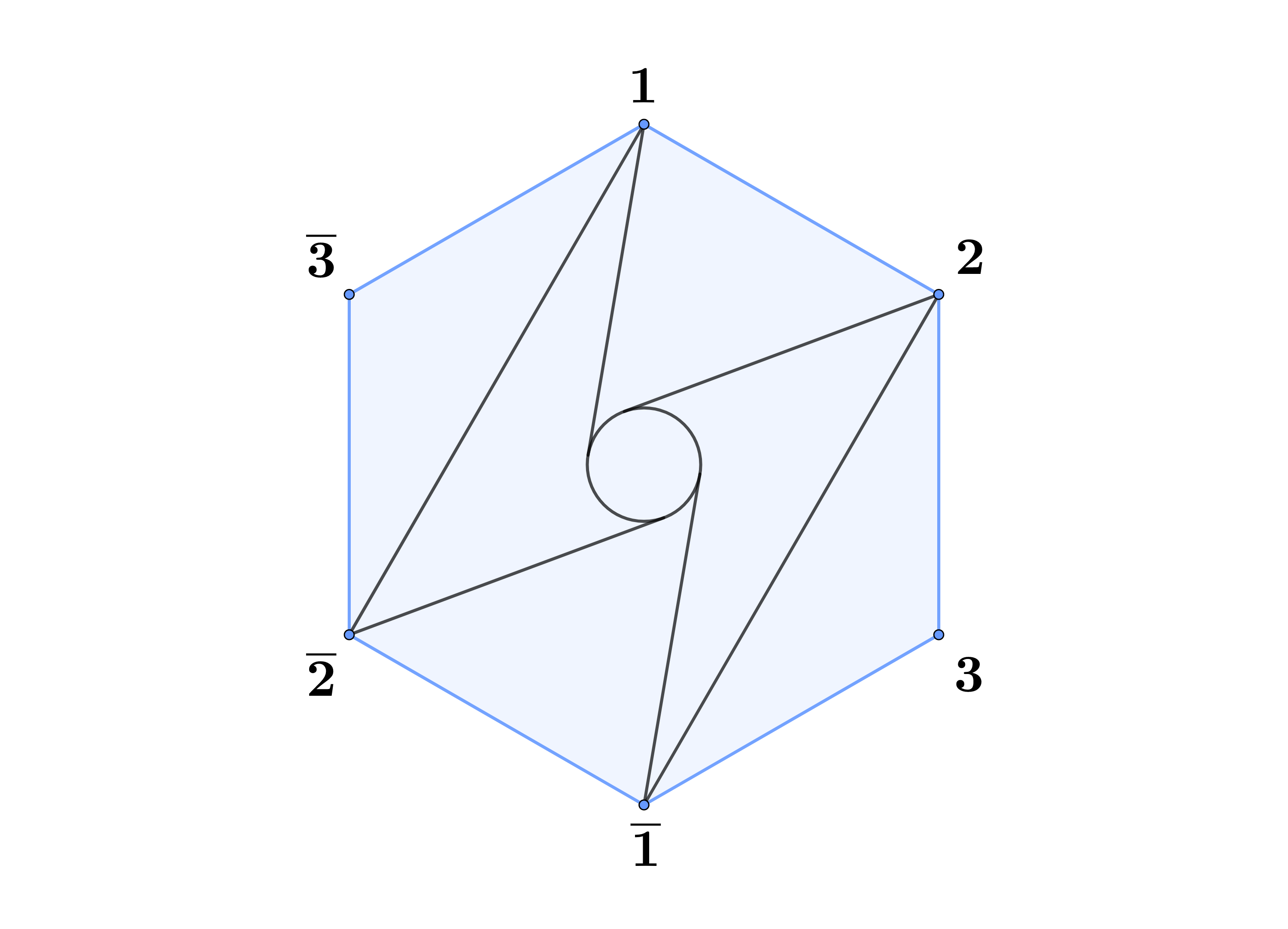}
 \caption{ $ \{1_{L}, \bar 1 _{L}\}, \{2_{L}, \bar 2 _{L}\}, \{ 1 \bar 2 , \bar 1 2 \}$}
\end{subfigure}
\begin{subfigure}{0.45\textwidth}
  \centering
\includegraphics[scale=1.45]{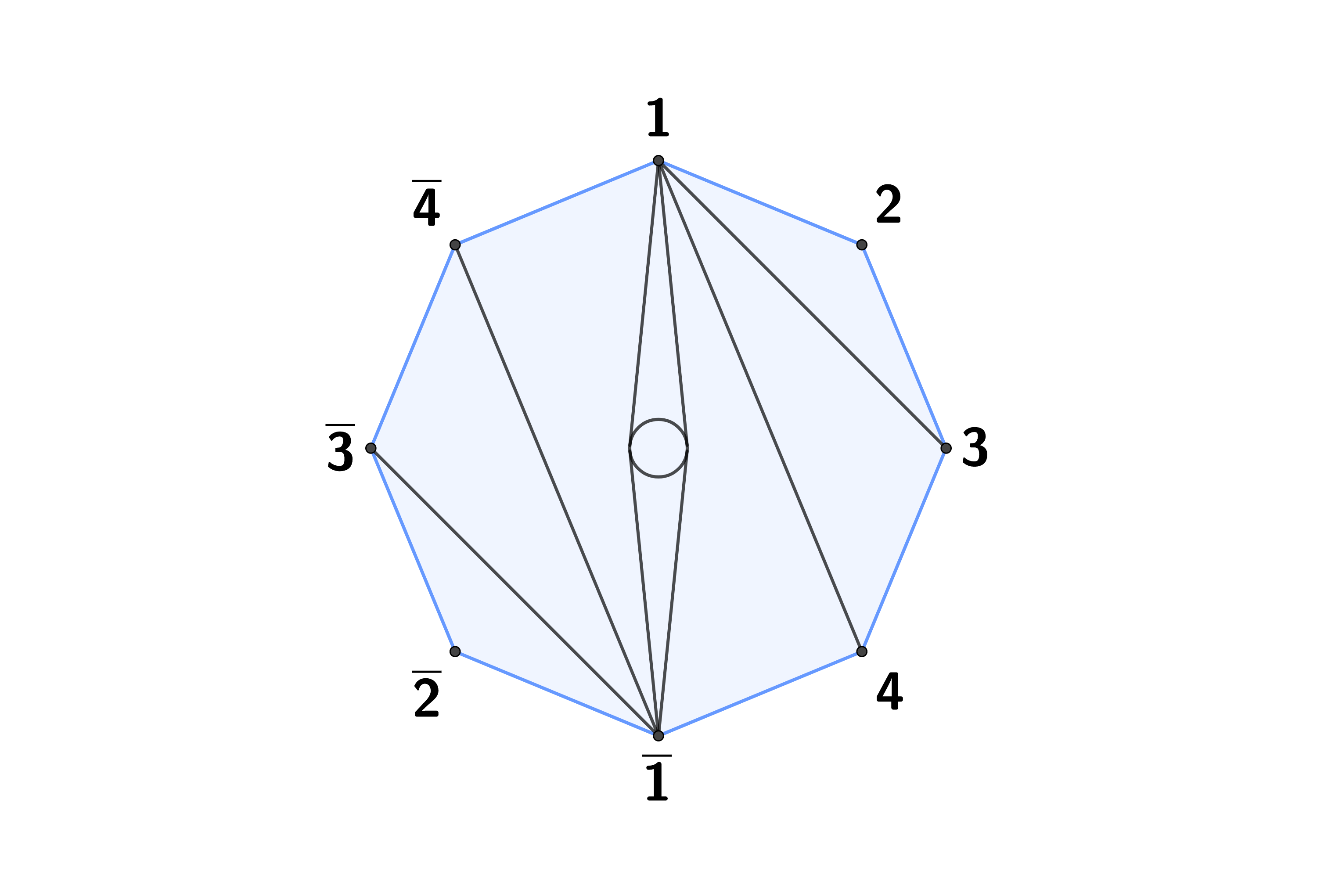}
 \caption{ $ \{1_{L}, \bar 1 _{L}\}, \{1_{R}, \bar 1 _{R}\}, \{ 1  3 , \bar 1 \bar 3 \}, \{ 1  4 , \bar 1 \bar 4 \}$}
\end{subfigure}
\caption{Examples of pseudo-triangulations.}
\label{Pseudo-triangulationexamples}
\end{figure}

A centrally symmetric pseudo-triangulation of $2n$-gon is a  centrally symmetric maximal crossing-free set of chords of the $2n$-gon. Each pseudo-triangulation contains $n$  centrally symmetric pairs of chords. Some examples of pseudo-triangulations are given in figure \ref{Pseudo-triangulationexamples}.  This forms a combinatorial model for type-D associahedra as was shown in \cite{ceballos-pilaud}. In order to describe this model, we need the notion of mutation. \cite{ceballos-pilaud}\\
Just as in the case of triangulations of a planar polygon, there exists an involution on set of pseudo-triangulations called mutation. Mutation can essentially be understood as follows. Given a pseudo-triangulation $T$, we consider a pseudo-quadrilateral formed by two cells that are adjacent to each other. The flip of the ``diagonal" of this pseudo-quadrilateral is an example of a  single mutation of $T$ in $T^{\prime}$. All the pseudo-triangulations of $T$ can be obtained  starting from any $T_{0}$  by a finite sequence of mutations.

\begin{figure}[H]
\centering
\begin{subfigure}{1.0\textwidth}
\centering
\begin{subfigure}{0.5\textwidth}
\centering
\begin{subfigure}{0.48\textwidth}
\includegraphics[scale=1.1]{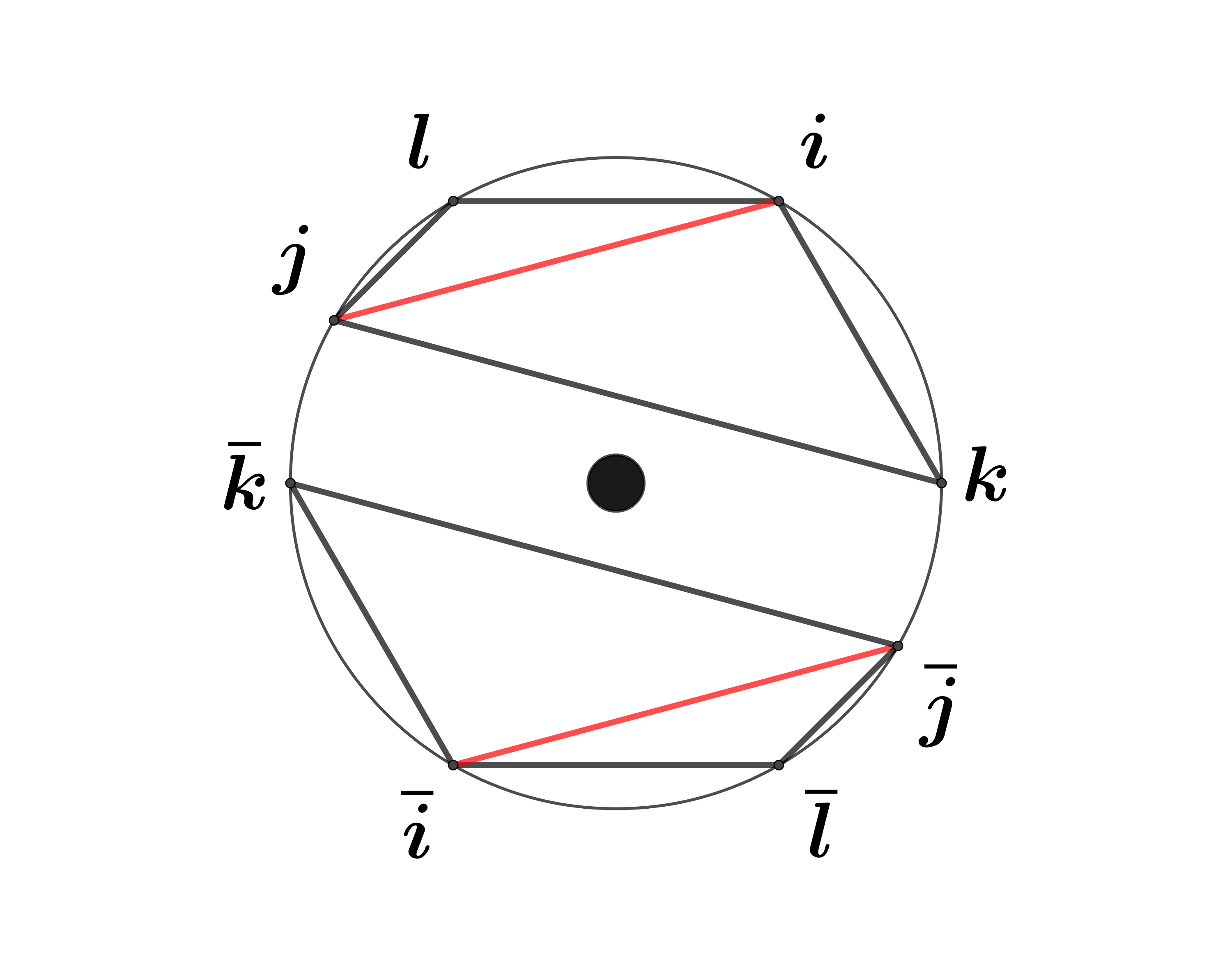}
\end{subfigure}
\begin{subfigure}{0.48\textwidth}
\includegraphics[scale=1.1]{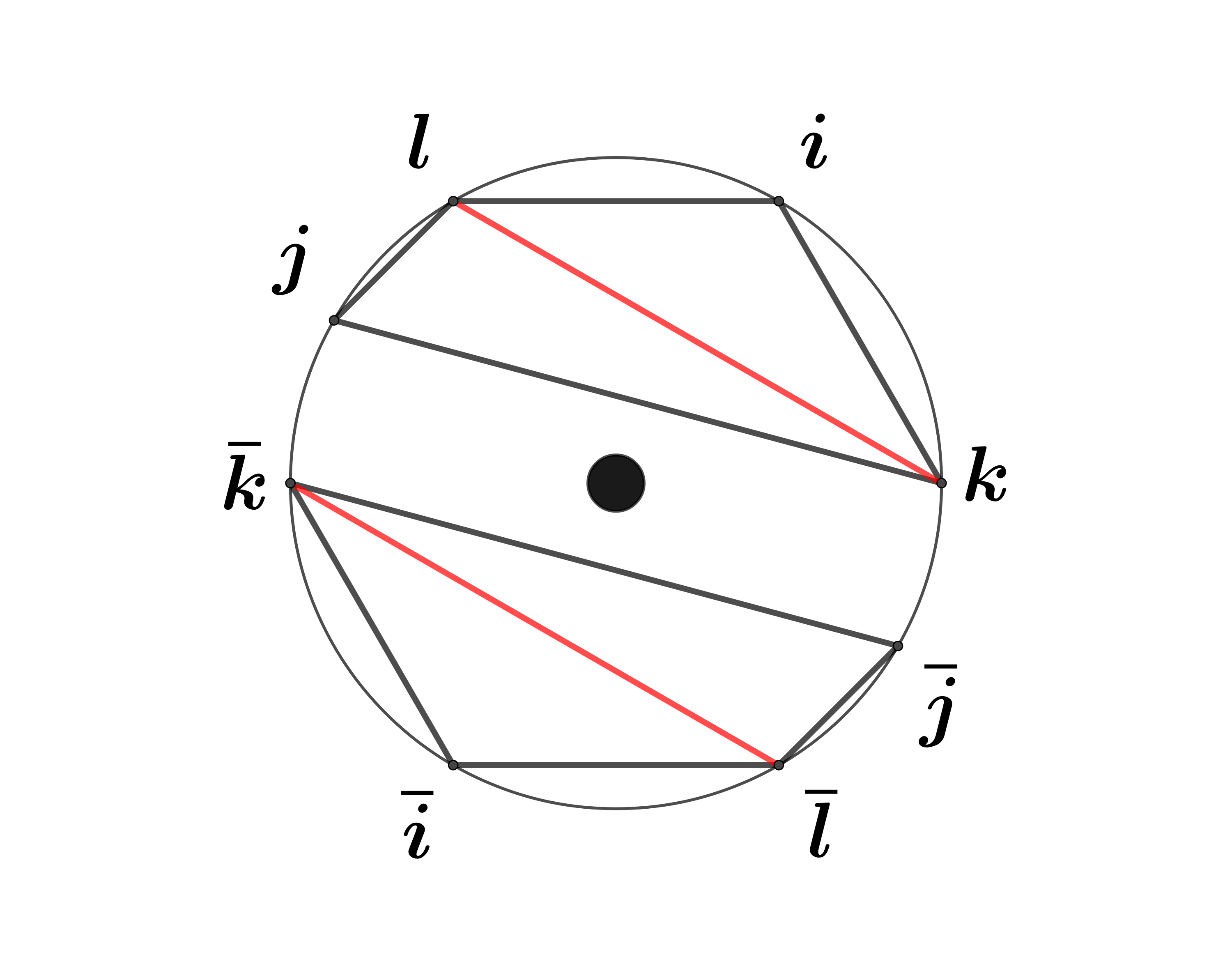}
\end{subfigure}
\caption{$\{ij,\bar i \bar j\} \leftrightarrow \{ k\ell, \bar k \bar \ell\}$}
\end{subfigure}%
\begin{subfigure}{0.5\textwidth}
\centering
\begin{subfigure}{0.48\textwidth}
\includegraphics[scale=1.1]{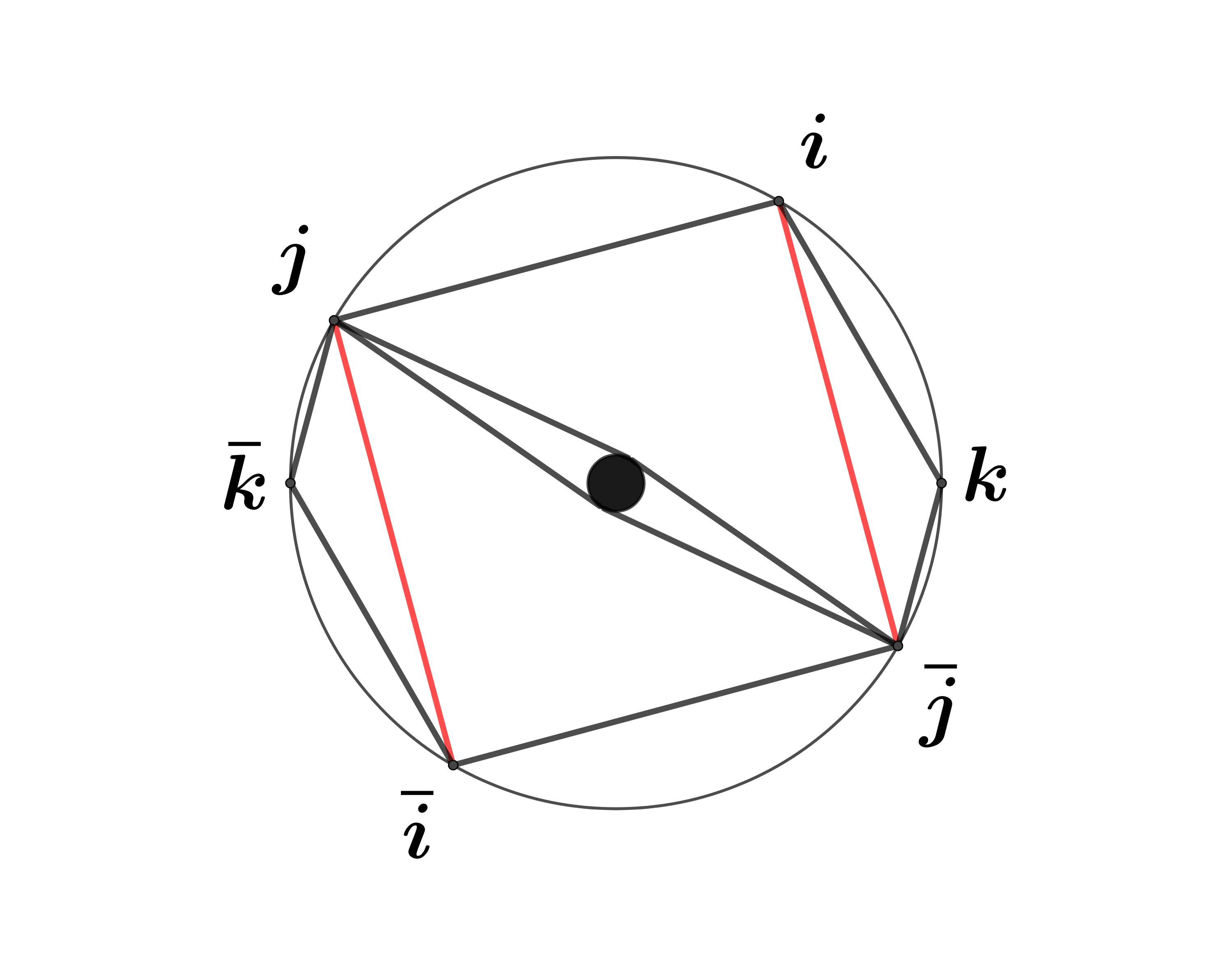}
\end{subfigure}
\begin{subfigure}{0.48\textwidth}
\includegraphics[scale=1.1]{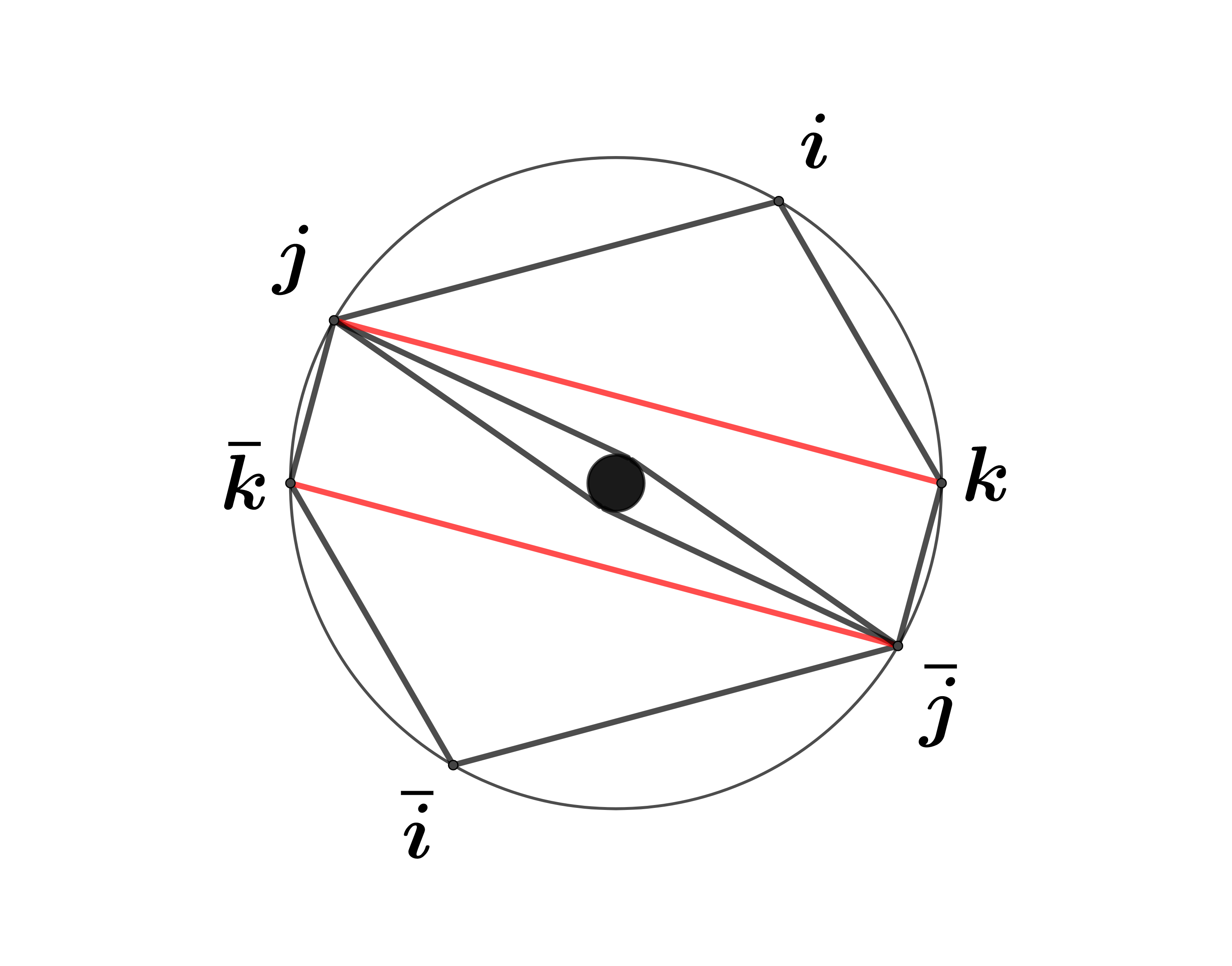}
\end{subfigure}
\caption{$\{i \bar j,\bar i  j\} \leftrightarrow \{ j k, \bar j \bar k\}$}
\end{subfigure}%
\end{subfigure}

\begin{subfigure}{1.0\textwidth}
\centering
\begin{subfigure}{0.5\textwidth}
\centering
\begin{subfigure}{0.48\textwidth}
\includegraphics[scale=1.1]{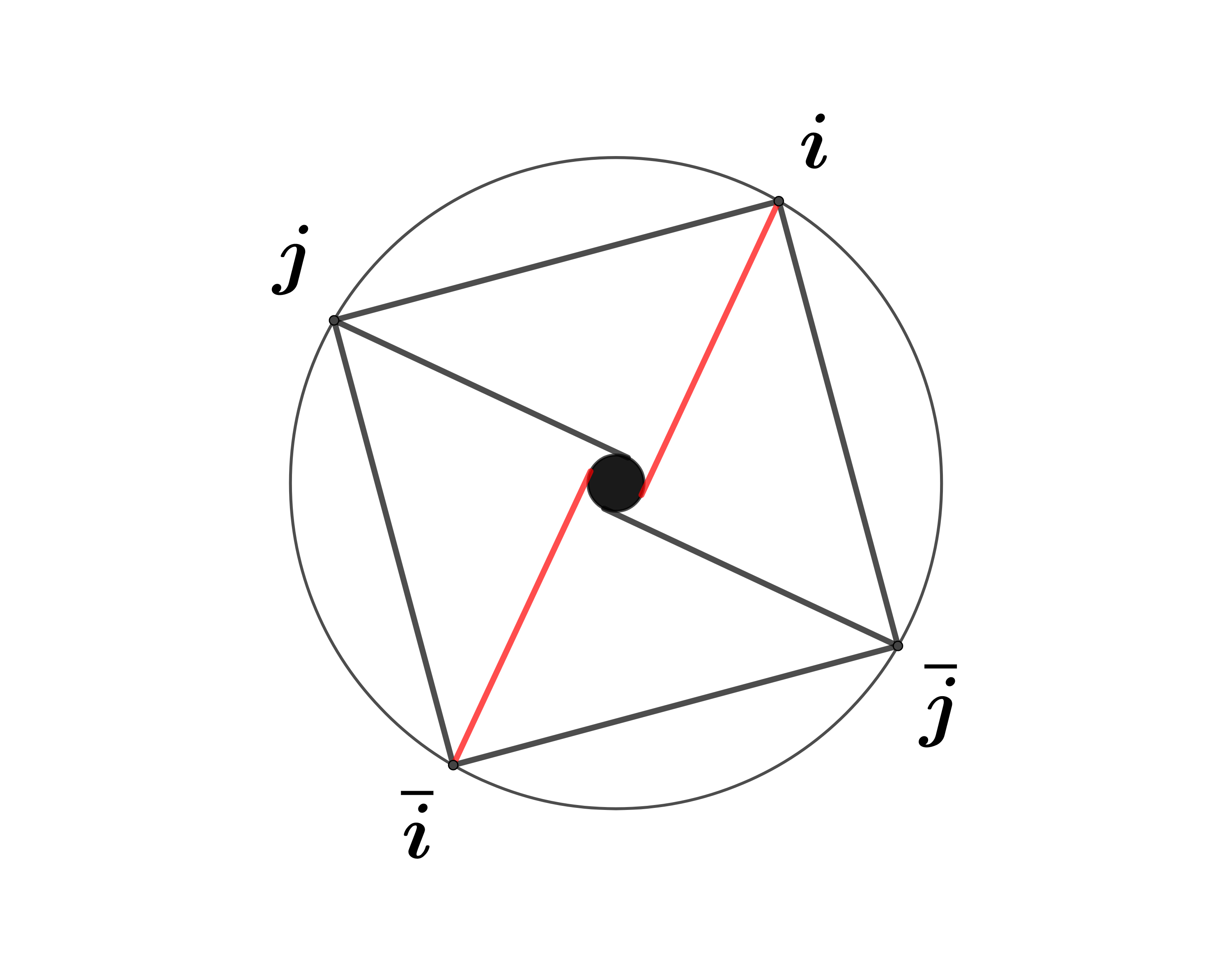}
\end{subfigure}
\begin{subfigure}{0.48\textwidth}
\includegraphics[scale=1.1]{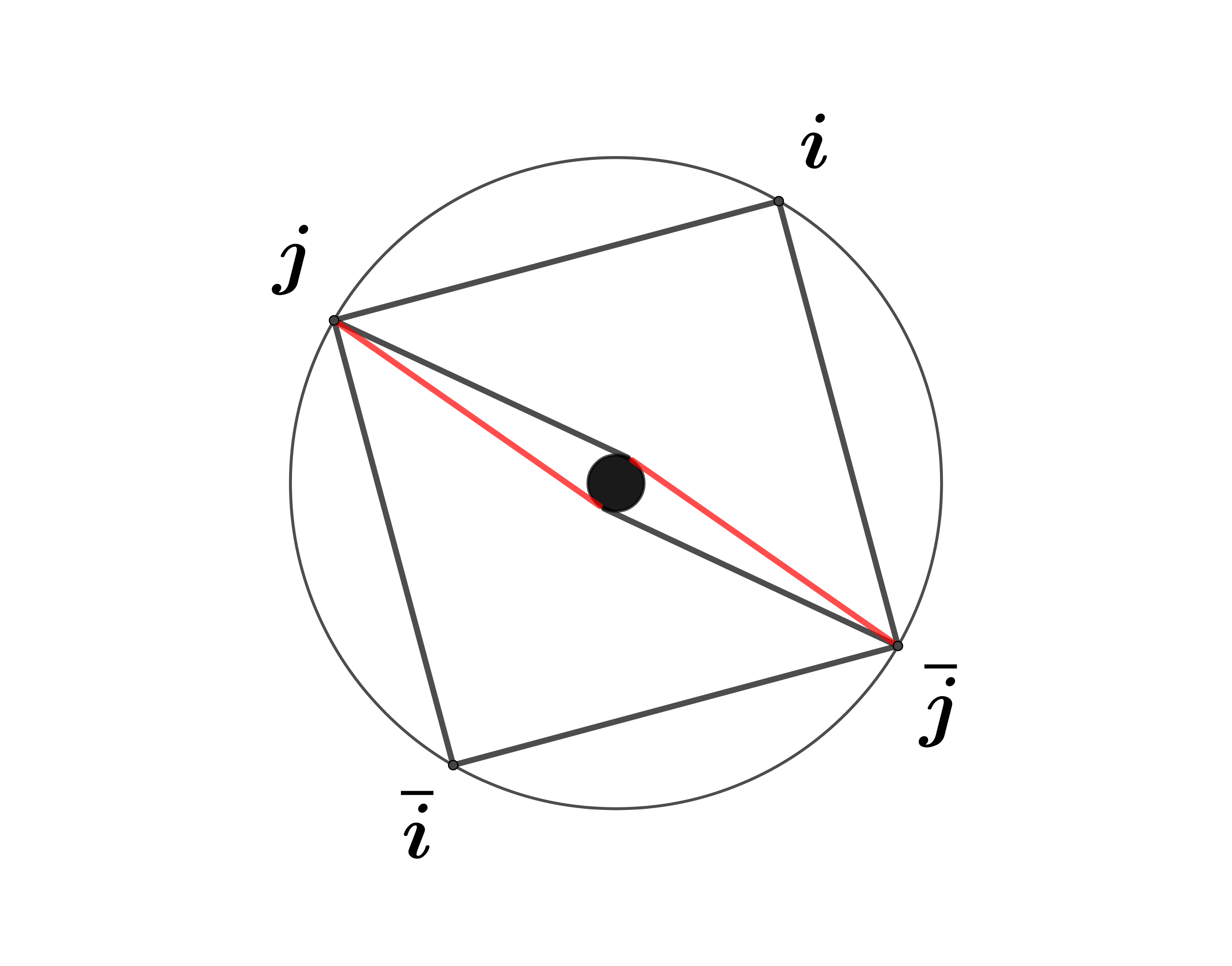}
\end{subfigure}
\caption{$\{i_{R},\bar i_{R}\} \leftrightarrow \{ i_{L},\bar i_{L}\}$}
\end{subfigure}%
\begin{subfigure}{0.5\textwidth}
\centering
\begin{subfigure}{0.48\textwidth}
\includegraphics[scale=1.1]{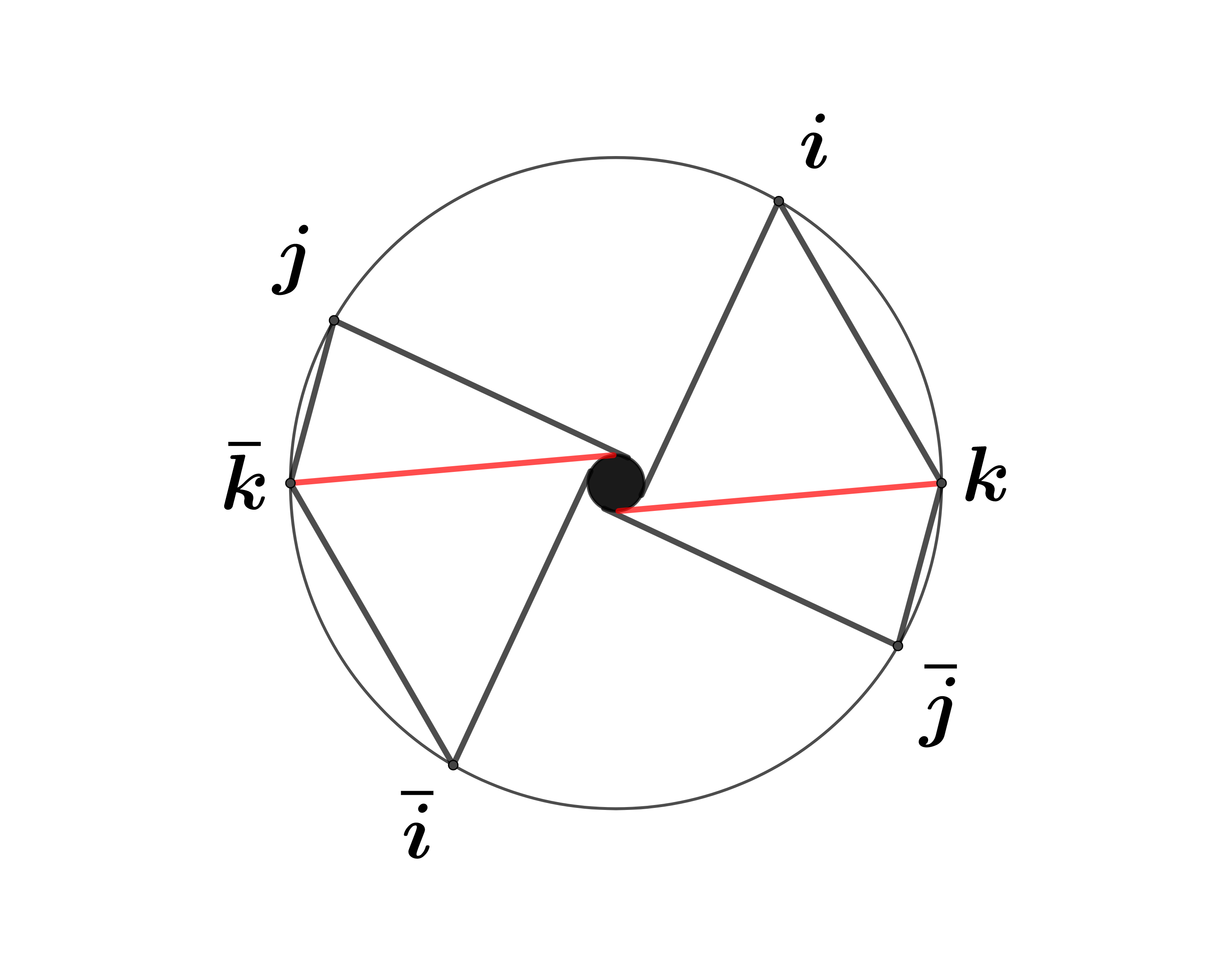}
\end{subfigure}
\begin{subfigure}{0.48\textwidth}
\includegraphics[scale=1.1]{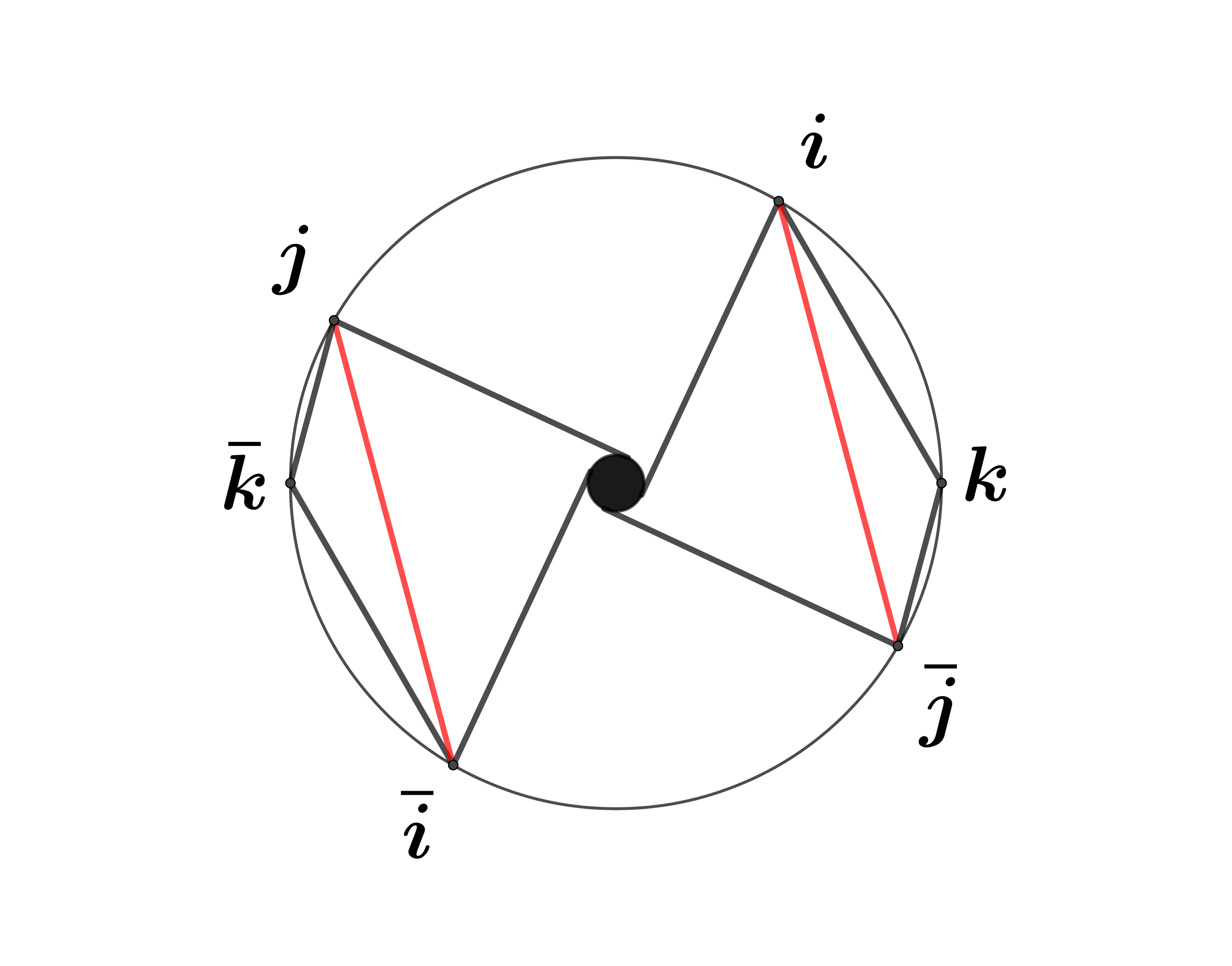}
\end{subfigure}
\caption{$\{k_{R},\bar k_{R}\} \leftrightarrow \{i \bar j,\bar i  j\}$}
\end{subfigure}%
\end{subfigure}

\caption{Different kinds of mutations.}
\label{Mutation1}
\end{figure}

We can now summarise the structure of the type-D associahedra as follows.
\vspace*{-0.1in}
\begin{itemize}
\item The set of centrally symmetric pseudo-triangulations of $2n$-gon are in bijection with the vertices of type-$D$ associahedra. 
\item The set of partial pseudo-triangulations with $n -r$ centrally symmetric pairs of chords gives the set of $r$ dimensional faces of type-$D$ associahedra. \footnote{More precisely, as was shown by Ceballos and Pilaud, Pseudo-triangulation model is a combinatorial model for the type-D cluster algebra due to the mutation which maps one pseudo-triangulation to a distinct pseudo-triangulation which corresponds to flips which map one seed in type-D cluster algebra.} 
\end{itemize}
We will also classify the pseudo-triangulations as follows. This classification is not preserved under mutation but it will serve an important role for us in section \ref{cg1}.\\
(1) Any pseudo-triangulation which contains precisely two central chords from $i$ to $\tp{i}$ which are tangential to left and right of the annulus will be called type-1 pseudotriangulations or $T_{1}$.\\
(2) A pseudo-triangulation which has no linear chord (and resembles spokes of a wheel) will be called type-2 pseudo triangulation, $T_{2}$.

\subsection{Scattering Forms in the Kinematic Space for 1-loop Integrands} 
In this section we review the construction of Kinematic space ${\cal KL}_{n}$ for 1-loop integrands and review the construction of a \emph{unique} scattering form obtained by demanding that it is odd under action of mutation.\\
For $n$-point tree-level scattering of massless particles  the kinematic space ${\cal K}_{n}$ is the $\frac{n(n-3)}{2}$ dimensional space of Mandelstam variables $\{ s_{ij} = 2 p_{i}\cdot p_{j} \}$ \cite{Arkani-Hamed:2017mur}. It is convenient to consider planar variables $X_{ij} = (\sum_{k=i}^{j-1} p_{k} )^{2}$ which form a basis of the kinematic space. 
\begin{equation}
s_{ij} = X_{i j+1} + X_{i+1 j} -X_{ij} -X_{i+1 j+1}. 
\end{equation}
Kinematic space ${\cal KL}_{n}$ for 1-loop integrands is an abstract $n^{2}$ dimensional space spanned by the variables,\\ 
(a) $X_{i,j}\, 1\, \leq\, i\, <\, j\, \leq\, n$, which are the planar variables spanning ${\cal K}_{n}$.\\ 
(b) $X_{i\overline{j}}\, =\, X_{\overline{i}j}$\\
(c) and $Y_{i}$, and $\tilde{Y}_{i}$.
\vspace*{-0.1in}
\begin{enumerate}
\item We emphasise that the $X_{i\overline{j}}$ variables do not have an a-priori physical interpretation and should be thought of as abstract variables co-ordinatising an (abstract) kinematic space. But in the final analysis when we obtain a loop integrand, these variables can be equated with $X_{i\overline{j}}\, =\, X_{\overline{i}j}\, =\, (p_{j}\, +\, \dots\, +\, p_{n}\, +\, p_{1}\, +\, \dots\, +\, p_{i-1})^{2}$. 
\item $2n$ $Y$-type variables are associated to propagators along the loop \cite{Arkani-Hamed:2019vag}, the loop propagator occurring just before external leg $i$ is denoted by $Y_{i}$ or $\tilde{Y}_{i}$. 
\end{enumerate}
These variables are associated with the pairs of centrally symmetric chords of $2n$-gon with a small disc at center from the Ceballos-Pilaud pseudo-triangulation model. 
\begin{itemize}
\item $X_{i j} = X_{j i}= X_{\bar i \bar j}= X_{\bar j \bar i}$ with $1 \leq i,j \leq n  $ is associated the pair $\{ ij, \bar i \bar j \}$.
\item $X_{i \bar j} = X_{j \bar i}= X_{\bar i  j}= X_{\bar j  i}$ with $1 \leq i,j \leq n  $ is associated the pair $\{ i \bar j, \bar i  j \}$.
\item $ Y_{i}= Y_{\bar i} $ with $1 \leq i,j \leq n  $ is associated the pair $\{ i_{L}, \bar i_{L} \}$.
\item $ \tilde{Y}_{i}= \tilde{Y}_{\bar i} $ with $1 \leq i,j \leq n  $ is associated the pair $\{ i_{R}, \bar i_{R} \}$.
\end{itemize}
Just as in the case of planar scattering form in ${\cal K}_{n}$, we can now use the mutations over pseudo-triangulations to write a $n$ form in the kinematic space ${\cal KL}_{n}$ which is projective \cite{Arkani-Hamed:2019vag}.  The \emph{unique} scattering form in the kinematic space ${\cal KL}_{n}$ is defined precisely as in the case of planar scattering form associated to tree-level amplitudes . 
\begin{flalign}\label{pflks}
\omega_{n}\, =\, \sum_{T_{n}}\, (-1)^{\sigma_{T_{n}}}\, \bigwedge_{I=1}^{n}\, d\ln X_{i_{I}j_{I}}
\end{flalign}
In the next section we will argue that this form descends to the canonical form on any geometric realisation of the type-D associahedra in (the positive region of) ${\cal KL}_{n}$ eventually leading to 1-loop integrand for planar $\phi^{3}$ amplitude.

\section{Geometric Realisations of Type-D Associahedron} 
As we reviewed in section \ref{rev}, all geometric realisations of associahedra can be understood from the gentle algebras associated to dissection quiver $Q_{T}$. The entire class of realisations are relevant for realising the scattering amplitude for planar $\phi^{4}$ interactions as lower forms on ${\cal A}_{T}$. Motivated by this idea, we propose a class of geometric realisation of type-D associahedra which are analogous to the geometric realisation of associahedra derived in \cite{1906ppp}. In order to motivate this proposal, we first \emph{define} the ``Mandelstam invariants" associated to 1-loop integrands. We will denote these variables as $s_{ij}$.
\begin{tcolorbox}[colback=black!5!white, colframe=black!75!black,arc=0mm]
\begin{align}
    -s_{ij}\, &= X_{ij} +\, X_{i+1  j+1} -\, X_{i j+1} - X_{i+1 j} &   1 \leq\, i\, <\, j\, \leq\, n\, \textrm{and}\, n-1 > j - i\, \geq\, 2,  & \endline 
    -s_{i\bar{j}} &= X_{i\bar{j}} + X_{i+1\overline{j+1}} - X_{i\overline{j+1}} - X_{i+1\overline{j}} &   1 \leq i, j \leq n-1, \vert j - i\vert \geq 2, & \endline 
    -s_{i\overline{i+1}} &= X_{i\overline{i+1}} + X_{i+1\overline{i+2}} - X_{i\overline{i+2}} - Y_{i+1} - \tilde{Y}_{i+1} &   1 \leq i \leq n, & \endline 
    - s_{i_{L}}\, &= Y_{i}\, +\, \tilde{Y}_{i+1} -\, X_{i \overline{i+1} }  &   1 \leq i \leq n, & \endline 
    - s_{i_{L}}\, &= \tilde{Y}_{i}\, +\, Y_{i+1} -\, X_{i \overline{i+1} } &   1 \leq i \leq n.&
\end{align}
\end{tcolorbox}
The third equation in the above system of equations include $i\, =\, n$, that is,
\begin{flalign}
-s_{n1}\, =\, X_{1n}\, +\, X_{\bar 1 2}\, -\, X_{2 n}\, -\, Y_{1}\, -\, \tp{Y}_{1}\, =\,  c_{n1}
\end{flalign} 
We now define the ``counter-clockwise cousin" $T^{c}$ of $T$. For this purpose it is useful to define the left and the right side of the annulus as two ``vertices" which form a closed configuration under translation by one unit : That is, $a_{L} - 1\, =\, a_{R},\, a_{R} -1\, =\, a_{L}$. Given a  centrally symmetric pseudo-triangulation $T$, $T^{c}$ is generated via following operations. 
\begin{flalign}
\begin{array}{lll}
(i, j)\, \rightarrow\, (i - 1, j-1)\, \forall\, (i,j)\, \in\, \{1,\, \dots,\, n,\, \tp{1},\, \dots,\, \tp{n}\, \}\\
(i, a_{R})\, \rightarrow\, (i-1, a_{L})\, \forall\, 1\, \leq\, i\, \leq\, n\\
(i, a_{L})\, \rightarrow\, (i-1, a_{R})\ \forall\, 1\, \leq\, i\, \leq\, n
\end{array}
\end{flalign}

We now claim the following.\\
\begin{tcolorbox}[colback=black!5!white, colframe=black!75!black,arc=0mm]
\begin{claim} \label{typedrealisation}
 Given any centrally symmetric pseudo triangulation $T$, a polytopal realisation of the type-D associahedra in the non-negative region of ${\cal KL}_{n}$ is given by a family of linear constraints, 
\begin{flalign}
\begin{array}{lll} 
s_{ij}\, =\, -\, c_{ij}\, \forall\, (ij)\, \notin\, T^{c}
\end{array}
\end{flalign}
\end{claim}
\end{tcolorbox}
In the next section we will try to derive the above constraints (which could be thought of as a ``physicist's proof"). Our derivation is based on the analysis in \cite{1906ppp} where a large class of polytopal realisations of  associahedra (which included ABHY associahedra)  was given using gentle algebra associated to the dissection quiver. 
But in this section, we elaborate on these constraints in terms of co-ordinates spanning ${\cal KL}_{n}$, and show that they include all the geometric realisations defined in \cite{Arkani-Hamed:2019vag}.
 These constraints can essentially be classified into three types.
\begin{equation}\label{treeconstraints}
s_{ij}  = -c_{ij}  \hspace{0.5cm}  \text{ for } | i-j | \neq 1 \text{ or } n , 
\end{equation}
\begin{align}\label{loopconstraints}
s_{i_{L} }\, = -c_{i_{L} }\, \textrm{and}\, s_{i_{R} }\, = -c_{i_{R} }
\end{align}
The constraints in equation \eqref{treeconstraints} are associated with the centrally symmetric pair of linear chords (equivalently tree variables) as follows
\begin{equation}
s_{ij}  = -c_{ij}  \hspace{0.5cm}  \text{ for } | i-j | \neq 1 \text{ or } n \hspace{0.5cm} \rightarrow  \{ i+1 j+1 , \overline{i+1} \overline{j+1} \} \sim X_{i+1  j+1}.
\end{equation}
 The constraints in equation \eqref{treeconstraints} are associated with the centrally symmetric pair of central chords as follows,
 \begin{align}
 s_{i_{L} }& = -c_{i_{L} }    \rightarrow \{ i+1_{R}, \bar i+1_{R} \} \sim \tilde{Y}_{i+1}, \endline
  s_{i_{R} }& = -c_{i_{R} }    \rightarrow \{ i+1_{L}, \bar i+1_{L} \} \sim Y_{i+1} .
 \end{align}
  
To get the geometric realisation of type D associahedra associated with the pseudo-triangulation $T$ we drop the constraints associated with diagonals of $T$ and consider all remaining constraints. We denote this geometric realisation by $\mathcal{D}_{T}$. For example, given reference pseudo-triangulation $\left( \{ 1_{L},\bar 1 _{L} \},  \{ 2_{L},\bar 2 _{L} \}, \{ 1 \bar 2, \bar 1 2 \}   \right)$ of a hexagon we drop the following three constraints
\begin{align}
c_{3_{R}} = -s_{3_{R}} &=  \tilde{Y}_{3} + Y_{1}- X_{1 3} \\
c_{1_{R}} = -s_{1_{R}} &= \tilde{Y}_{1} + Y_{2}- X_{1 \bar 2} \\
c_{1 3} = -s_{13} & = X_{1 3} +X_{1 \overline{2}} - Y_{1} - \tilde{Y}_{1} .
\end{align}
The geometric realisation is given by $X_{ij}, Y_{i}, \tilde{Y}_{i} \geq 0$ and the following six constraints,
\begin{align}\label{wq123}
c_{2_{R}} = - s_{2_{R}} & =  \tilde{Y}_{2} + Y_{3}- X_{2 \bar 3} \\
c_{1_{L}} = - s_{1_{L}} & =  Y_{1} + \tilde{Y}_{2}- X_{1 \bar 2} \\
c_{2_{L}} = - s_{2_{L}} & =  Y_{2} + \tilde{Y}_{3}- X_{3 \bar 3} \\
c_{3_{L}} = - s_{3_{L}} & =  Y_{3} + \tilde{Y}_{1}- X_{1  3} \\
c_{1 \bar 2} = -s_{1 \bar 2 } &= X_{1 \bar 2} +X_{2 \overline{3}} - Y_{2} - \tilde{Y}_{2}  \\
c_{2 \bar 3} = -s_{2 \bar 3 } &= X_{2 \bar 3} +X_{1 3} - Y_{3} - \tilde{Y}_{3}. 
\end{align}
Several remarks are in order.
\begin{itemize}
\item Just as in the case of associahedron,  our proposal for geometric realisation of type D polytopes includes the geometric realisation obtained from causal diamonds \cite{Arkani-Hamed:2019vag}. That is, whenever $T$ can be thought of as specifying ``initial data" on a null slice, the geometric realisation we obtain precisely matches with the one defined in \cite{Arkani-Hamed:2019vag}. 
\item In section \ref{rev}, we reviewed how the geometric realisations of associahedra obtained via $s_{ij}\, =\, -\, c_{ij}\, \forall\, T\, \notin\, T^{c}$  included the ABHY associahedra.  In the same vein, the geometric realisations defined in {\bf claim-1} for type-D associahedra are defined for \emph{any} pseudo-triangulation and we suspect that they include those realisations that come from starting with acyclic seeds in type-D cluster algebra. \cite{2003InMat.15463F, Arkani-Hamed:2019vag}
\end{itemize}
We now review the construction of 1-loop integrand for planar $\phi^{3}$ amplitude using the canonical top form on ${\cal D}_{T}$. As an illustrative example which is distinct from the ones given in \cite{Arkani-Hamed:2019vag}, we consider $T\, =\, \left( \{ 1_{L},\bar 1 _{L} \},  \{ 2_{L},\bar 2 _{L} \}, \{ 3_{L} \bar 3_{L}\}   \right)$.
 The projective form on ${\cal KL}_{3}$ as defined in eq.\eqref{pflks} can be explicitly written as,
\begin{flalign}
\begin{array}{lll}
\omega_{n=3}\, =\\[0.4em]
d\ln Y_{1}\, \wedge\, d\ln Y_{2}\, \wedge\, d\ln Y_{3} + d\ln \tp{Y}_{1}\, \wedge\, d\ln \tp{Y}_{2}\, \wedge\, d\ln \tp{Y}_{3} \\[0.4em]
+\, d\ln X_{2\bar 3}\, [\, d\ln Y_{2}\, \wedge\, d\ln \tp{Y}_{2}\, -\, d\ln Y_{3}\, \wedge\, d\ln \tp{Y}_{3}\,  -\, d\ln Y_{2}\, \wedge\, d\ln  Y_{3}\,  -\, d\ln \tp{Y}_{2}\, \wedge\, d\ln \tp{Y}_{3}\, ]\\  +\, d\ln X_{2\bar 1} [\, d\ln Y_{1}\, \wedge\, d\ln \tp{Y}_{1}\, -\, d\ln Y_{2}\, \wedge\, d\ln \tp{Y}_{2}\, -\, d\ln Y_{1}\, \wedge\, d\ln Y_{2}\,  -\, d\ln \tp{Y}_{1}\, \wedge\, d\ln \tp{Y}_{2}\, ]\\ +\, d\ln X_{13}\, [\, d\ln Y_{2}\, \wedge\, d\ln \tp{Y}_{2}\, -\, d\ln Y_{3}\, \wedge\, d\ln \tp{Y}_{3}\, -\, d\ln Y_{2}\, \wedge\, d\ln Y_{3}\,   -\, d\ln \tp{Y}_{2}\, \wedge\, d\ln \tp{Y}_{3}\,  ]
\end{array}
\end{flalign}

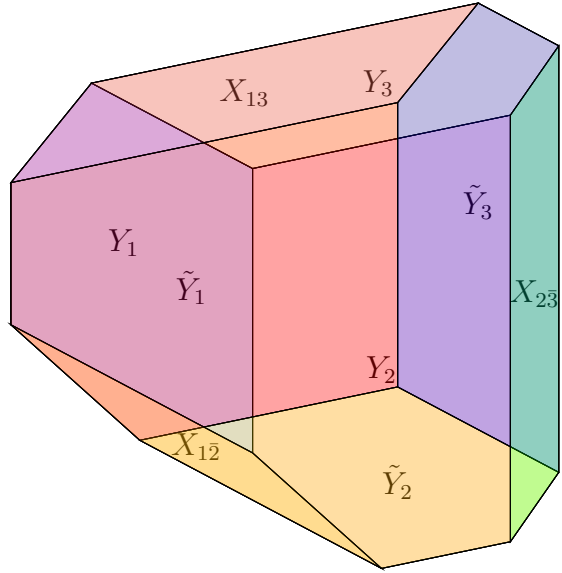
\begin{figure}[h]
\centering
\tdplotsetmaincoords{-70.5}{-32}
\begin{tikzpicture}[tdplot_main_coords,scale=2.]

\coordinate (Y1Y2Y3) at (0,0,0);
\coordinate (Y1Y2X12) at (0,0,2);
\coordinate (Y1Y3X13) at (0,2,0);
\coordinate (Y2Y3X23) at (2,0,0);
\coordinate (Y1Yt1X12) at (0,3,2);
\coordinate (Y1Yt1X13) at (0,3,1);
\coordinate (Y3Yt3X13) at (3,2,0);
\coordinate (Y3Yt3X23) at (3,1,0);
\coordinate (Y2Yt2X23) at (2,0,3);
\coordinate (Y2Yt2X12) at (1,0,3);
\coordinate (Yt1Yt2X12) at (1,3,3);
\coordinate (Yt1Yt3X13) at (3,3,1);
\coordinate (Yt2Yt3X23) at (3,1,3);
\coordinate (Yt1Yt2Yt3) at (3,3,3);

\draw (Y1Y2Y3) -- (Y1Y2X12) -- (Y1Yt1X12) --(Y1Yt1X13)--(Y1Y3X13) -- cycle;
\draw (Y1Y2Y3) -- (Y1Y2X12) -- (Y2Yt2X12) --(Y2Yt2X23)--(Y2Y3X23) -- cycle;
\draw (Y1Y2Y3) -- (Y2Y3X23) -- (Y3Yt3X23) --(Y3Yt3X13)--(Y1Y3X13) -- cycle;

\draw (Yt1Yt2Yt3) -- (Yt1Yt2X12) -- (Y1Yt1X12) --(Y1Yt1X13)--(Yt1Yt3X13) -- cycle;
\draw (Yt1Yt2Yt3) -- (Yt1Yt2X12) -- (Y2Yt2X12) --(Y2Yt2X23)--(Yt2Yt3X23) -- cycle;
\draw (Yt1Yt2Yt3) -- (Yt2Yt3X23) -- (Y3Yt3X23) --(Y3Yt3X13)--(Yt1Yt3X13) -- cycle;

\draw (Y1Y2X12) -- (Y1Yt1X12) -- (Yt1Yt2X12) --(Y2Yt2X12) -- cycle;
\draw (Y1Y3X13) -- (Y1Yt1X13) -- (Yt1Yt3X13) --(Y3Yt3X13) -- cycle;
\draw (Y2Y3X23) -- (Y2Yt2X23) -- (Yt2Yt3X23) --(Y3Yt3X23) -- cycle;

\draw (0,1.6,1) node{$ Y_{1} $};
\draw (1,0,1.6) node{$ Y_{2} $};
\draw (1.6,1,0) node{$ Y_{3} $};

\draw (1.4,3,2) node{$ \tilde{Y}_{1} $};
\draw (2,1.4,3) node{$ \tilde{Y}_{2} $};
\draw (3,2,1.4) node{$ \tilde{Y}_{3} $};

\draw (0.5,1.5,2.5) node{$ X_{1 \bar 2}$};
\draw (1.5,2.5,0.5) node{$ X_{1 3}$};
\draw (2.5,0.5,1.5) node{$ X_{2 \bar 3}$};

\draw [fill opacity=0.15,fill=blue] (Y1Y2Y3) -- (Y1Y2X12) -- (Y1Yt1X12) --(Y1Yt1X13)--(Y1Y3X13) -- cycle;
\draw [fill opacity=0.15,fill=red](Y1Y2Y3) -- (Y1Y2X12) -- (Y2Yt2X12) --(Y2Yt2X23)--(Y2Y3X23) -- cycle;
\draw [fill opacity=0.15,fill=yellow](Y1Y2Y3) -- (Y2Y3X23) -- (Y3Yt3X23) --(Y3Yt3X13)--(Y1Y3X13) -- cycle;

\draw [fill opacity=0.25,fill=red] (Yt1Yt2Yt3) -- (Yt1Yt2X12) -- (Y1Yt1X12) --(Y1Yt1X13)--(Yt1Yt3X13) -- cycle;
\draw [fill opacity=0.25,fill=yellow] (Yt1Yt2Yt3) -- (Yt1Yt2X12) -- (Y2Yt2X12) --(Y2Yt2X23)--(Yt2Yt3X23) -- cycle;
\draw[fill opacity=0.25,fill=blue] (Yt1Yt2Yt3) -- (Yt2Yt3X23) -- (Y3Yt3X23) --(Y3Yt3X13)--(Yt1Yt3X13) -- cycle;

\draw[fill opacity=0.25,fill=orange] (Y1Y2X12) -- (Y1Yt1X12) -- (Yt1Yt2X12) --(Y2Yt2X12) -- cycle;
\draw[fill opacity=0.25,fill=mycolor1](Y1Y3X13) -- (Y1Yt1X13) -- (Yt1Yt3X13) --(Y3Yt3X13) -- cycle;
\draw[fill opacity=0.25,fill=green](Y2Y3X23) -- (Y2Yt2X23) -- (Yt2Yt3X23) --(Y3Yt3X23) -- cycle;
\end{tikzpicture}
\caption{Geometric realisation of Type-D associahedra associated with the reference $ \left( \{ 1_{L},\bar 1 _{L} \},  \{ 2_{L},\bar 2 _{L} \}, \{ 3_{L} \bar 3_{L}\}   \right)$ }
\label{geometricy1y2y3}
\end{figure}

The geometric realisation associated with the reference  $T\, =\, \left( \{ 1_{L},\bar 1 _{L} \},  \{ 2_{L},\bar 2 _{L} \}, \{ 3_{L} \bar 3_{L}\}   \right)$ is given by $X_{ij}, Y_{i}, \tilde{Y}_{i}>0$ and the following set of constraints, 
\begin{align} \label{cons123}
c_{1_{L}}  & =  Y_{1} + \tilde{Y}_{2}- X_{1 \bar 2} &  c_{1 \bar 2} &= X_{1 \bar 2} +X_{2 \overline{3}} - Y_{2} - \tilde{Y}_{2}   \endline
c_{2_{L}} & =  Y_{2} + \tilde{Y}_{3}- X_{3 \bar 3}  &  c_{2 \bar 3}  &= X_{2 \bar 3} +X_{1 3} - Y_{3} - \tilde{Y}_{3}  \endline
c_{3_{L}} & =  Y_{3} + \tilde{Y}_{1}- X_{1  3}  & c_{1 3}  &= X_{1 3} +X_{1 \overline{2}} - Y_{1} - \tilde{Y}_{1}.
\end{align}
This geometric realisation is given in figure \ref{geometricy1y2y3}

We can  use eqn. \eqref{cons123} to project $\omega_{n=3}$ on ${\cal D}_{T}$ and obtain the canonical top form on the ${\cal D}_{T}$
\begin{flalign}
\begin{array}{lll}
\omega_{n=3}\vert_{{\cal D}_{T}}\, =\, m^{\textrm{1-loop}}_{3}\, d Y_{1}\, \wedge\, d Y_{2}\, \wedge\, d Y_{3}
\end{array}
\end{flalign}
where 
\begin{flalign}\label{m3d3}
\begin{array}{lll}
m^{\textrm{1-loop}}_{3}\, =\, \frac{1}{Y_{1} Y_{2} Y_{3}}\,+ \frac{1}{\tp{Y}_{1} \tp{Y}_{2} \tp{Y}_{3}}\, +\, \sum_{i=1}^{3}\, \{\, \frac{1}{X_{i,i+2}Y_{i}Y_{i+2}}\, + \frac{1}{X_{i,i+2}\tp{Y}_{i}\tp{Y}_{i+2}}\, +\, \frac{1}{X_{i,i+2}Y_{i}\tp{Y}_{i}}\, +\, \frac{1}{X_{i,i+2}Y_{i+2}\tp{Y}_{i+2}}\,  \}
\end{array}
\end{flalign}
We now state the general result.
\begin{claim}
The pull-back of the kinematic form $\omega_{n}$ on ${\cal D}_{n}^{T}$ is,
\begin{flalign}
\omega_{n}\vert_{{\cal D}_{n}}\, =\, m_{n}^{\textrm{1-loop}}\, \wedge_{(ij) \in T} d X_{ij}
\end{flalign}
\end{claim}
\begin{claimproof}
The proof is precisely analogous to the proof in section \ref{cl1} and we do not repeat it here. This is due to the fact that the geometric realisations of associahedra in ${\cal K}_{n}^{+}$ and type-${\cal D}$ polytopes in ${\cal KL}_{n}^{+}$ are obtained via $s_{ij}\, =\, -\, c_{ij}\, \forall\, (ij)\, \notin\, T^{c}$ for $T$ being a triangulation or a pseudo-triangulation respectively. Thus the proof of the above claim simply amounts to changing the range of $i$ and $j$ from vertex set of a polygon to a vertex set of a doubled polygon with a marked annulus. \footnote{Hence the only caveat to be kept in mind is that the telescopic sum $\sum_{p,q} s_{p,q}$ will involve vertex set that may include $0_{L}, 0_{R}$. However this introduces no additional complications in this proof.}
\end{claimproof}\\
As was argued in \cite{Arkani-Hamed:2019vag}, the canonical form on ${\cal D}_{3}$ has all the singularities of the 1-loop integrand but it counts  those poles which are not related to the tadpole diagrams twice. This is due to the centrally symmetric terms in eqn. \eqref{m3d3}. Hence it is not immediately clear as to how one can use this canonical form to write the loop integrand for $\phi^{3}$ amplitudes. \\
It was also beautifully shown in \cite{Arkani-Hamed:2019vag} that  if we start with the doubled quiver\\ 
(e.g. $\{\, (1_{L}, \bar{1}_{L}), (1_{R}, \bar{1}_{R}),\, 2\bar{1}\, \}$) then the corresponding geometric realisation has a symmetry under $Y_{i}\, \leftrightarrow\, \tp{Y}_{i}$. That is, one could consistently impose the constraint $Y_{i}\, -\, \tp{Y}_{i}\, =\, Y_{0}$ and dissect the polytope in two halves separated by $Y_{0}\, =\, 0$ plane. The resulting half polytope obtained by considering $Y_{0}\, \geq\, 0$ plane was called a $\overline{{\cal D}}_{n}$ polytope and captured the singularities of the 1-loop integrand faithfully.  As the eqns. \eqref{cons123} are not symmetric under $Y_{i}\, \leftrightarrow\, \tp{Y}_{i}$, an analogous ``split" of the polytope into two halves which separates singularities and their centrally symmetric avatars is not possible in this case. Hence a ``reduction" to the $\overline{{\cal D}}_{n}$ polytope for any geometric realisation \emph{apart from} the one where the reference pseudo-triangulation is associated to the degenerate quiver is not known. We leave this as an interesting open question for future investigation. Fortunately, in the case of quartic interactions, we will only require access to the ${\cal D}_{n}$ polytopes. 
\subsection{ From Pseudo-triangulation to Gentle Algebra}\label{cg1}
\begin{figure}
\centering
  \includegraphics[scale=2.5]{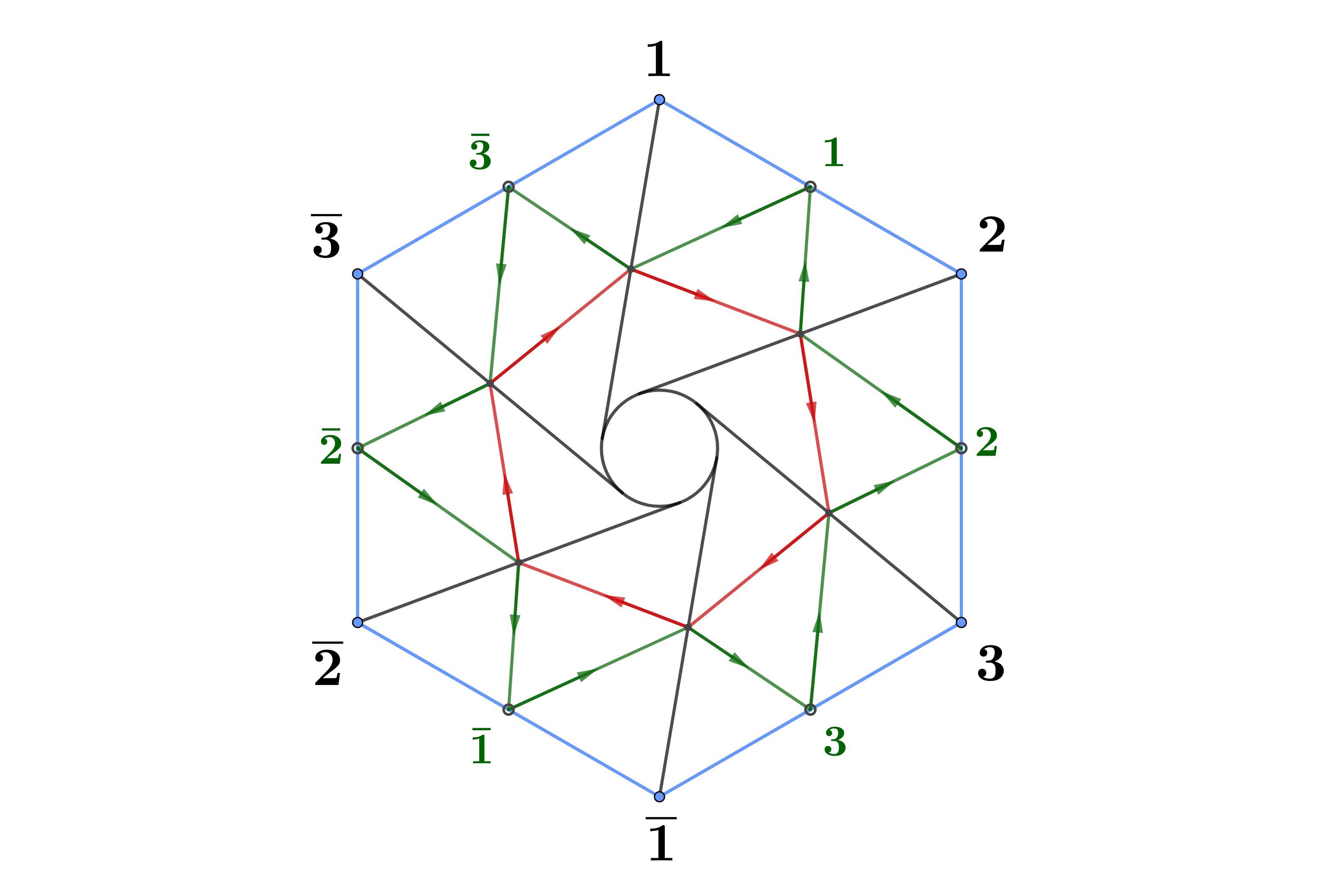}
\caption{Quiver for the pseudo-triangulation $( \{1_{L},\bar 1_{L}\}, \{2_{L},\bar 2_{L}\},\{3_{L},\bar 3_{L}\} )$.}
\label{Wheel Quiver}
\end{figure}

In this section we attempt to derive the constraints in eqn.\eqref{treeconstraints}, \eqref{loopconstraints} from ``first principles". Our ideas should be thought as a natural extension of the remarkable construction of geometric associahedra in \cite{1906ppp} (for a ``gentle" introduction to these ideas from the perspective of physicists, we refer the reader to  \cite{Aneesh:2019cvt}).\\
 That is,  starting with a  reference $T$, we consider the corresponding ``doubled  dissection quiver" $\tp{Q}(T)$ as defined in \cite{ceballos-pilaud}.  Rather then reviewing the definition of $\tp{Q}(T)$, we illustrate it with an example shown in figure \ref{Wheel Quiver}.\\
 We then consider a ``color"  gentle algebra which is a quotient of a colored path algebra modulo the two-sided ideal generated by all paths of length 2, which belong to the same cell. The path algebra is defined in the same way as in \cite{1906ppp} but with additional degree associated to each path.  More in detail, each walk which begins and ends in $i$ and  in any vertex $j$ is assigned  a degree $\{0,\ \pm 1\}$.  Hence two paths which begin and end in a vertex and it's centrally symmetric partner are not necessarily isomorphic in the resulting gentle algebra. This is why we refer the algebra as colored gentle algebra.
We will argue that just as the linear relations among elements of  gentle algebra can be used to obtain all possible geometric realisations of associahedron  \cite{1906ppp}, the colored gentle algebras leads to all possible geometric realisations of type-D associahedra.\linebreak
However unlike \cite{1906ppp}, we do not prove our assertion and merely give some evidence in their favour by means of certain examples.  We would like to emphasize that the literature on cluster algebras (and their relationship with triangulations of marked surfaces) is rather vast. It is plausible that a more precise version of our idea is already known to mathematicians. However to the best of our knowledge, there has been no extension of the relationship between gentle algebras and geometric realisations of polytopes to pseudo-dissections and ours is a ``physicist" attempt to such an extension. 
We first recall that a gentle algebra associated to a dissection quiver is the quotient of the path algebra generated by the paths of the quiver modulo the two-sided ideal generated by paths of length two which belong to the same cell.  
For pseudo-triangulation of a $2n$-gon with an annulus at the center, the path algebra is generated by all the  (oriented) paths that belong to the dissection quiver. However the difference from the path algebra corresponding to dissection quiver of a polygon is that we color the paths that begin and end in a vertex $i$ and it's centrally symmetric partner $\bar{i}$.\\
The rules for the covering are as follows. To begin with, we put hollow vertices on all the dissections in $T$. We then assign a degree to each dissection as follows.
\begin{itemize}
 \item If a hollow vertex $v$ is on a ``central chord $i_{R}$" ($1\, \leq\, i\, \leq\, \bar{n}$), then the degree of a walk $w$ at $v$ is $\pm 1$ depending on whether $w$ has a peak or dip at $v$.  
 \item  If a hollow vertex $v$ is on a ``central chord $i_{L}$", then the degree of a walk $w$ at $v$ is $\pm 1$ depending on whether $w$ has a peak or dip at $v$. 
\item If the hollow vertex is on the full arc $i_{L}\bar{i}_{L}$, then the degree is zero. 
 \end{itemize}
 Degree of a walk from $i$ to $\bar{i}$ is obtained by adding the degree of all of it's hollow vertices which are on diagonals. 
 Two walks from $i$ to $\bar{i}$ are inequivalent if they have even or odd degrees respectively. The colored gentle algebra is now obtained by quotienting the path algebra modulo the two sided ideal generated by all the paths of length two which belong to the same cell.\footnote{There is a cavaeat for $T$ with degenerate pseudo-triangles. Namely, as each degenerate pseudo-triangle has only one path in it's interior, the ideal is only generated by paths of length two in all except degenerate pseudo-triangles.} 
In order to obtain geometric realisation from a reference pseudo-triangulation, we generalise the algorithm devised in \cite{1906ppp} for planar dissections with two important distinctions. Before stating the algorithm, we introduce an important concept called reduced degree of a walk in $\tp{Q}(T)$.\\
Consider an internal path $p$ with respect to which we write a particular constraint.  
\begin{itemize}
\item We associate to any walk from $i$ to $\overline{i}$ a  reduced degree $d_{r}(i,\bar{i})$ which is the total degree of the walk minus the degree of the hollow vertices adjacent to $p$ modulo 2.
\end{itemize}
For any given path,  if an exchangeable pair is given by the walks $w(i,\, \bar{i}),\, w(j,\bar{j})$ then we conjecture the following.\\
(1), Either $d_{r}(i,\bar{i}),\, d_{r}(j, \bar{j})$ are both zero or\\
(2)  If $d_{r}(i,\bar{i})\, \pm 1$  then $d_{r}(j,\bar{j})\, =\, \mp\, 1$.\\
Although a proof of the above statements is missing, we have verified it in a number of examples. 
We now give the set of prescription for obtaining geometric realisation of type-D associahedra from the dissection quiver $\tp{Q}(T)$. 
\begin{itemize}
 \item Given an internal path $p$, if paths $X_{i\overline{j}}$ and $X_{\overline{i}j}$ form an exchangeable pair, then we  consider the walk which is degree preserving : that is, if $X_{i\bar{i}}$, $X_{j\bar{j}}$ are the non-kissing walks through $p$, and if $d_{r}(i,\bar{i})\, =\, d_{r}(j,\bar{j})\, =\, 0$ then $X_{i,\bar{j}},\, X_{\bar{i}, j}$ both form the kissing pairs. 
 \item But in the complimentary case, only one of $X_{i,\bar{j}},\, X_{\bar{i},j}$ form a kissing walk, chosen by the condition : $X_{i\bar{i}}$ has a peak/dip away from $p$ on a central chord, and   $X_{j\bar{j}}$  has a dip/peak away from $p$ on another central chord, then we choose either $X_{i\bar{j}}$ or $X_{\bar{i}j}$ or both depending on which of these walks have $d_{i},\, d_{j}$.
 \item If a walk from $i$ to $i+1$ circumscribes the annulus then we define it as $X_{i\overline{i+1}}$. 
\end{itemize}
We now define a correspondence between a walk and a kinematic variable.
\begin{itemize}
\item The kinematic variable associated to even degree walk from $i$ to $\overline{i}$ is, $Y_{i}\, +\, \tilde{Y}_{i}$ and
\item the kinematic variable associated to a positive odd degree walk is $Y_{i}$ and  $\tilde{Y}_{i}$ if the degree is negative odd. 
\end{itemize}
More in detail, we assign 
\begin{flalign}
X_{i\bar{i}}^{+1}\, =\, Y_{i}\, \textrm{and}\, X_{i\bar{i}}^{-1}\, =\, \tilde{Y}_{i}
\end{flalign}
In appendix \ref{appA}, we use the color gentle algebra and the resulting constraints in several examples. Namely, for $T_{1}$, $T_{2}$ family of pseudo-triangulations,  the colored gentle algebras generate geometric realisations consistent with our proposal. 

\section{Pseudo-accordiohedron : A Polytope from Pseudo-quadrangulations.}\label{fi41l}
\begin{figure}[H]
    \centering
    \includegraphics[scale=0.23]{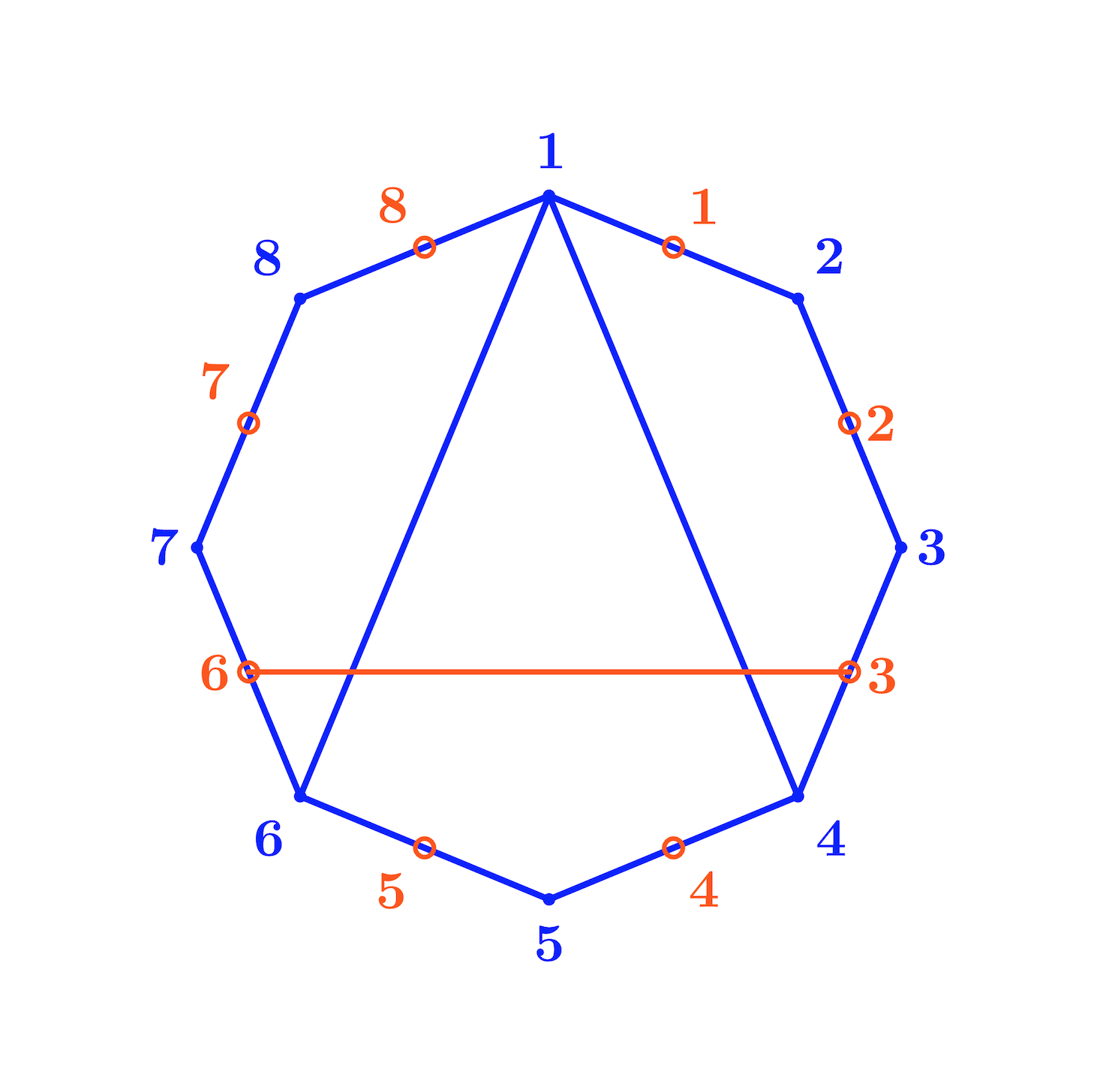}
    \caption{Quadrangulation $(14,16)$ of the octogon and an accordion diagonal $3^{\circ}6^{\circ}$}
    \label{8ptquadrangulation}
\end{figure}

In this section we introduce a new (class of) combinatorial polytopes which arise from considering pseudo-dissections of the $2n$-gon with an annulus where each cell is a pseudo p-gulation with $p\, >\, 3$. For concreteness we consider pseudo-quadrangulations but the definition is applicable even when $p\, >\, 4$. These polytopes are a natural generalisation of the accordiohedra which are defined using arbitrary dissections of a planar polygon. We will call them (for a lack of a better name) pseudo-accordiohedra. We do not validate the existence of pseudo-accordiohedron  by scrutinising the definition at the level of mathematical rigour, as has been done for accordiohedron \cite{accoref}, but in section \ref{crpa} ,we discover a geometric realisation of these polytopes in the interior of type-D associahedra, thus validating the combinatorial definition indirectly. A detailed scrutiny of these polytopes is desirable, but beyond the scope of this paper.\\
We first start by reviewing the definition of accordiohedron \cite{accoref}.\\
Let $Q$ be a quadrangulation of an $n$-gon such that the union of all the external edges and the elements of $Q$ is $\overline{Q}$. Let  ${\cal Q}^{\circ}$ be a dual quadrangulation of the dual $n$-gon. We color the vertices of the initial $n$-gon as solid vertices, $\{i^{\bullet}\, \vert\, i\, =\, 1,\, \dots,\, n \}$ and vertices of the dual polygon as hollow vertices $\{i^{\circ}\, \vert\, i\, =\, 1,\, \dots,\, n \}$. (See figure \ref{8ptquadrangulation}) We now consider any diagonal, say $i^{\circ} j^{\circ}$ of the dual polygon whose intersection with $\overline{Q}$ satisfies two conditions.
\vspace*{-0.1in}
\begin{enumerate}\itemsep0em 
\item  $i^{\circ} j^{\circ}$ does not enter and exit any cell of $\overline{Q}$ from non-incident sides.\footnote{Two sides of a cell are called non-incident if they do not meet in a common solid vertex.} 
\item The intersection set of $i^{\circ} j^{\circ}$ with elements of $\overline{Q}$ is a connected set. 
\end{enumerate}
\vspace*{-0.1in}

Such a diagonal is called an accordion diagonal. We now consider set of all possible accordion diagonals and separate sub-sets  of mutually non-intersecting accordion diagonals, such that each of these sub-sets is a  either partial or complete quadrangulation of a dual polygon.\\
A set of solid diagonals is compatible with ${\cal Q}$ if each diagonal in the set is an accordion. We call this a compatibilty condition between diagonals of the solid polygon and diagonals of the hollow polygon. 
An accordiohedron polytope is a combinatorial polytope whose vertices are in bijection with the subsets corresponding to complete quadrangulations and whose co-dimension $k$ facets is in 1-1 correspondence with subsets corresponding to $k$ partial quadrangulations.\\
The compatibility condition which defines an accordion in the case of a planar polygon can be used to define a ``pseudo-accordion" even in the case of the $2n$-gon with an annulus.
\pagebreak
\subsubsection*{Pseudo-Accordion and Pseudo-Accordiohedron}
\begin{figure}[H]
    \centering
    \includegraphics[scale=0.23]{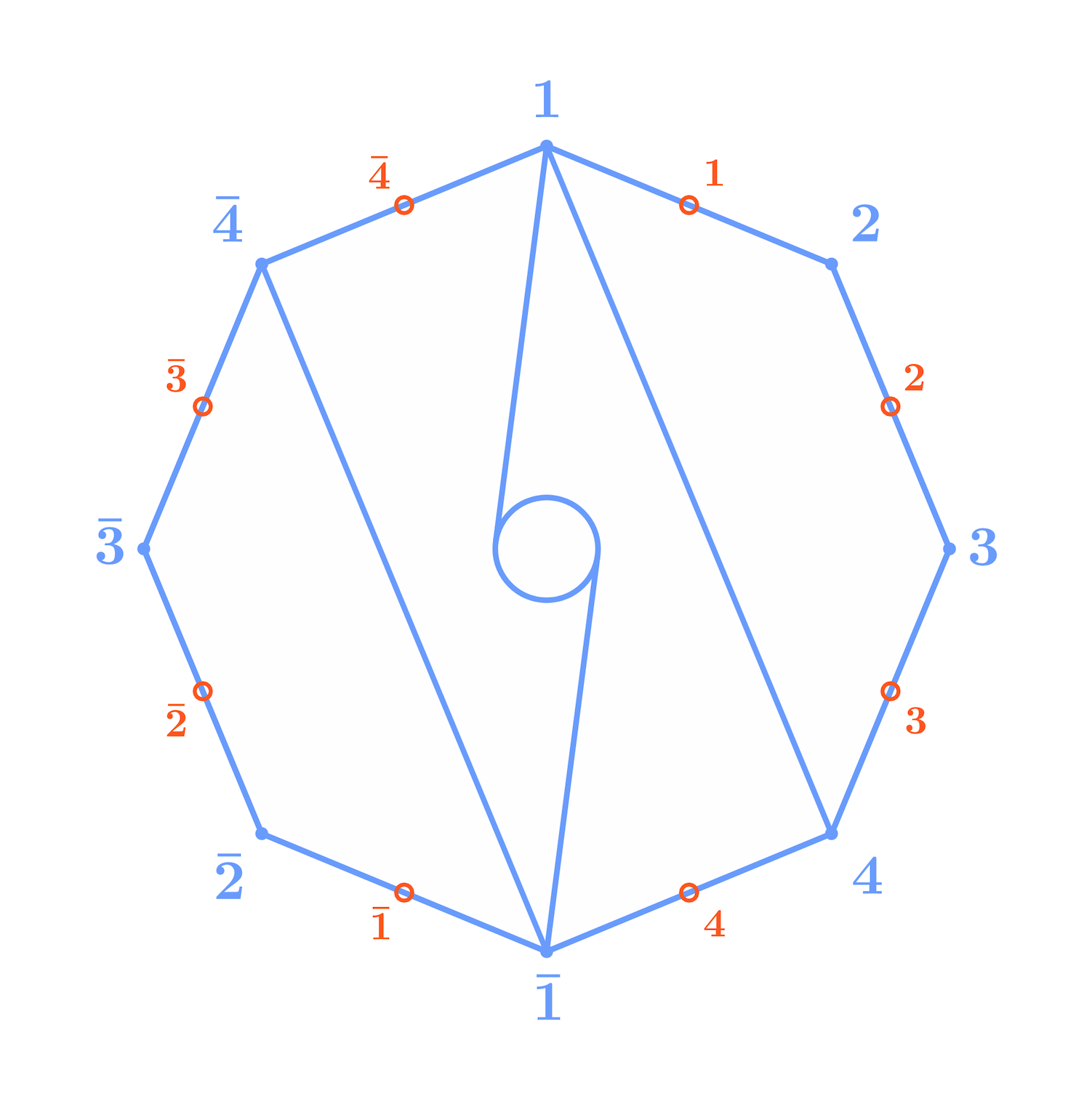}
    \caption{Pseudo-quadrangualtion ${\cal Q}\ =\, \{\, 14,\, 1_{L},\, \tp{1}_{L},\, \bar{1}  \bar{4} \}$}
    \label{pseudoquadrangulationy1x14}
\end{figure}

{\bf First definition} :\\
We consider a ``solid" $2n$-gon with vertices $\{ 1^{\bullet},\, \dots,\, n^{\bullet},\, \bar{1}^{\bullet},\, \dots,\, \bar{n}^{\bullet}\, \}$ and the dual ``hollow" 2n-gon with vertices $\{ 1^{\circ},\, \dots,\, n^{\circ},\, \bar{1}^{\circ},\, \dots,\, \bar{n}^{\circ}\, \}$ (see figure \ref{pseudoquadrangulationy1x14} for details).
Let ${\cal Q}$ be a reference pseudo-quadrangulation of the solid $2n$-gon.  ${\cal Q}$ divides the $2n$-gon into pseudo-quadrilaterals. A pseudo-quadrilateral is defined as follows. 
\begin{enumerate}\itemsep0em 
\item Any polygon with at least 4 convex corners is a pseudo-quadrilateral. Hence any regular quadrilateral is a pseudo-quadrilateral. Similarly, in the $2n$-gon with annulus, e.g. $\{1_{L},1 2, 2 3, 3_{L}\}$ is a pseudo-quadrilateral. 
 \item A polygon with vertex configuration $\{ (v_{1})_{L}, v_{1} v_{2}, v_{2} \bar{v}_{1},  (\bar{v}_{1})_{L}\, \}$  is also defined to be a pseudo-quadrilateral. 
\end{enumerate} 
\vspace*{-0.2in}

\begin{figure}
\centering

\begin{subfigure}{0.24\textwidth}
\centering
\includegraphics[scale=0.15]{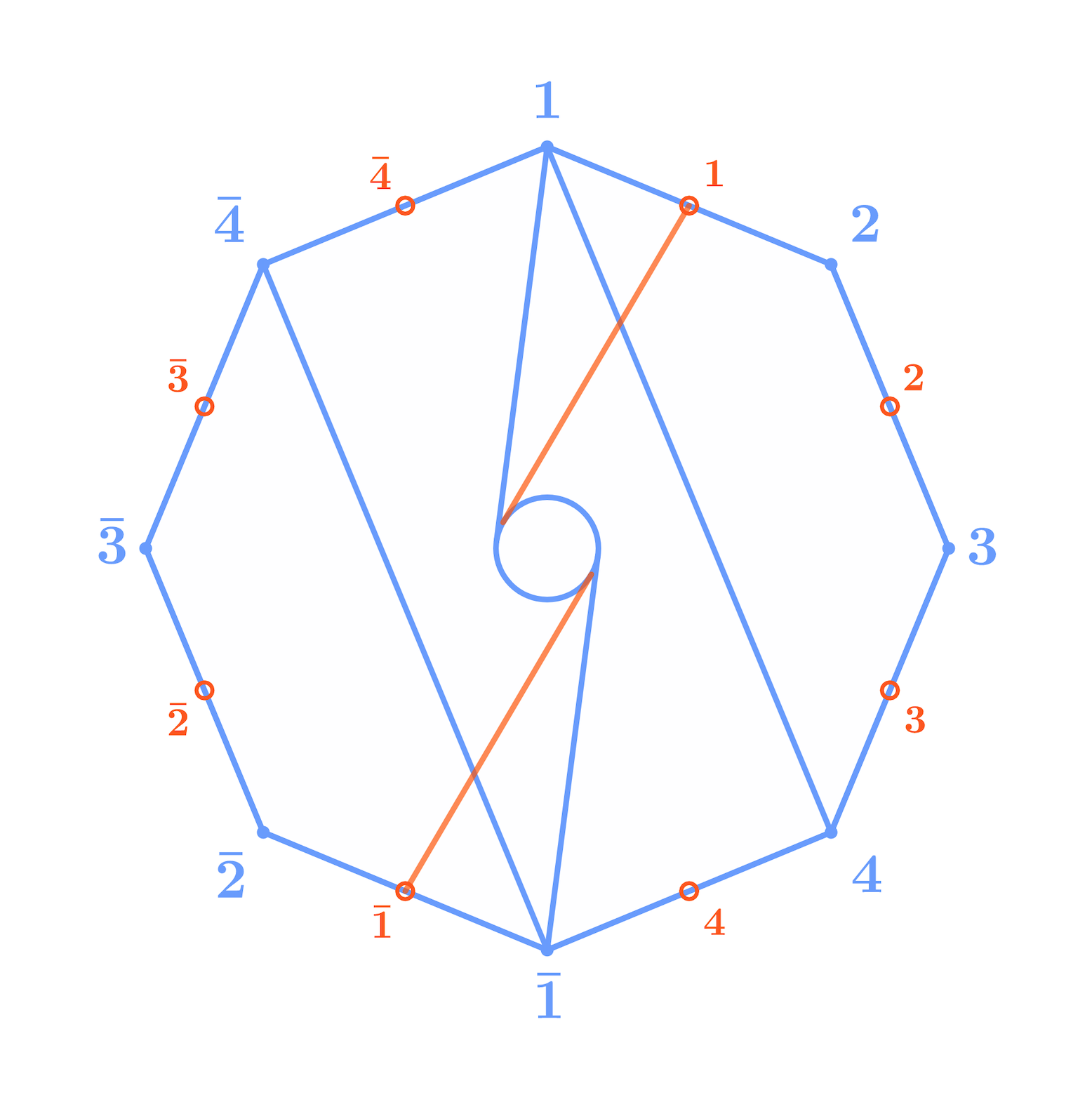}
\end{subfigure}
\begin{subfigure}{0.24\textwidth}
\centering
\includegraphics[scale=0.15]{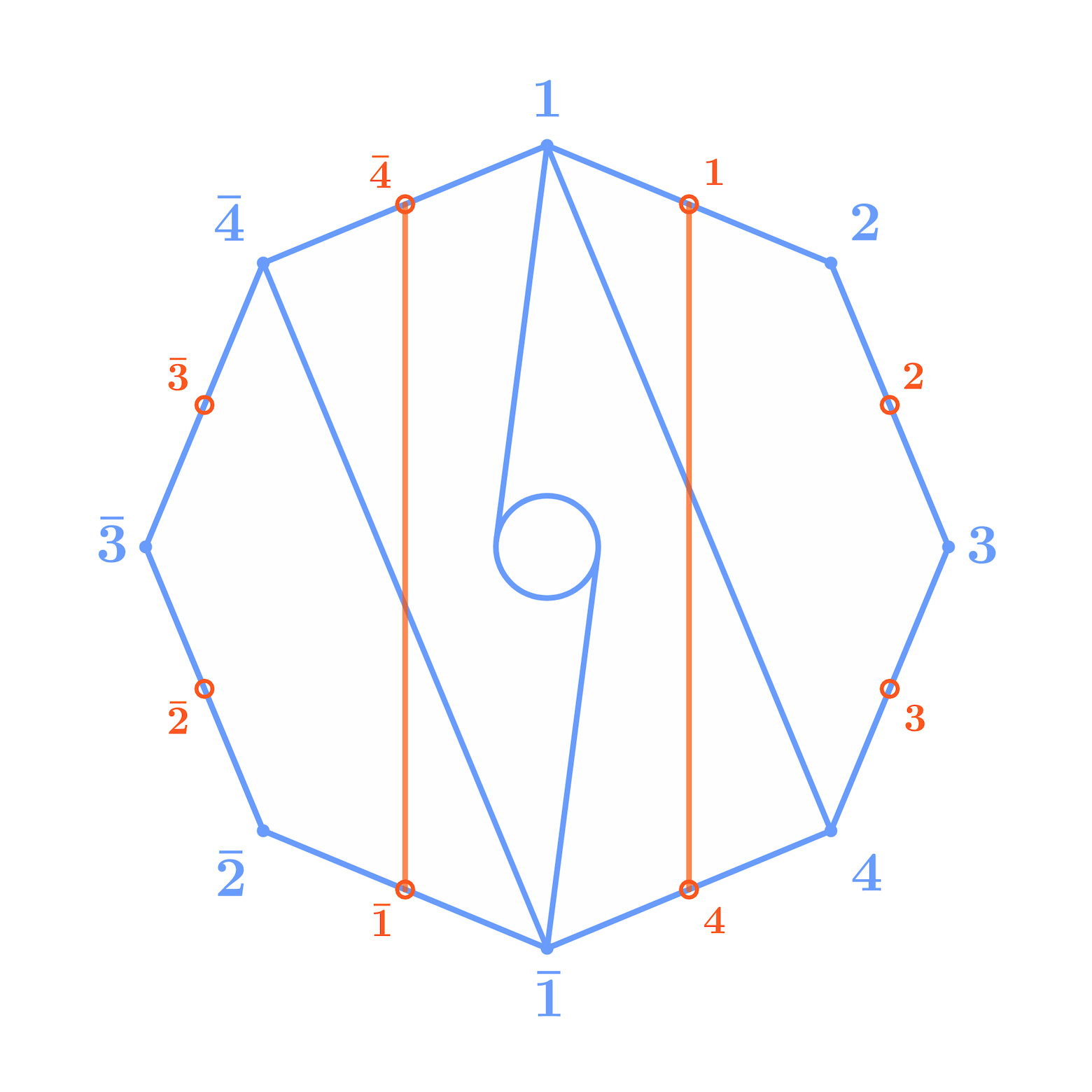}
\end{subfigure}
\begin{subfigure}{0.24\textwidth}
\centering
\includegraphics[scale=0.15]{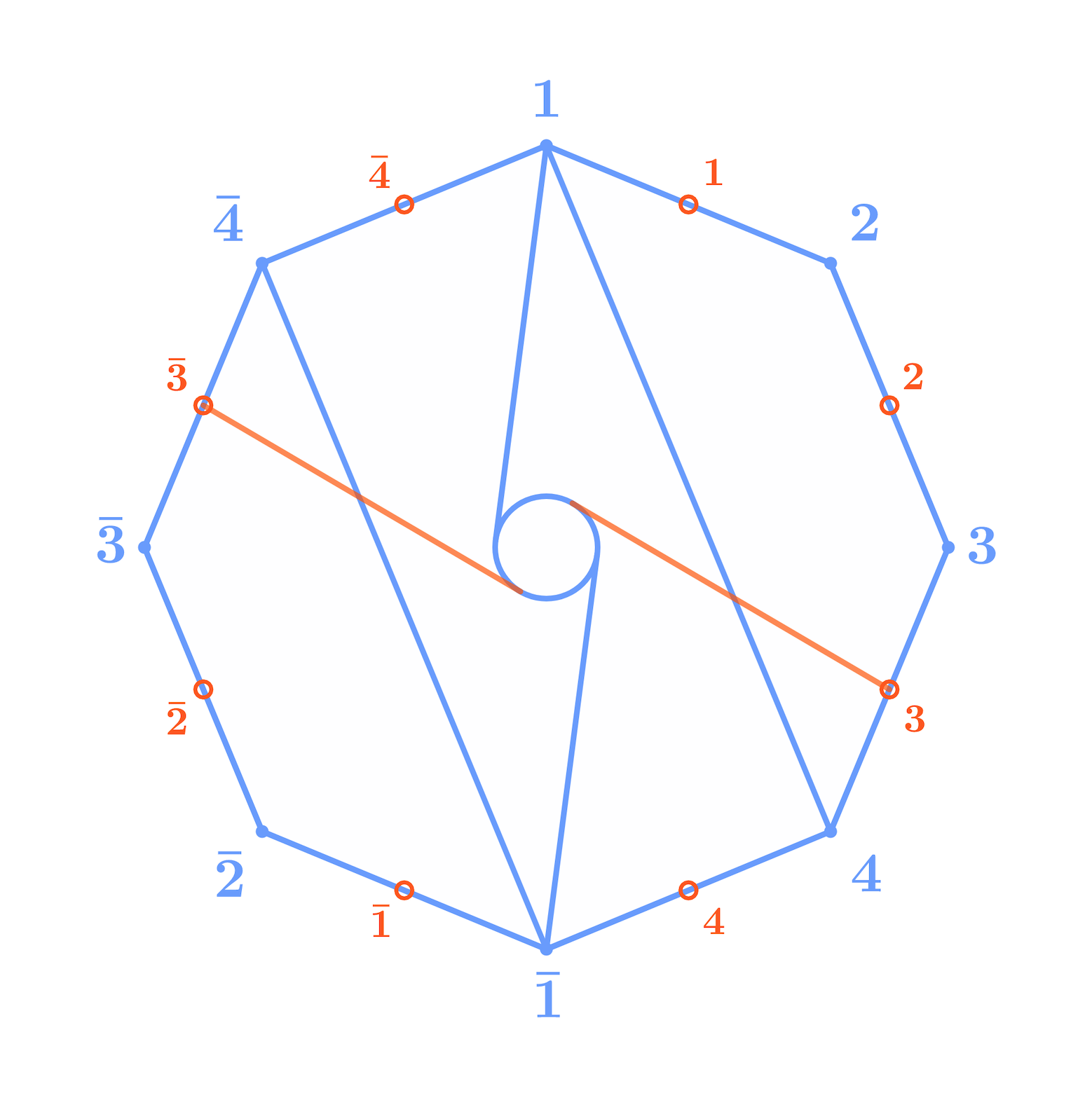}
\end{subfigure}
\begin{subfigure}{0.24\textwidth}
\centering
\includegraphics[scale=0.15]{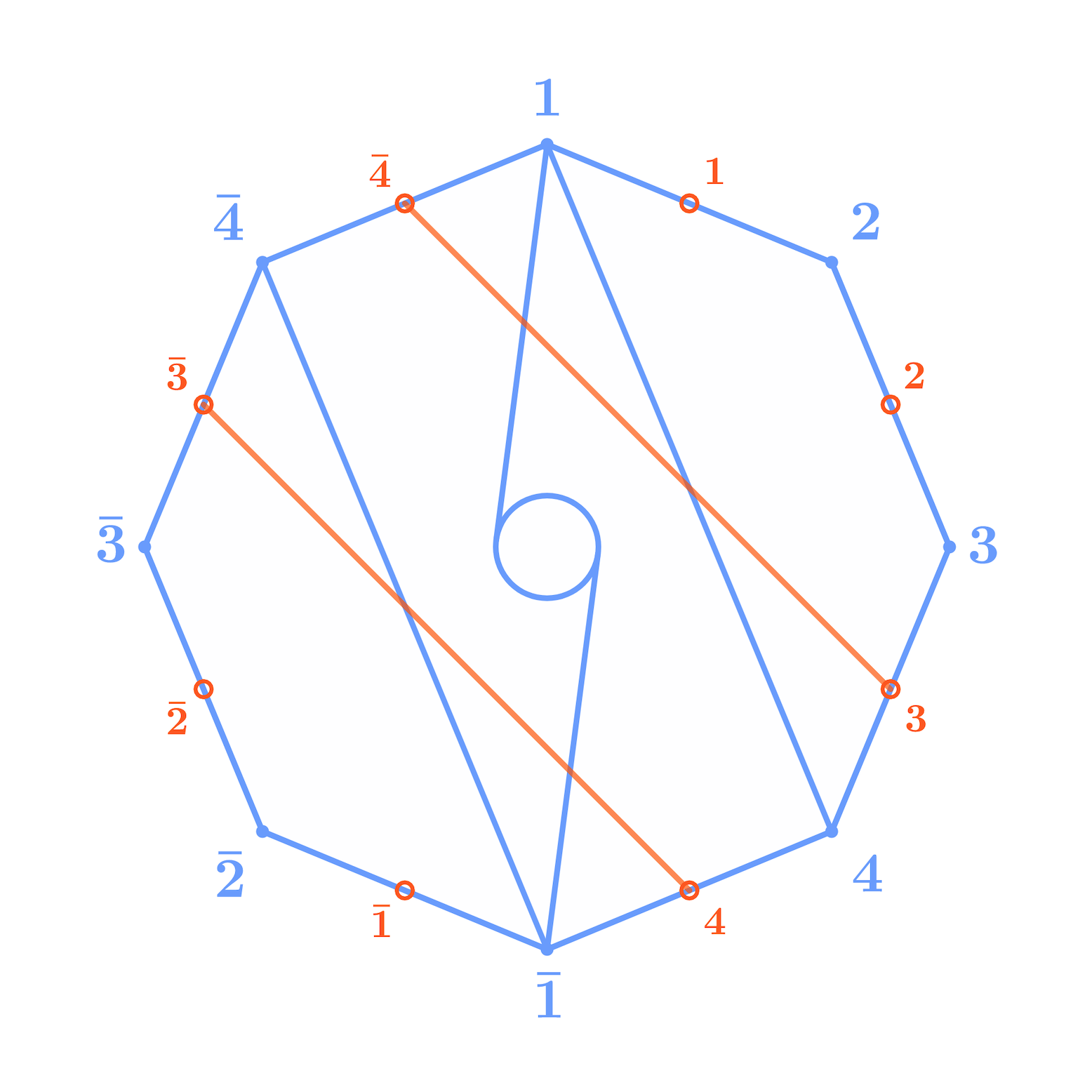}
\end{subfigure}

\begin{subfigure}{0.24\textwidth}
\centering
\includegraphics[scale=0.15]{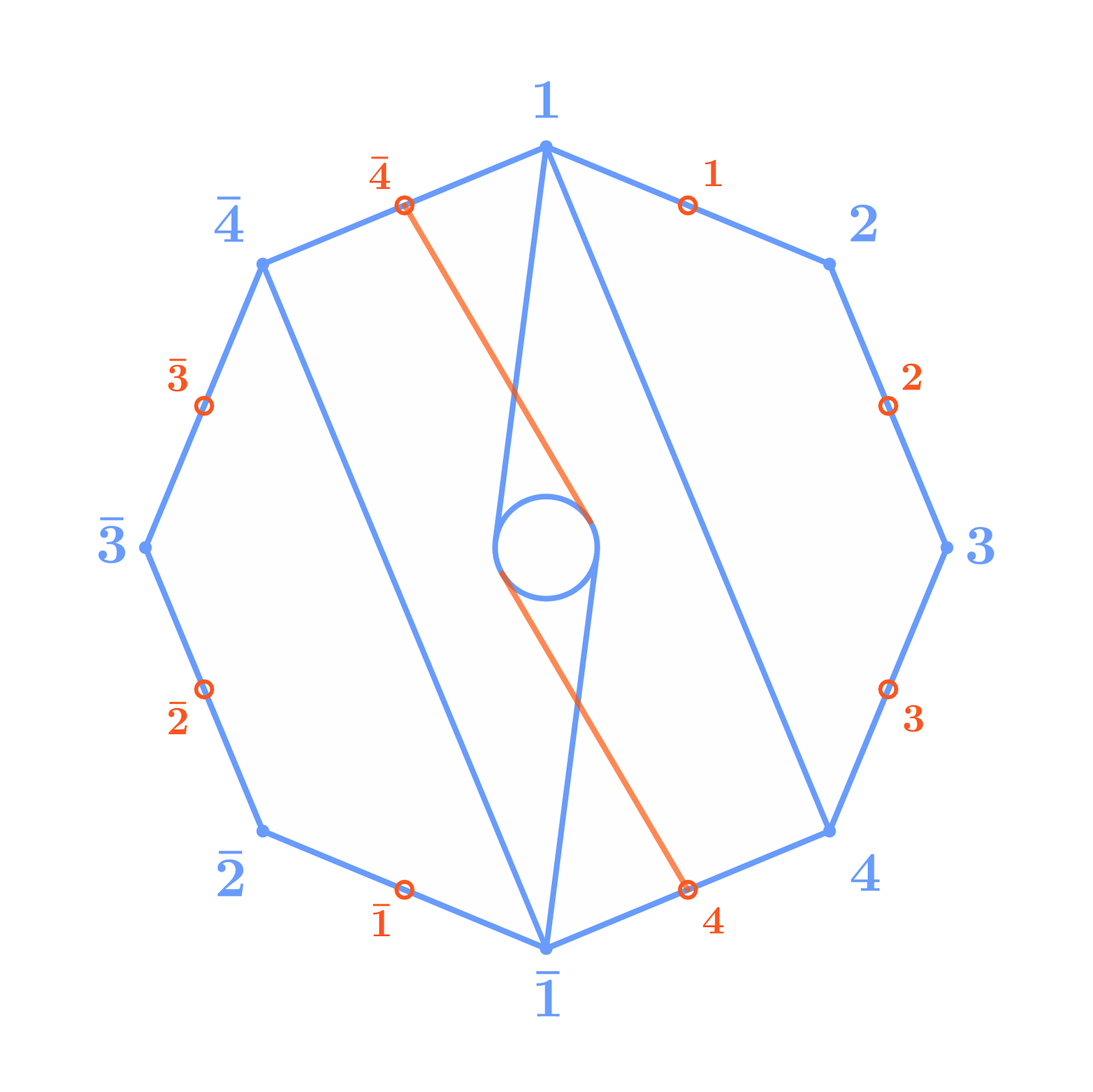}
\end{subfigure}
\begin{subfigure}{0.24\textwidth}
\centering
\includegraphics[scale=0.15]{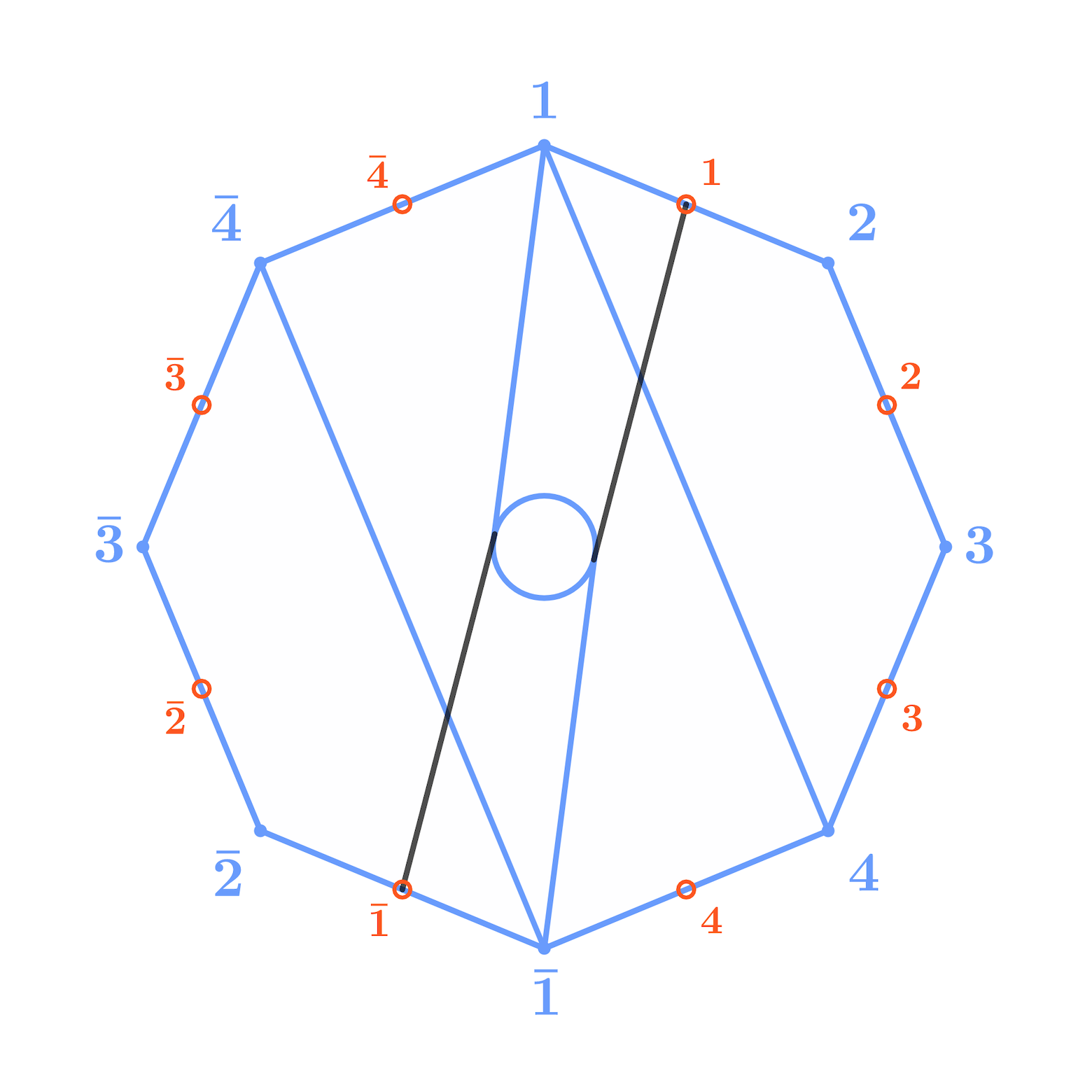}
\end{subfigure}
\begin{subfigure}{0.24\textwidth}
\centering
\includegraphics[scale=0.15]{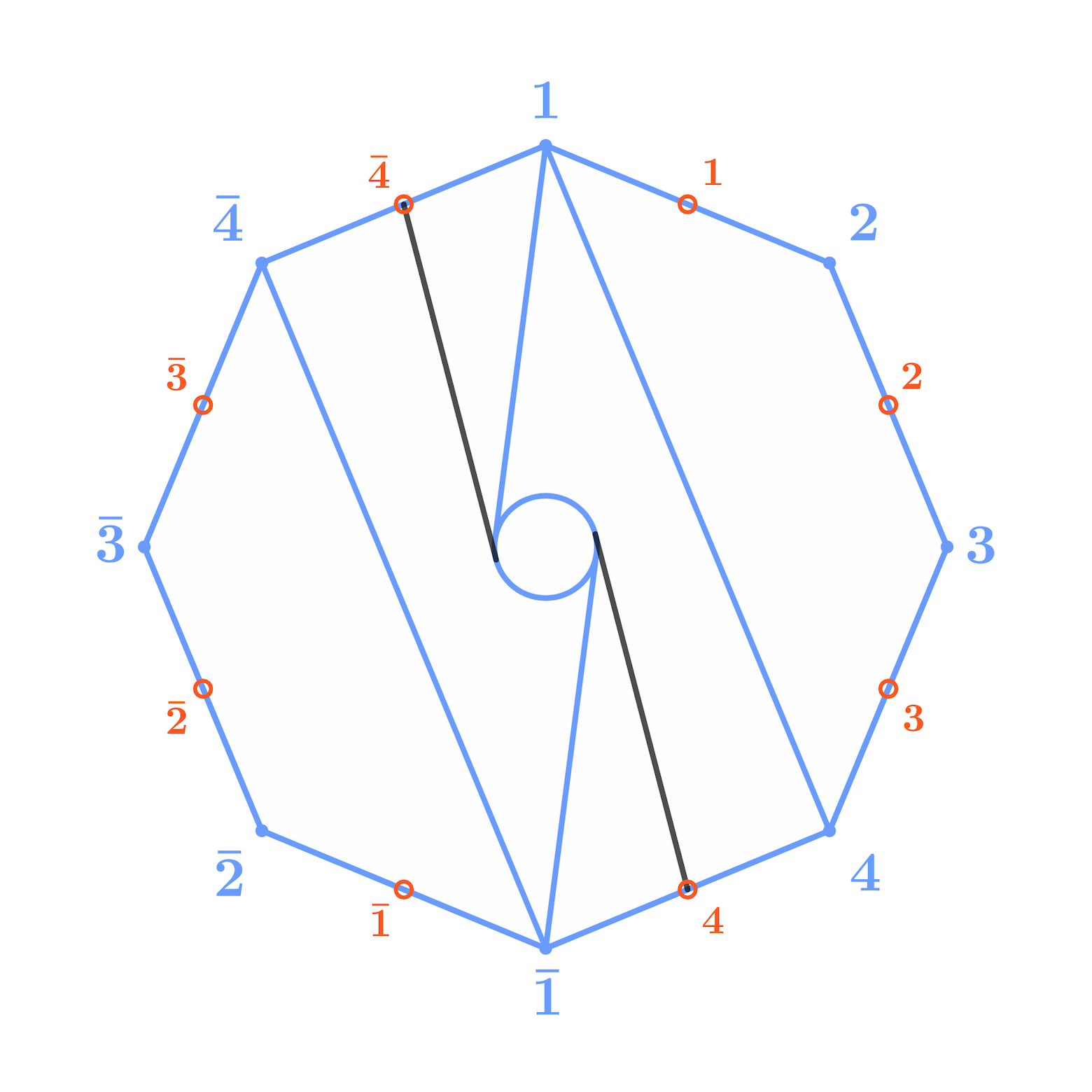}
\end{subfigure}
\begin{subfigure}{0.24\textwidth}
\centering
\includegraphics[scale=0.15]{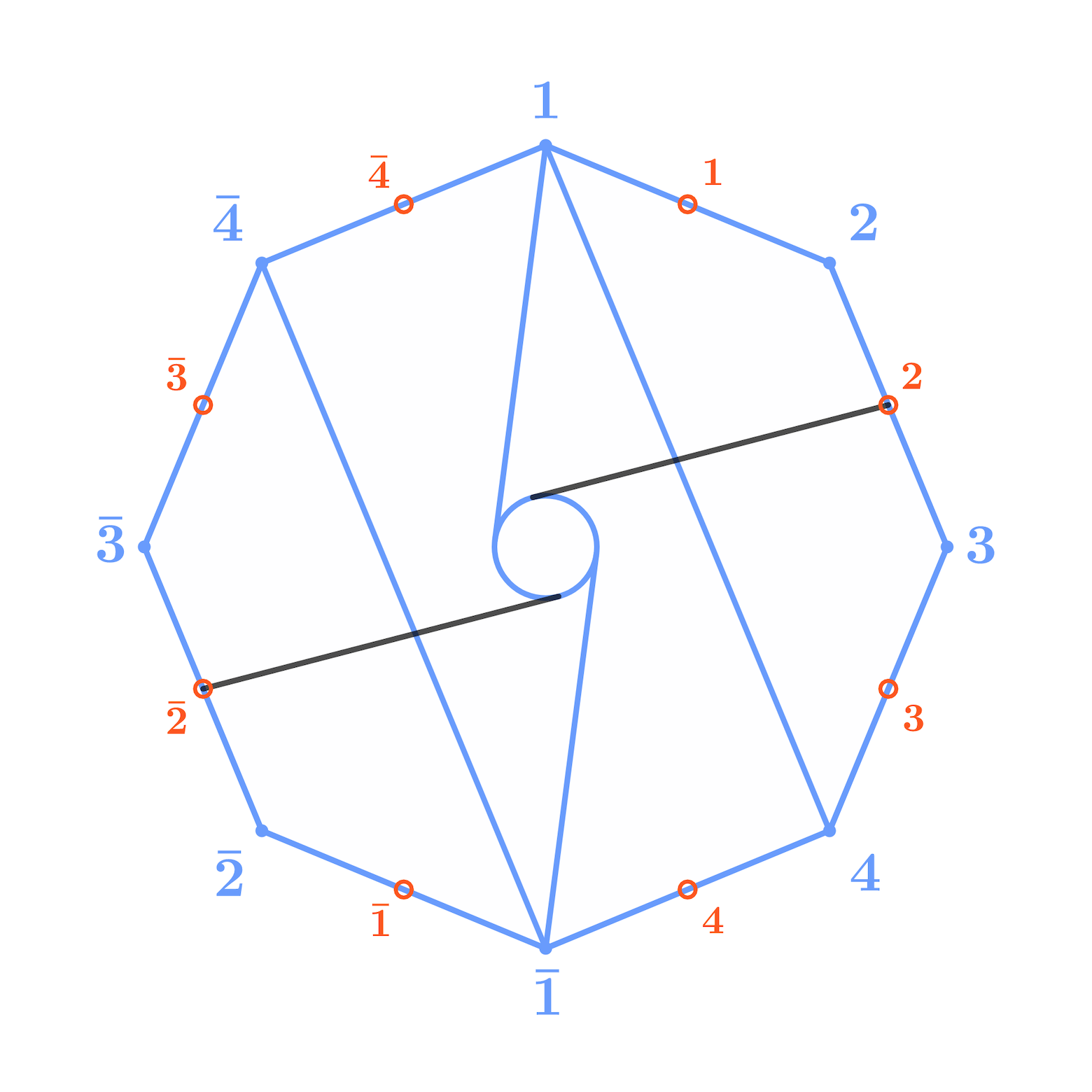}
\end{subfigure}

\begin{subfigure}{0.24\textwidth}
\centering
\includegraphics[scale=0.15]{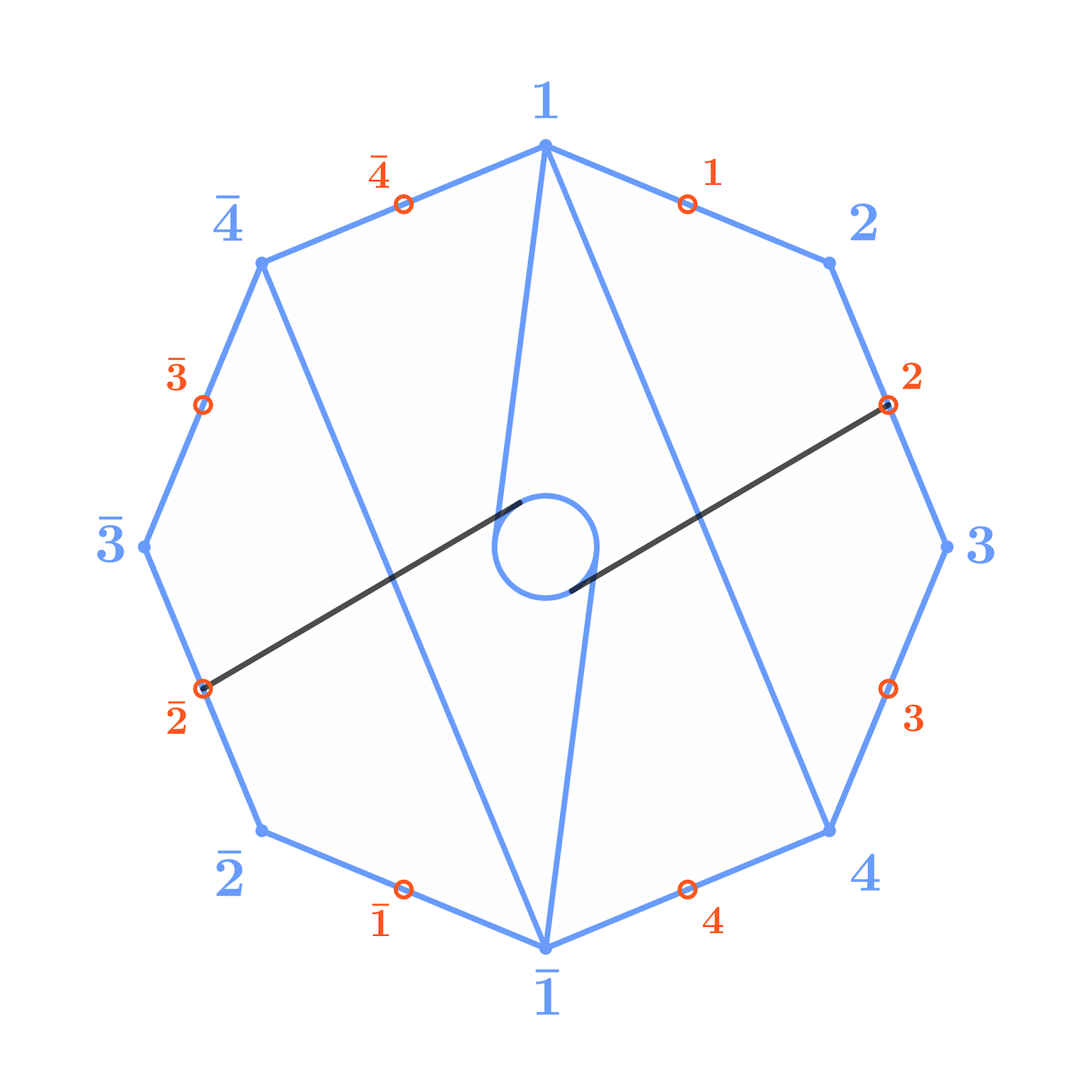}
\end{subfigure}
\begin{subfigure}{0.24\textwidth}
\centering
\includegraphics[scale=0.15]{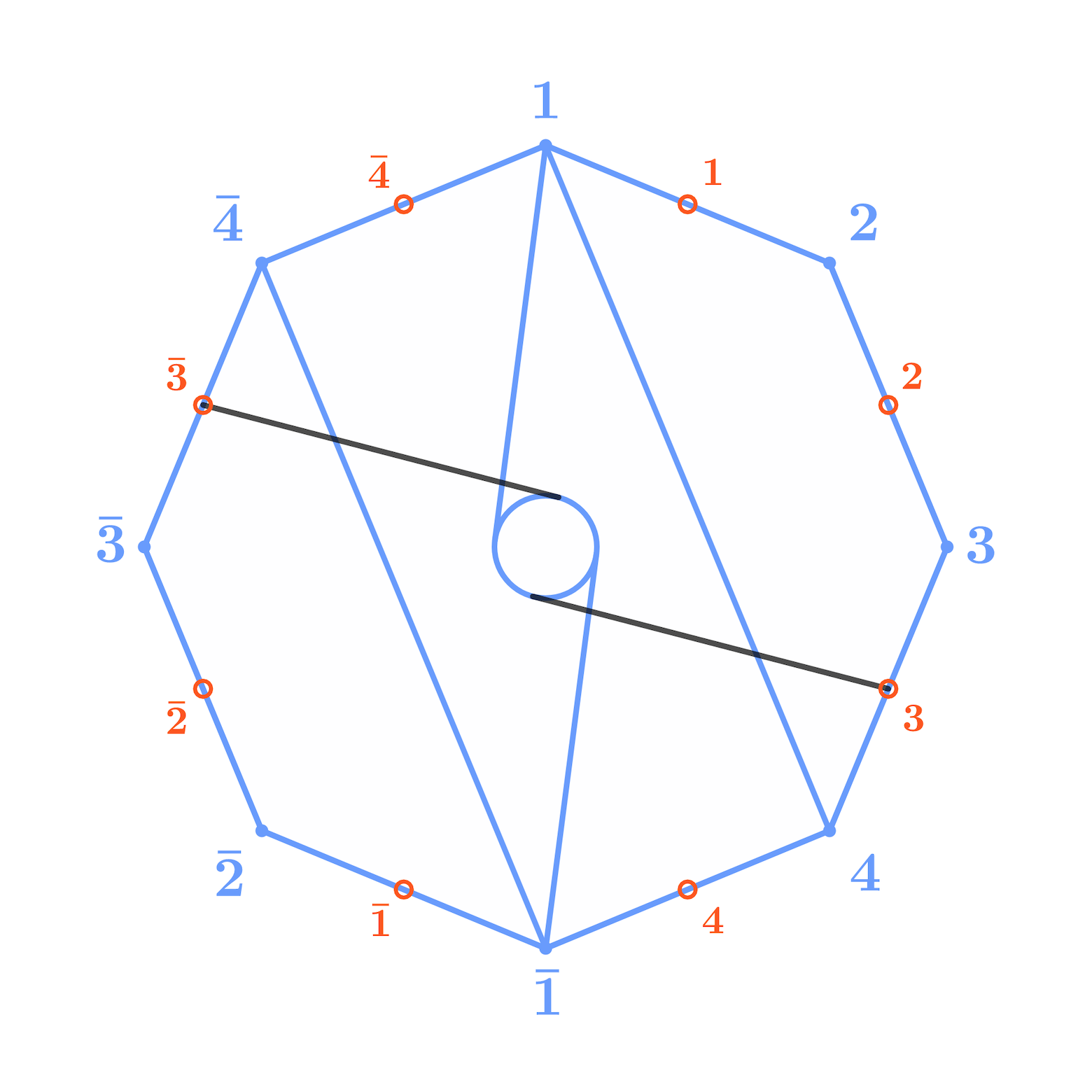}
\end{subfigure}
\begin{subfigure}{0.24\textwidth}
\centering
\includegraphics[scale=0.15]{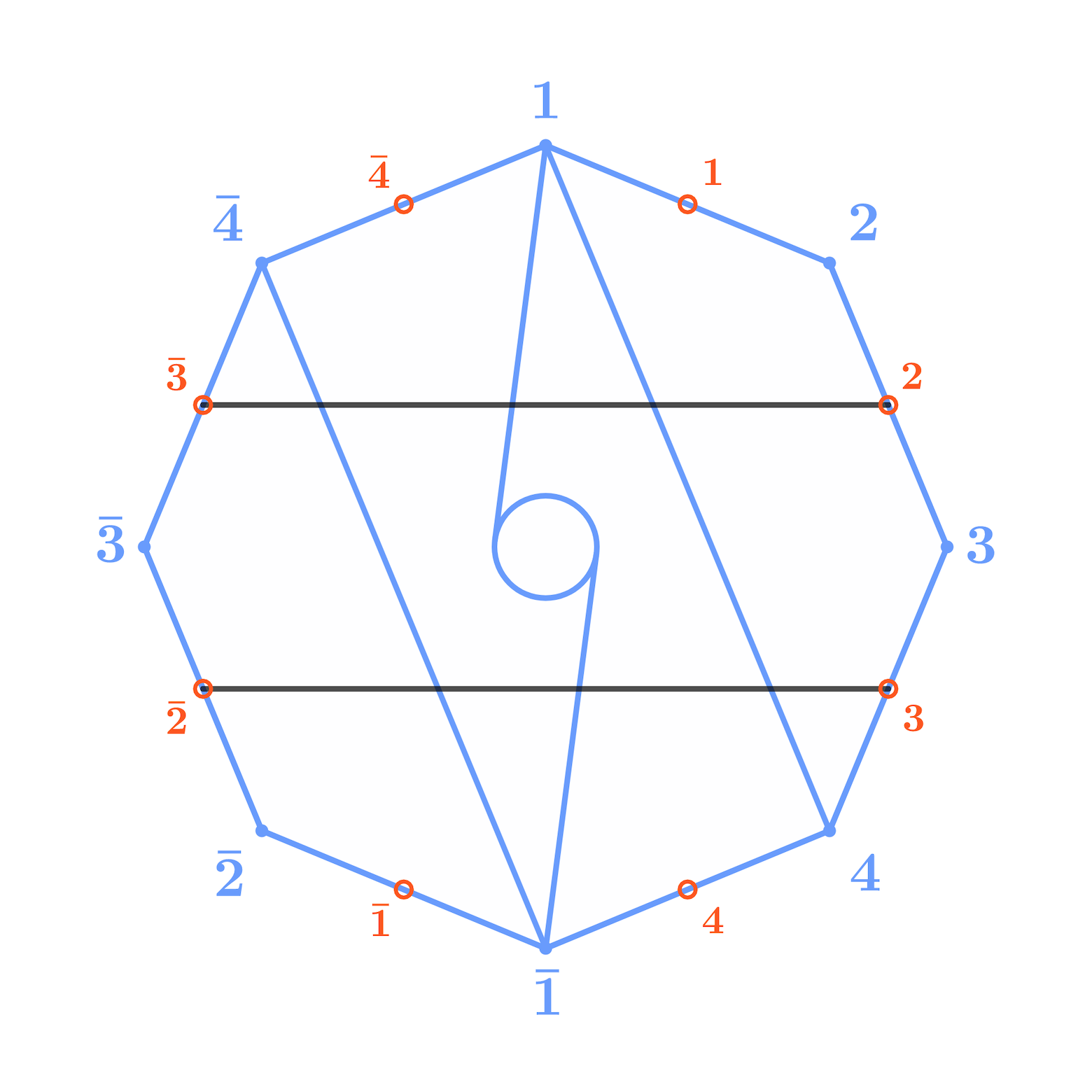}
\end{subfigure}
\begin{subfigure}{0.24\textwidth}
\centering
\includegraphics[scale=0.15]{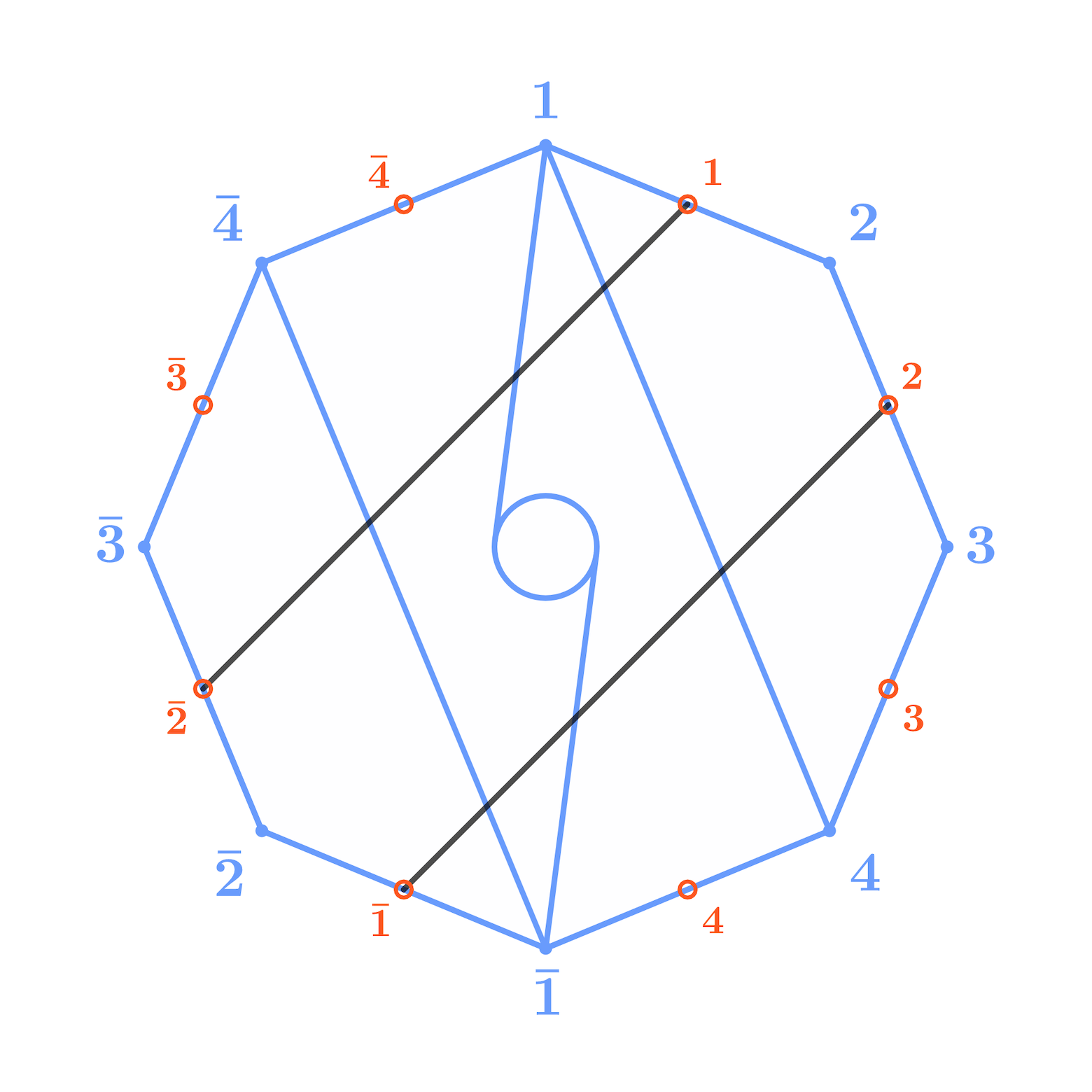}
\end{subfigure}

\begin{subfigure}{0.24\textwidth}
\centering
\includegraphics[scale=0.15]{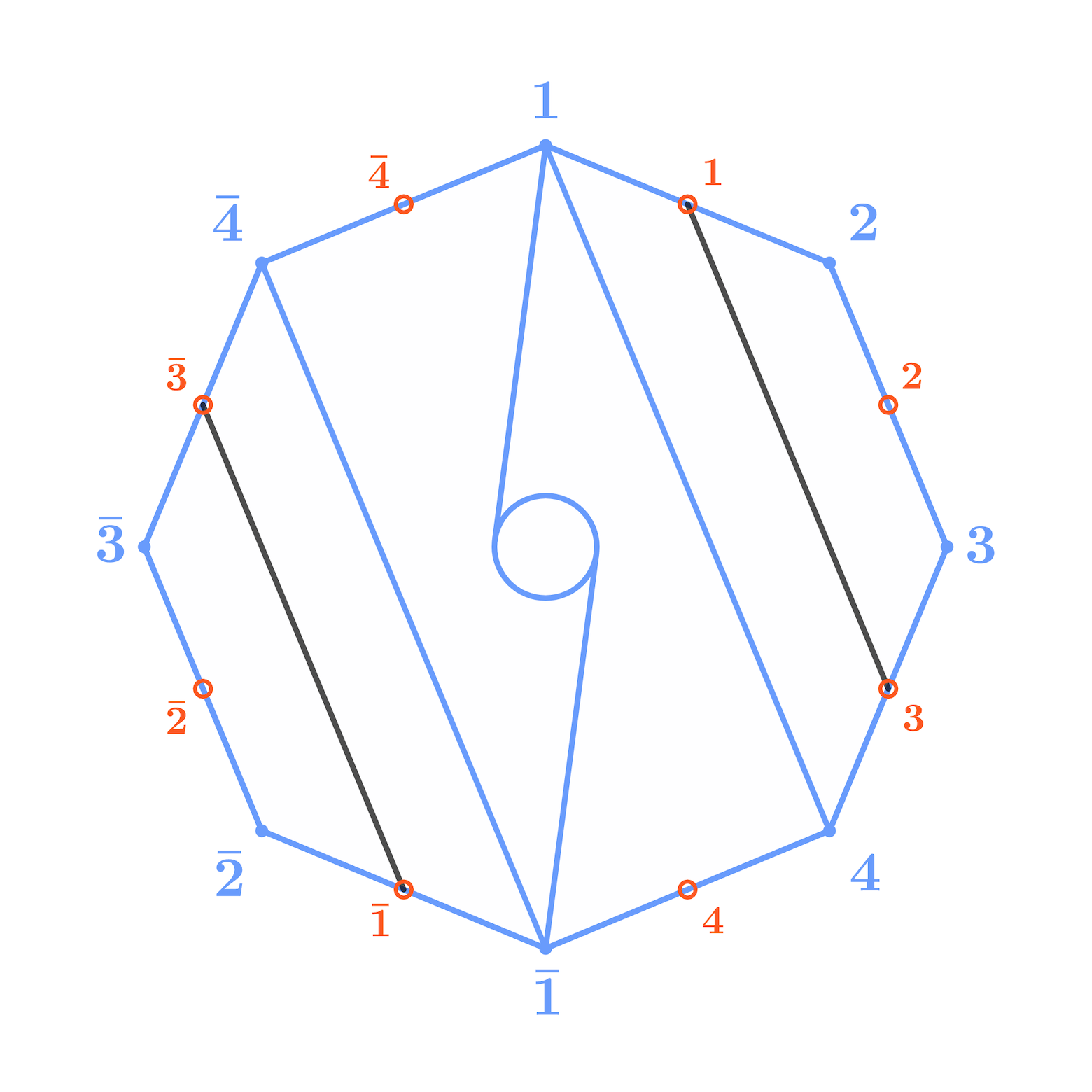}
\end{subfigure}
\begin{subfigure}{0.24\textwidth}
\centering
\includegraphics[scale=0.15]{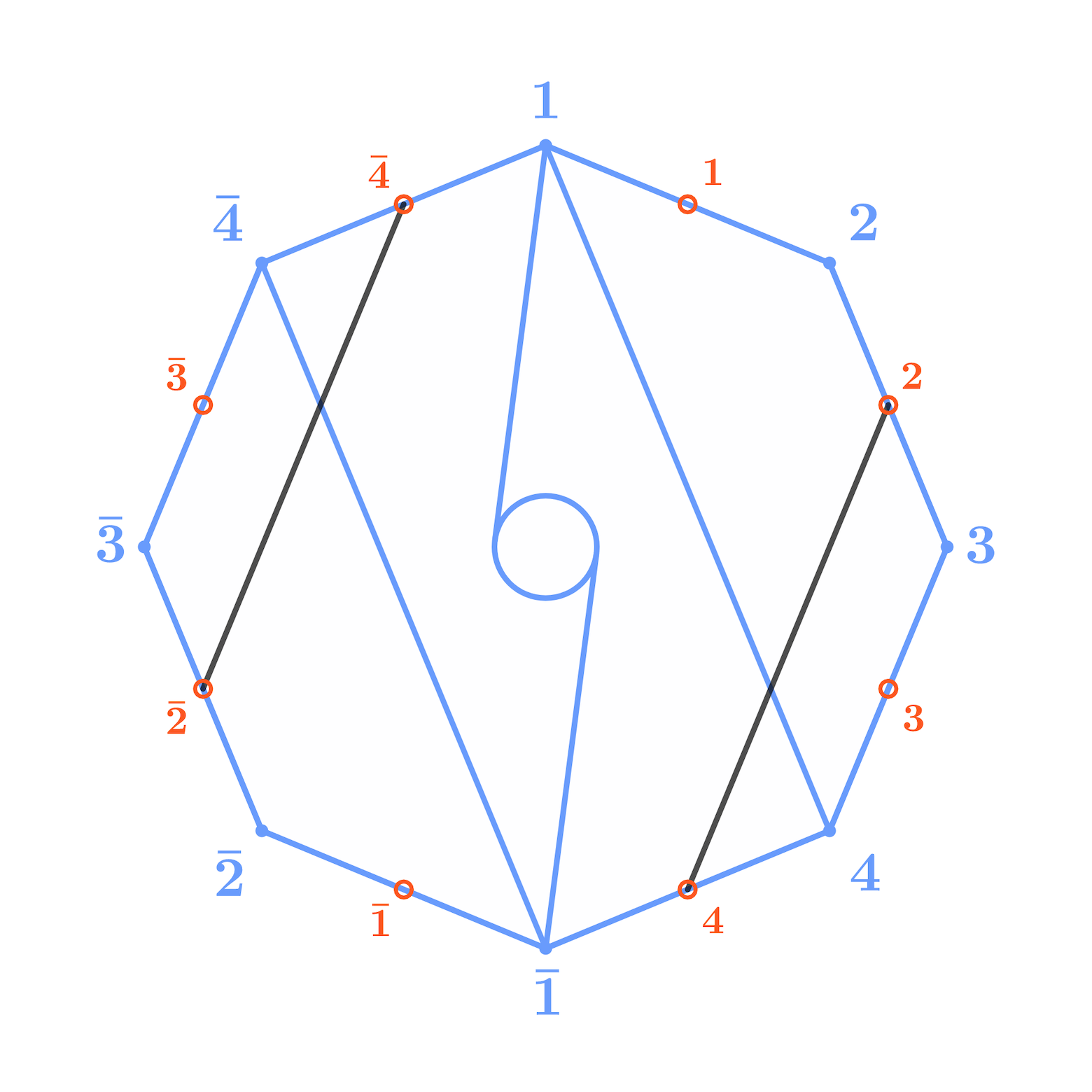}
\end{subfigure}
\begin{subfigure}{0.24\textwidth}
\centering
\includegraphics[scale=0.15]{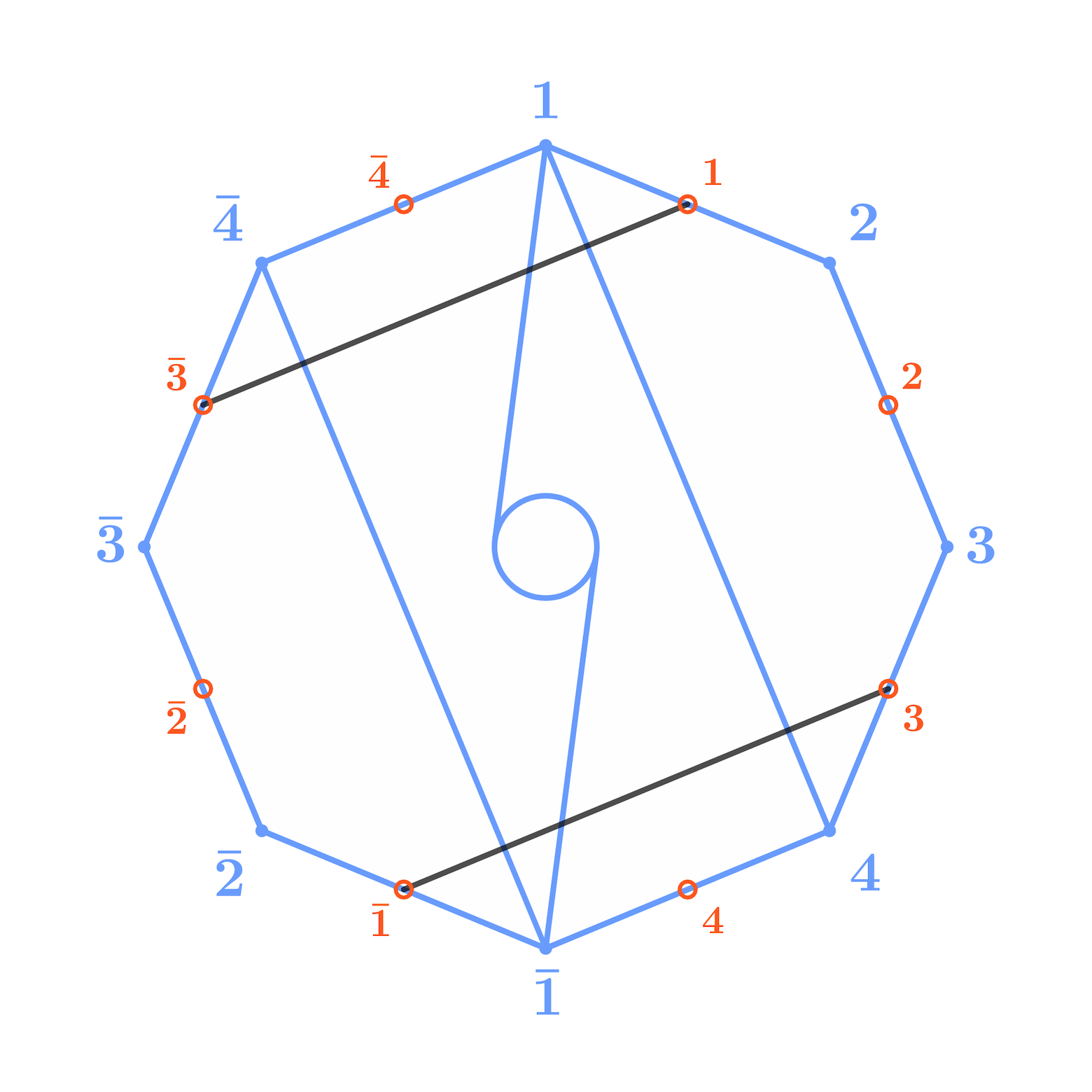}
\end{subfigure}
\begin{subfigure}{0.24\textwidth}
\centering
\includegraphics[scale=0.15]{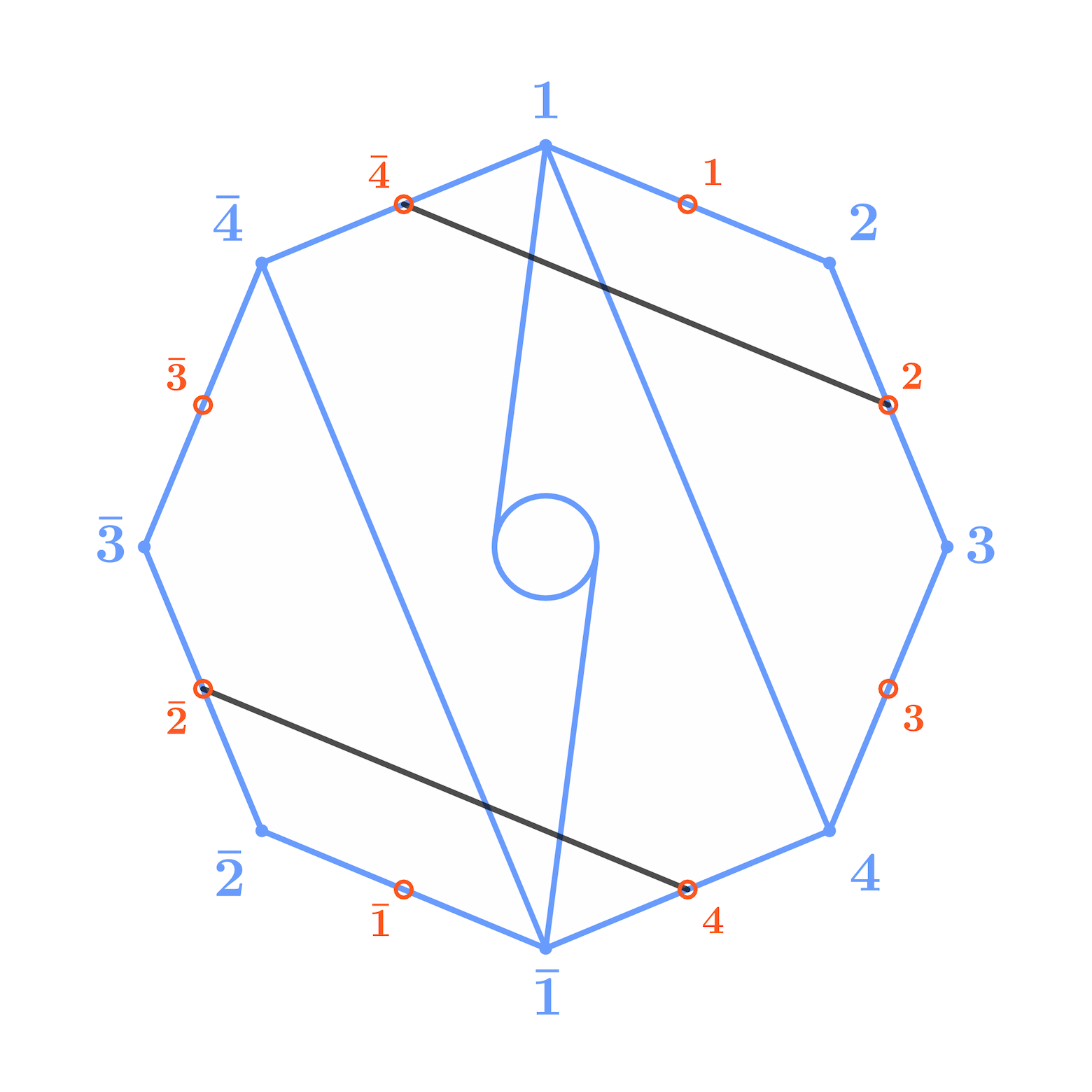}
\end{subfigure}
\caption{Chords Compatible with the reference are in orange while chords incompatible with the reference are in black.}
\label{compaincompay1x14}
\end{figure}

Hence ${\cal Q}$ splits the $2n$-gon in pseudo-quadrilaterals. We consider any two edges of a given pseudo-quadrilateral as incident if they meet in a given vertex. 
We define a hollow diagonal  $i^{\circ} j^{\circ}$  to be \emph{compatible} with ${\cal Q}$ only if,
\vspace*{-1.2em}
\begin{itemize}
\itemsep0em 
\item It does not enter and exit the same cell in ${\cal Q}$ from non-incident edges, and 
\item The set of chords (including the external edges) that $d_{i,j}$ intersects in $\overline{{\cal Q}}$ forms a connected set. (see figure \ref{pseudoquadrangulationy1x14Compatiblequadrangulation}).  
\begin{figure}[H]
    \centering
    \includegraphics[scale=0.23]{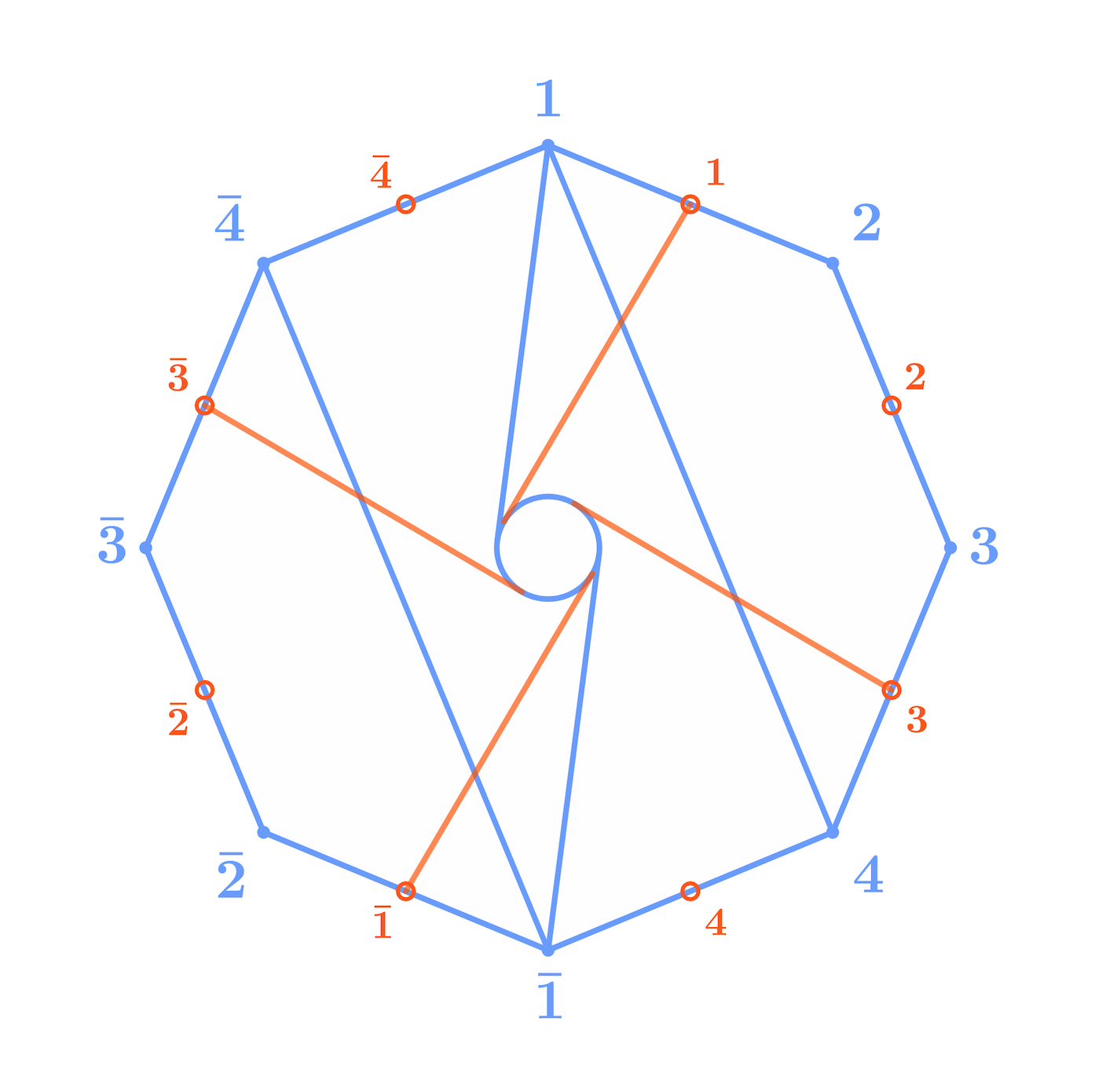}
    \caption{Reference pseudo-quadrangulation ${\cal Q}_{1} =\, \{\, 14,\, 1_{L},\, \bar{1}_{L},\, \bar{1}    \bar{4}\, \}$ and the quadrangulation ${\cal Q}^{\prime}\ =\, \{\, 1_{L},\, 3_{L},\, \overline{1}_{L},\, \overline{3}_{L} \}$ compatible with ${\cal Q}$.}
    \label{pseudoquadrangulationy1x14Compatiblequadrangulation}
\end{figure}

\end{itemize}
\vspace*{-0.8em}
We note that a compatible hollow diagonal begins and ends on the external edges of the solid polygon. We will also distinguish between two hollow diagonals which are tangent to the annulus at anti-podal positions. That is, we distinguish between $(v^{\circ}_{L},\, \bar{v}^{\circ}_{R})$ from $(v^{\circ}_{R},\, \bar{v}^{\circ}_{L})$.
 First point  of the above definition requires some clarification. Let us consider an example of pseudo-quadrangulation for an $8$-gon. 
\begin{flalign}\nonumber\\
{\cal Q}_{1} =\, \{\, 14,\, 1_{L},\, \bar{1}_{L},\, \bar{1}    \bar{4}\, \}
\end{flalign}
Following are some of the non-trivial examples of compatible hollow diagonals.
\begin{flalign}
\textrm{Compatible}\, =\, \{ (1^{\circ}_{L}, \bar{1}^{\circ}_{L} ) ,\, ( 3^{\circ}_{L}, \bar{3}^{\circ}_{L} )\, \}
\end{flalign}
$(1^{\circ}_{L}, \bar{1}^{\circ}_{L} )$ and $(3^{\circ}_{L}, \bar{3}^{\circ}_{L})$ can be thought of as entering and exiting the quadrilateral $[14,\, 1_{L},\, \bar{1}_{L}]$ from $14$ and $1_{L}$ respectively.\\
Some examples of incompatible hollow diagonals are 
\begin{flalign}
\textrm{Incompatible}\, =\, \{ (1^{\circ}_{R}, \bar{1}^{\circ}_{R} ) ,\, (4^{\circ}_{L}, \bar{4}^{\circ}_{L})\, \}
\end{flalign}
$(1^{\circ}_{R}, \bar{1}^{\circ}_{R} )$ enters and exists from $14$ and $\bar{1}_{L}$ respectively. $(4^{\circ}_{L}, \bar{4}^{\circ}_{L})$ enters and exists from $4,\bar{1}$ and $1_{L}$ respectively. For other compatible and incompatible diagonals see figure \ref{compaincompay1x14}\\
We refer to a compatible hollow diagonal as a pseudo-accordion.\\
A pseudo-quadrangulation of the solid $2n$-gon, ${\cal Q}^{\prime}$ is compatible with ${\cal Q}$ if it consists of mutually non-intersecting pseudo-accordions (see figure \ref{pseudoquadrangulationy1x14Compatiblequadrangulation}).\\
This definition sounds a mouthful and we now give an \emph{equivalent} definition which is perhaps easier to work with and is in fact closer to the ``original" definition of the compatible quadrangulations given by Barshnykov \cite{Baryshnikov}.\\
{\bf Second definition} :\\
 We consider the $2n$-gon with vertices $\{1,\, 2,\, \dots\, \bar{n},\, 0_{L},\, 0_{R}\, \}$. In this vertex set, we have included the two halves of the annulus as (abstract) vertices. 
\vspace*{-0.2in}
\begin{itemize}
\item Given a pseudo-quadrangulation ${\cal Q}$, consider  a (pseudo-hexagon) formed by the external edges and internal chords. As ${\cal Q}$ is a quadrangulation, the hexagon is dissected by one of the chords, say $c_{i,j}$, such that if we label the vertices of this hexagon clockwise as $1,\, \dots,\, 6$, $(i,\, j)$ is either $(1 4),\, (2 5),\, \textrm{or},\, (3, 6)$. We can obtain a compatible pseudo-quadrangulation by flipping $c_{1,4}$ to $c_{3,6}$, $c_{3,6}$ to $c_{2,5}$ and $c_{2,5}$ to $c_{1,4}$ respectively.
\item If within a pseudo-quadrangulation we have a configuration where the annulus is bounded by two parallel linear chords so as to form a square. Let the square be labelled by external vertices, $i_{1}, j_{1}, \bar{i}_{1}, \bar{j}_{1}$. If there is a diagonal $i_{L}, \bar{i}_{L}$ inside this square, then we flip this diagonal to $j_{R},\, \bar{j}_{R}$. 
\end{itemize}
We thus say that a pseudo-quadrangulation ${\cal Q}^{\prime}$ is compatible with ${\cal Q}$ only if, diagonals in ${\cal Q}^{\prime}$ are obtained from the flips of one or more diagonals belonging to ${\cal Q}$.\\
The above rules are perhaps simpler to understand as they do not refer to a polygon and it's dual. But they are rather specific to pseudo-quadrangulations. The definition of compatible ${\cal Q}$ given previously is in fact applicable to any pseudo-dissections and generates a family of combinatorial polytopes.\\
We end this section with a few remarks.
\vspace*{-0.2in}
\begin{enumerate}\itemsep0em 
\item To the best of our knowledge, the combinatorial polytopes whose vertices are in $1-1$ correspondence with pseudo-quadrangulations have not been studied in the mathematics literature before. However in a striking paper \cite{konk-palu}, the authors define a large class of polytopes based on dissections of a compact Riemann surface with marked points and possble boundaries. These polytopes are referred to as accordion complex in \cite{konk-palu}. We strongly believe that the pseudo-accordiohedra defined in this paper are accordion complexes. However an explicit proof is required.   
\item We expect that many of the remarkable properties that accordiohedra satisfies, such as it's pseudo-manifoldness, it's dimensionality given by number of hollow diagonals will continue to remain valid for pseudo-accordiohedra. However this requires a detailed analysis far beyond the level of mathematical rigour in this paper.
\item Starting with any reference pseudo-triangulation,  mutations lead to a unique type-D associahedra. However (just as in the case of accordiohedron), this is not the case for pseudo-quadrangulations and the polytope retains the memory of the reference (pseudo)-dissection ${\cal Q}$. 
\end{enumerate}
\vspace*{-0.2in}
\subsection{Projective Forms on ${\cal KL}_{n}$}
\vspace*{-0.2in}
We now use the flips of pseudo-quandrangulations to define a family of projective $\frac{n}{2}$ forms $\Omega_{{\cal Q}}^{n}$ on ${\cal KL}_{n}$. That is, if $\{\, {\cal Q}_{1},\, \dots,\, {\cal Q}_{N({\cal Q})}\, \}$ is the set of compatible pseudo-quadrangulations then \footnote{It is an important open question to derive $N({\cal Q})$ as was done for the Stokes polytopes by F. Chapoton in \cite{Chapoton}.} 
\begin{flalign}
\Omega_{{\cal Q}}^{n}\, =\, \sum_{I=1}^{N(n)}\, (-1)^{\sigma({\cal Q}_{I}, {\cal Q})}\, \bigwedge{k=1}^{\frac{n}{2}} d\ln X_{i_{k} j_{k}}
\end{flalign}
where 
\begin{flalign}
\sigma({\cal Q}_{I},\, {\cal Q})\, =\, \pm 1
\end{flalign}
depending on the number of flips required to reach ${\cal Q}_{I}$ from ${\cal Q}$.  For each ${\cal Q}$ this form is projective,\footnote{This is in precise analogy with the tree-level  projective scattering forms  on the planar kinematic space generated by Accordiohedra.} We explain this construction with a simple example of a 2 particle case. 
Four possible pseudo-quadrangulations are,
\begin{flalign}
\begin{array}{lll}
{\cal Q}^{2}_{1}\, =\, \{\, 1_{L}, \bar{1}_{L}\, \}\\
{\cal Q}^{2}_{1^{\prime}}\, =\, \{\, 1_{R}, \bar{1}_{R}\, \}\\
{\cal Q}^{2}_{2}\, =\, \{\, 2_{L}, \bar{2}_{L}\, \}\\
{\cal Q}^{2}_{2}\, =\, \{\, 2_{R}, \bar{2}_{R}\, \}\\
\end{array}
\end{flalign}
If we start with reference $Q$ as either $\{1_{L},\, \bar{1}_{L}\, \}$ or $\{\, 1_{R},\, \bar{1}_{R}\, \}$ then the corresponding one dimensional polytopes have vertices associated with  $\{\{ 1_{L},\bar{1}_{L}\},\{ 2_{R},\bar{2}_{R}\} \}$ and $\{\{ 1_{R},\bar{1}_{R}\},\{ 2_{L},\bar{2}_{L}\} \}$ respectively.\\
On ${\cal KL}_{2}$, it is easy to write the projective forms. For ${\cal Q}_{1}$ we have, 
\begin{flalign}
\Omega_{{\cal Q}_{1}} =  \pm d\ln\frac{Y_{1}}{\tp{Y}_{2}}
\end{flalign}
As slightly more involved examples, we consider $n\, =\, 4$ with the reference pseudo-quadrangulations being, 
\begin{flalign}
{\cal Q}_{1}\, &=\, \{1_{L},\, 14,\,   \overline{1}_{L}\, \bar{1}    \bar{4},\, \}  &  {\cal Q}_{2}\, &=\, \{1_{L},\, 1\bar 2,\,   \overline{1}_{L}\, \overline{1}2,\, \} &  {\cal Q}_{3}\, &=\, \{1_{L},\, 3_{L},\,   \overline{1}_{L}\, \overline{3}_{L},\, \}. 
\end{flalign}

\begin{figure}[H]
\centering 
\begin{subfigure}{0.45 \textwidth}
\centering 
\includegraphics[scale=0.25]{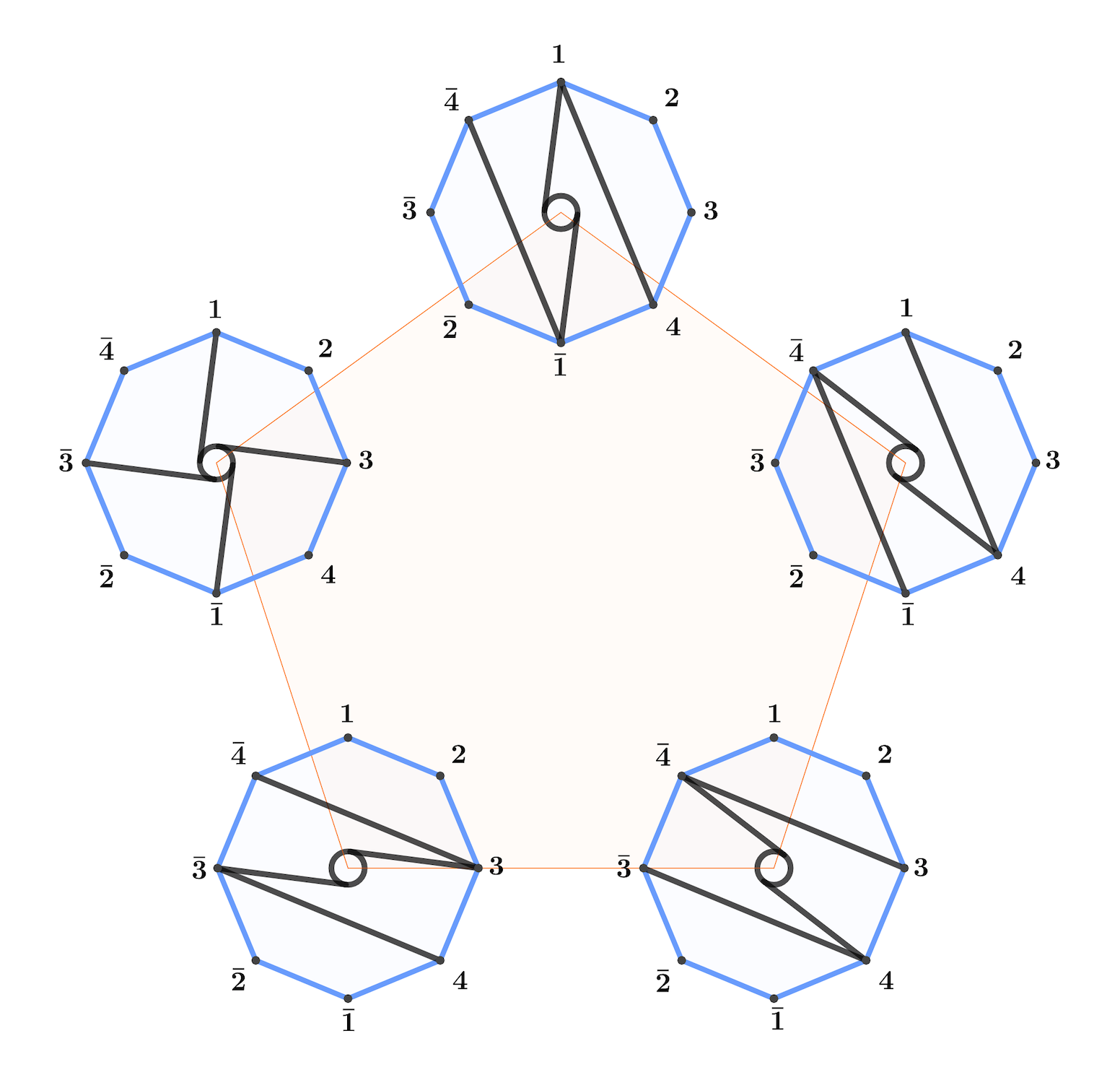}
\caption{$\mathcal{PAC}(Q_{1})$}
\end{subfigure}
\begin{subfigure}{0.45 \textwidth}
\centering 
\includegraphics[scale=0.25]{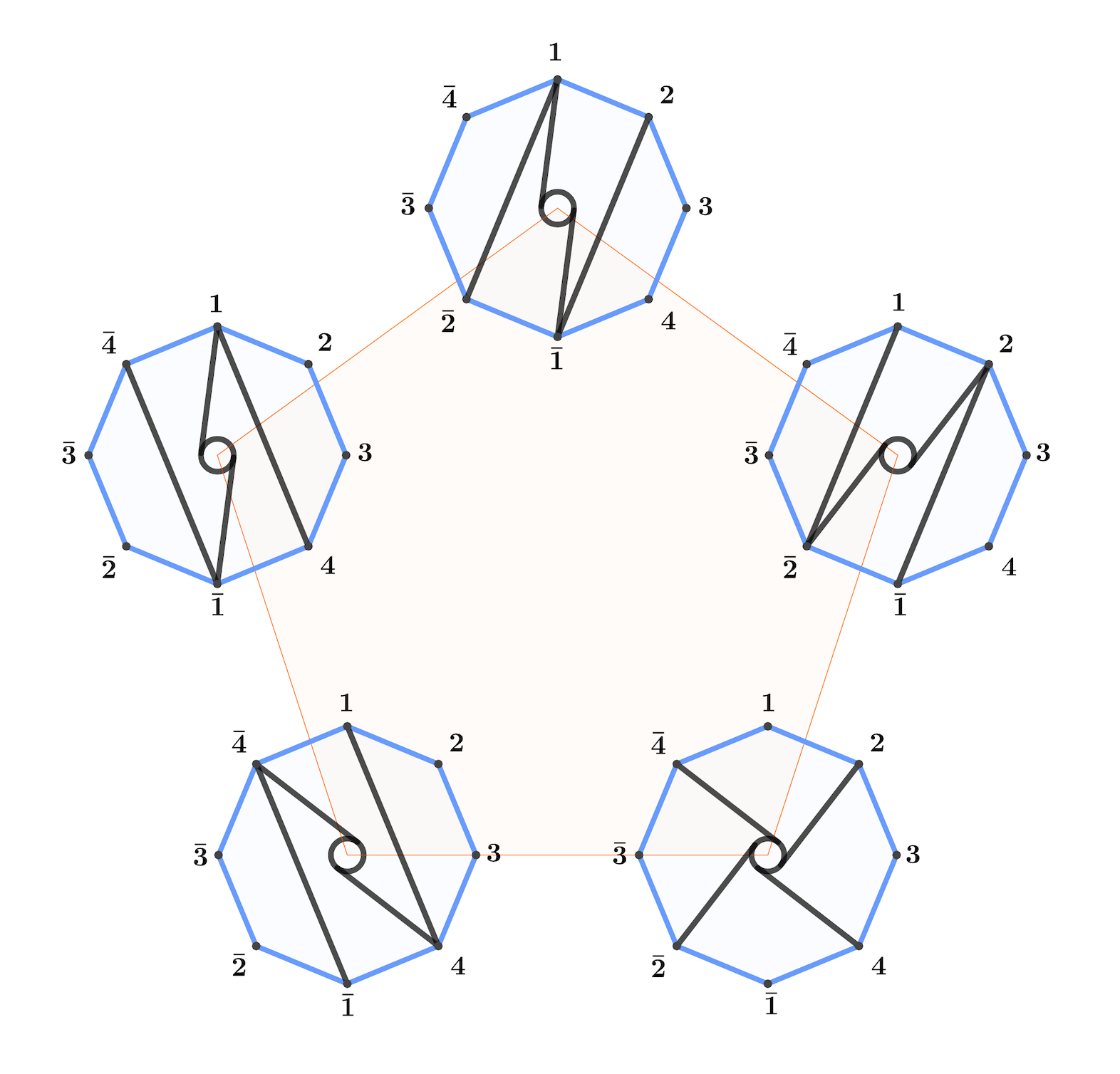}
\caption{$\mathcal{PAC}(Q_{2})$}
\end{subfigure}
\begin{subfigure}{0.45 \textwidth}
\centering 
\includegraphics[scale=0.25]{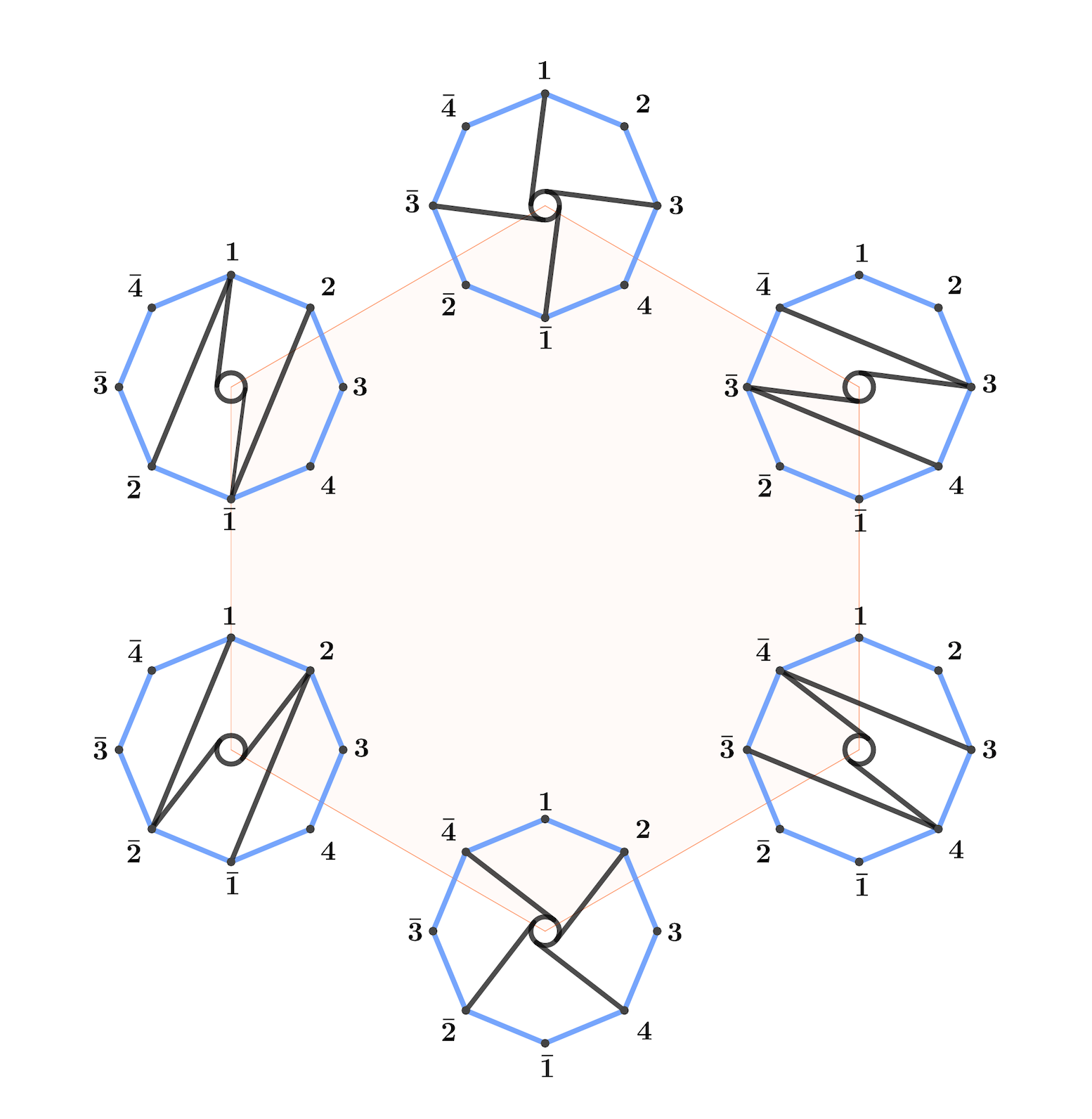}
\caption{$\mathcal{PAC}(Q_{3})$}
\end{subfigure}
\caption{Pseudo-accordiohedra }
\label{pseudo-accordiohedra}
\end{figure}

In these cases, the Pseudo-accordiohedra have the following structure (see figure \ref{pseudo-accordiohedra}) and  the projective 2-forms on ${\cal KL}^{(1)}_{4}$ are given by
\begin{flalign}
\Omega_{{\cal Q}_{1}}^{4}\, &=\,    d\ln \frac{Y_{1}}{\tp{Y}_{4}}\,  \wedge\,   d\ln \frac{X_{14}}{Y_{3}}\,  +\, d\ln \frac{\tp{Y}_{4}}{Y_{3}}  \wedge\,  d\ln \frac{X_{3\bar 4}}{Y_{3}}\,  \endline
\Omega_{{\cal Q}_{2}}^{4}\, &=\,    d\ln \frac{Y_{1}}{\tp{Y}_{2}}\,  \wedge\,   d\ln \frac{X_{1\bar 2}}{X_{14}}\,  +\, d\ln \frac{\tp{Y}_{2}}{X_{14}}  \wedge\,  d\ln \frac{\tp{Y}_{4}}{X_{14}}\, \endline
\Omega_{{\cal Q}_{3}}^{4}\, &=\,    d\ln \frac{Y_{1}}{X_{3 \bar 4}}\,  \wedge\,   d\ln \frac{Y_{3}}{X_{1\bar 2}}\,  +\, d\ln \frac{\tp{Y}_{2}}{X_{3 \bar 4}}  \wedge\,  d\ln \frac{X_{1 \bar 2 }}{\tp{Y}_{4}}\,.
\end{flalign}
\subsection{From Projective Forms in ${\cal KL}_{n}$ to Lower Forms on ${\cal D}_{T}$}
In this section, we restrict the $\frac{n}{2}$ forms labelled by any pseudo-quadrangulation to one of the (geometric realisations of) ${\cal D}_{n}$ to obtain a form which has simple poles only on the faces of ${\cal D}_{n}$ that correspond to poles in the loop integrand of  planar $\phi^{4}$ amplitudes. As we show below, each such form generates a partial contribution to the loop integrand associated to quartic interactions. A weighted sum over all the pseudo-quadrangulations finally produces the $n$-point loop integrand for $\phi^{4}$ theory. 
Our main claim in this section is the following. 

\begin{tcolorbox}[colback=black!5!white, colframe=black!75!black,arc=0mm]
\begin{claim}
Let ${\cal Q}$ be a pseudo-quadrangulation of a $2n$-gon with an annulus. Let $T$ be any pseudo-triangulation which containts ${\cal Q}$ as a proper subset. Then, on the geometric realisation of the type-D associahedra which is associated to $T$ the scattering form $\Omega^{n}_{{\cal Q}}$ projects to,
\begin{flalign}\label{ctfopa}
\omega^{n}_{{\cal Q}}\vert_{{\cal D}_{T}}\, =\, \left(\, \sum_{v\, \in\, {\cal PAC}_{\mathcal{Q}}} \prod\, \frac{1}{X_{i_{v}j_{v}}}\, \right)\, \bigwedge_{c\, \in\, {\cal Q}}\, d X_{c}
\end{flalign}
\end{claim}
\end{tcolorbox}
where the sum $\sum_{v\, \in\, {\cal PAC}_{\mathcal{Q}}}$ is over all the vertices of the pseudo-accordiohedron. 
\begin{claimproof}
Our proof goes along the same lines as the proof of claim \ref{cl2}. However we need to analyse the mutations involving central chords carefully. Let's consider an example where pseudo-quadrangulation $\mathcal{Q}_{1}$ mutates to $\mathcal{Q}_{2}=  \mathcal{Q}_{1}  \setminus \{i j, \bar{i} \bar{j} \}   \cup \{i_{L},\bar i_{L}\} $. Suppose the flip occurs in a hexagon whose two sides are $i_{L}$ and $kj$. Then the compatibility rules imply there are no chords of the reference quadrangulation with one end in $\{i+1,\ldots, k\}$ and other end is in $\{ j+1, \ldots, \overline{k}\}$. Therefore we have 
\begin{equation}
X_{i j} + Y_{k} - X_{k j } - Y_{i} =   \sum_{p=i}^{k-1} \left( \sum_{q= j}^{\overline{i-1}} c_{pq} + \sum_{q = \overline{p+1} }^{\overline{k-1}} c_{pq} + c_{p_{R}}  \right).
\end{equation}
Therefore we have 
\begin{flalign}\label{qproofpw}
\bigwedge_{(ij) \in \mathcal{Q}_{1}}\, d X_{ij}\, =\, -\, \bigwedge_{(mn) \in \mathcal{Q}_{2}}\, d X_{mn}. 
\end{flalign} 
 Similarly we can show \eqref{qproofpw} for other mutations as well.
\end{claimproof}

The restriction of $\Omega_{\mathcal{Q}_{1}}^{4}$ on geometric realisation of type-D associahedra $\mathcal{PA}_{T}^{4}$ associated with $\mathcal{T}=  \{1_{L},1_{R} , 14 , 24,  \overline{1}_{L}, \overline{1}_{R} \bar{1}    \bar{4}, \bar{2} \bar{4} \} $ is given by 
\begin{equation}\label{q1restriction}
    \Omega_{\mathcal{Q}_{1}}^{4} \vert_{\mathcal{PA}_{T}^{4}} = m^{\mathcal{Q}_{1}}_{4} = \frac{1}{X_{14} Y_{1}} + \frac{1}{X_{14}\tilde{Y}_{4}} + \frac{1}{X_{3 \bar 4}\tilde{Y}_{4}} + \frac{1}{X_{3 \bar 4} Y_{3}} + \frac{1}{Y_{1}Y_{3}}. 
\end{equation}

\subsection{Geometric Realisation of Pseudo-accordiohedron}\label{crpa}
In this section, we will obtain the geometric realisation of pseudo-accordiohedra with reference $Q \subset T $ by projecting the geometric realisation of type-D associahedra. 

The coordinates on the $n^{2}$ dimensional kinematic space ${\cal KL}_{n}$ are the $n^{2}$ kinematic variables $X_{ij}, Y_{i}, \tilde{Y}_{i}$. The geometric realisation $\mathcal{D}_{T}$ of type-D associahedra associated with the pseudo-triangulation $T$, sitting inside the kinematic space ${\cal KL}_{n}$ is given by $X_{ij}, Y_{i}, \tilde{Y}_{i} >0$ and
\begin{equation}\label{Dsijcij}
s_{ij} = - c_{ij} \hspace{1cm} \forall (ij) \notin T^{c}.
\end{equation}
These $n(n-1)$ constraints \eqref{Dsijcij} cut out a $n$ dimensional hyper-plane ${\cal K}_{T}$ in ${\cal KL}_{n}$. This $n$-dimensional hyper-plane is parametrized by the kinematic variables associated with the chords of the triangulation $T$. That is, given any point of ${\cal KL}_{n}$ sitting in this space, any coordinate $z$ of that point can be expressed as 
\begin{equation}
z = \sum_{x \in T} g_{z x}  x+ c_{z},
\end{equation} 
where $g_{z x}$\footnote{$g_{z x}$ is the $x$th component of the $g$-vector of $z$. } and $c_{z}$ are some constants. Thus the kinematic variables associated with the chords of $T$ form a coordinate system on ${\cal K}_{T}$.  On this hyper-plane the geometric realisation of type-D associahedra is given by 
\begin{equation}
AD_{T} = \{(x_{1},\ldots, x_{n}) \in {\cal K}_{T}   \vert z = \sum_{x_{i} \in T} g_{z x_{i}}  x_{i} + c_{z} \geq 0 \hspace{0.2cm} \forall  \text{  kinematic variables } z \}.
\end{equation} 
The geometric realisation of the pseudo-accordiohedra $\mathcal{PAC}(Q)$ can be given by projecting the geometric realisation $ AD_{T}$ on the $\frac{n}{2}$ dimensional hyper-plane $\mathcal{K}_{Q}$ spanned by the variables $x \in Q$. Suppose $ P : \mathcal{K}_{T} \rightarrow \mathcal{K}_{Q} $ is the projection map. Then 
\begin{equation}
\mathcal{PAC}(Q) = \{ (x_{1}, \ldots x_{\frac{n}{2}})  \in \mathcal{K}_{Q}   \vert \exists  (y_{1}, \ldots , y_{n}) \text{ such that } P(y_{1}, \ldots , y_{n}) = (x_{1}, \ldots x_{\frac{n}{2}}   )  \}.
\end{equation}

Given a pseudo-quadrangulation $Q $ which is a subset of pseudo-triangulation $T$, a chord $z$ is compatible with the reference quadrangulation $Q$ if and only if the constants $g_{z x}$ appearing in $z = \sum_{x \in T} g_{z x}  x+ c_{z},$ are zero for all $x \notin Q$. Therefore on the hyper-plane the geometric realisation of $\mathcal{PAC}(Q)$ is given by 
\begin{equation}
  \mathcal{PAC}(Q) =    \{ (x_{1}, \ldots x_{\frac{n}{2}}  ) \in \mathcal{K}_{Q} \vert z = \sum_{x_{i} \in Q} g_{z x_{i}}  x_{i} + c_{z} \geq 0 \hspace{0.2cm} \forall  \text{ compatible variables } z \}.
\end{equation}

Let's look at an example with $\mathcal{Q}_{1} = \{1_{L},\, 14,\,   \overline{1}_{L}\, \bar{1}    \bar{4},\, \} \subset \mathcal{T} =  \{1_{L},1_{R} , 14 , 24,  \overline{1}_{L}, \overline{1}_{R} \bar{1}    \bar{4}, \bar{2}    \bar{4} \} $.
The geometric realisation is given by the following in-equations
\begin{align}
Y_{1}& \geq 0,  & X_{14}  & \geq 0, &  Y_{3} = Y_{1}- X_{14} + c_{Y_{3}} &\geq 0,  & 
\tilde{Y}_{4}  = - Y_{1} + c_{\tilde{Y}_{4}} & \geq 0, & X_{3 \bar 4} = -X_{14} + c_{X_{3 \bar 4}} &\geq 0. 
\end{align}
Equivalently 
\begin{align}
0 \leq &  Y_{1} \leq   c_{\tilde{Y}_{4}} , &  0 \leq &  X_{14} \leq c_{X_{3 \bar 4}} ,  & X_{14} \leq Y_{1} + c_{Y_{3}}  .
\end{align}
Where the constants $c_{Y_{3}},c_{\tilde{Y}_{4}}$ and $c_{X_{3 \bar 4}} $ are
\begin{align*}
c_{Y_{3}} &= c_{14} + c_{24} + c_{1_{R} } + c_{1 \bar 2} + c_{2_{R}}  \endline
 c_{\tilde{Y}_{4}} &= c_{1 \bar 2} + c_{1 \bar 3 } + c_{2 \bar 3} + c_{1_{L}} +c_{2_{L}} + c_{3_{L}} \endline
 c_{X_{3 \bar 4}}&=  c_ {1 4} + c_ {1 _ {L}} + c_ {1 _ {R}} + c_ {1\bar 2} + c_ {1 \bar 3} + c_ {2 4} + c_ { 2 \bar 1} + c_ {2 _ {L}} + c_ {2 _ {R}}  + c_ {2\bar 3}.
\end{align*}

\subsection{Factorisation Properties of Pseudo-accordiohedra}
Type-D associahedra have striking factorisation properties just as associahedron. As was reviewed in \cite{Arkani-Hamed:2019vag}, Any facet (boundary of co-dimension one) of a ${\cal D}_{n}$  is  either a type-D associahedra or a product of type-D associahedra and associahedra of lower dimension. The situation is analogous for pseudo-accordiohedron $\mathcal{PAC}_{\mathcal{Q}}$. That is, every facet of the pseudo-accordiohedron is either a pseudo-accordiohedron or a direct product of a pseudo-accordiohedron with a stokes poset. That is, we claim that, if (1) $j\, >\, i$ and (2) ${\cal SP}_{Q^{\prime}}$ is a Stokes polytope with reference dissection $Q^{\prime}$ then,
\begin{tcolorbox}[colback=black!5!white, colframe=black!75!black,arc=0mm]
\begin{flalign}
\begin{array}{lll}
\mathcal{PAC}^{n}_{\mathcal{Q}}\xrightarrow{X_{ij}\, \rightarrow\, 0}\, {\cal SP}^{n - (j-i) - 1}_{\mathcal{Q}_{L}}\, \times\, \mathcal{PAC}^{j-1}_{\mathcal{Q}_{R}}\\[0.2em]
\mathcal{PAC}^{n}_{\mathcal{Q}} \xrightarrow{Y_{i}\, \rightarrow\, 0}\, {\cal SP}^{n+2}_{\mathcal{Q} \setminus i\cup\tp{i}}\\[0.2em]
\mathcal{PAC}^{n}_{\mathcal{Q}} \xrightarrow{X_{i\tp{j}}\, \rightarrow\, 0}\, {\cal SP}^{i + 1 + (n - j)}_{\mathcal{Q}_{L}}\, \times\, \mathcal{PAC}^{n-(j - i)}_{\mathcal{Q}_{R}}
\end{array}
\end{flalign}
\end{tcolorbox}
where $\mathcal{Q}_{L}$ and $\mathcal{Q}_{R}$ are pseudo-quadrangulations to the left and the right of the diagonal which is set to zero.  We suspect that this structure is rather general and extend to pseudo p-gulations for $p\, >\, 4$.

\subsection{Intgrands for $\phi^{4}$ amplitudes}
In this section,  we  relate the pull-back of the $\frac{n}{2}$ forms $\omega_{n}^{Q}$ on ${\cal D}_{n}^{T}$(or equivalently, the canonical form associated to ${\cal S}_{n}^{Q}$) to the integrands of $\phi^{4}$ 1-loop amplitudes in a ``forward limit" representation.\footnote{A more detailed relationship between the canonical forms on pseudo-accordiohedra and the Feynman integrands along the lines of \cite{Arkani-Hamed:2019vag, Yang:2019esm} is left for the future.} 

We start the $n\, =\, 4$ example and choose reference pseudo-quadrangulation as ${\cal Q}_{1}\, =\, \{1_{L},\, 14,\,   \overline{1}_{L}\, \bar{1}    \bar{4},\, \}$.\\ 
For ${\cal Q}_{1}$ we can write the canonical form on ${\cal S}_{{\cal Q}_{1}}^{4}$ ( defined in eqn. \eqref{ctfopa})  as,\begin{flalign}
\begin{array}{lll}
\omega^{n}_{{\cal Q}_{1}}&=\, \left(\, \frac{1}{Y_{1}}\, (\, \frac{1}{Y_{3}}\, +\, \frac{1}{X_{14}}\, )\, +\, \frac{1}{\tp{Y}_{4}}\, (\, \frac{1}{X_{14}}\, +\, \frac{1}{X_{3\bar{4}}}\, )\, +\, \frac{1}{Y_{3}} \frac{1}{X_{3\bar{4}}}\, \right)\, dY_{1}\, \wedge\, d X_{14}\\[0.1em]
&=:\, m_{4}^{{\cal Q}_{1}}\, d Y_{1}\, \wedge\, d X_{14}
\end{array}
\end{flalign}
We thus see that,
\begin{flalign}
\lim_{Y_{1}\rightarrow\, 0}\, Y_{1}\, m_{4}^{{\cal Q}_{1}}\, =\, \frac{1}{Y_{3}}\, +\, \frac{1}{X_{14}}\\
\end{flalign}
This residue is in fact a $6$ point forward scattering (partial) amplitude associated to reference quadrangulation $Q_{t}$. 
\begin{flalign}
m_{6}^{Q_{t}\, =\, (14)}(1,\dots, 4, -, +)\, =\, \frac{1}{Y_{3}}\, +\, \frac{1}{X_{14}}
\end{flalign}
where we identify $k_{+}\, =\, -k_{-}\, =\, l$ and hence  we identify $Y_{3}$ with $Y_{3}\, =\, 2 p_{3}\, \cdot l$

where $m_{6}^{Q_{t}\, =\, (14)}(1,\dots, 4, -, +)$ is the tree-level scattering amplitude in the forward limit with  $X_{3+}\, =:\, Y_{3}$. Similarly, if we choose ${\cal Q}_{2}$, as a reference
\begin{flalign}
{\cal Q}_{2}\, =\,  \{2_{L},\, 4_{L},\, \overline{2}_{L},\, \overline{4}_{L}\, \}\\
\end{flalign}
then it can be easily verified that, 
\begin{flalign}
\begin{array}{lll}
\lim_{Y_{2}\, \rightarrow\, 0}\, Y_{2} m_{4}^{{\cal Q}_{2}}\, =\, m_{6}^{Q_{t}\, =\, (4+)}(1,\dots, 1,\dots, -, +)\\[0.2em]
m_{6}^{Q_{t}\, =\, (4+)}(2,\, \dots,\, 1,\, \dots,\,  -,\,  +)\, =\, \frac{1}{Y_{4}}\, +\, \frac{1}{X_{3\bar{2}}}
\end{array}
\end{flalign}
This result is true in general. That is, given any reference pseudo-quadrangulation $Q\, =\, Q_{t}\, \cup\, i_{L/R}$, if the variable associated to the central chord is $y_{i}$ ($y_{i}$ can be $Y_{i}$ or $\tp{Y}_{i}$ depending on the orientation of the central chord) then,
\begin{flalign}
\lim_{y_{i}\rightarrow 0}\, y_{i}\, m_{n}^{Q_{t}\, \cup\, i_{L/R}}\, =\, m_{n+2}^{Q_{t}}(i+1,\, \dots,\, n,\, 1,\, -\, +)
\end{flalign}

\section{Primitives and Weights}
Just as in the case of tree-level amplitudes for quartic interactions, a single polytope of a given dimension does not encompass all the singularities of the $n$ point one loop integrand. The full integrand is obtained after summing over all possible pseudo accordiohedra of a given dimension with each term in the sum weighted by a factor. In literature these factors are referred to as weights and are determined by the combinatorics of the (pseudo) quadrangulations. For a detailed discussion on the weights in generic $\phi^{p\, >\, 3}$ theories, we refer the reader to \cite{Raman:2019utu, Aneesh:2019ddi}. Recently it has been proved that for tree-level amplitudes, these weights are uniquely determined using certain recursion relations \cite{Kojima:2020tox} and have a nice relationship with BCFW recursion relations in $\phi^{4}$ theory \cite{Srivastava:2020dly, Feng:2009ei}. 

In this section, we derive the formula for weights associated to pseudo-accordihedra.\\
Upon identifying centrally symmetric points of a $2n$-gon with a disc at center, a centrally symmetric pair of chords reduces to a chord of an $n$-gon. Thus, there is a natural action of dihedral group $\mathrm{Dih}_{n}$ on the space of centrally symmetric pairs chords of $2n$-gon. Along with the $\mathrm{Dih}_{n}$ action there is a $\mathbb{Z}_{2}$ action on the space of centrally symmetric pairs chords of $2n$-gon which takes $\{i_{L},\bar i_{L}\}$ to $\{i_{R},\bar i_{R}\}$  and $\{i_{R},\bar i_{R}\}$  to $\{i_{L},\bar i_{L}\}$. These  $\mathrm{Dih}_{n}$ and $\mathbb{Z}_{2}$ group actions induce $\mathrm{Dih}_{n}$ and $\mathbb{Z}_{2}$ group actions on the space of centrally symmetric pseudo-quadrangulations and the space of pseudo-accordiohedra. We call the representatives of the orbits under combined $\mathrm{Dih}_{n}$ and $\mathbb{Z}_{2}$ action on the space of centrally symmetric pseudo-quadrangulations \textbf{primitives}. We denote the set of all primitive pseudo-quadrangulations of $2n$-gon by $\mathcal{P}_{4,n}$. The map which takes reference pseudo-quadrangulation to its pseudo-accordiohedra is $\mathrm{Dih}_{n}$ and $\mathbb{Z}_{2}$ equivariant map.  Therefore once we know the pseudo-accordiohedron associated with a primitive we know pseudo-accordiohedra associated with all other elements of the orbit of that primitive. 

Using the fact that the map which takes dissections of $n$-gon to accordiohedra associated with them is $\mathbb{Z}_{n}$ equivariant it was argued in \cite{Aneesh:2019ddi} that the weights of accordiohedra associated with dissections of $n$-gon in the orbit of $\mathbb{Z}_{n}$ action are same. These arguments can be extended to weights of pseudo-accordiohedra. Thus the weights of pseudo-accordiohedra associated pseudo-quadrangulations of $2n$-gon in the orbit of combined  $\mathrm{Dih}_{n}$ and $\mathbb{Z}_{2}$ action are same.

To compute weights it would be useful to define the functions $\delta(D_{i},D_{j})$ and $M(D_{i},D_{j})$.  $\delta(D_{i},D_{j})$ tells you wether the pseudo-quadrangulation $D_{j}$ occurs in $\mathcal{PAC}(D_{i})$, the pseudo-accordiohedra of $D_{i}$ or not. That is, 
\begin{equation}
\delta(D_{i},D_{j})=
 \begin{cases} 
1 \text{ if $D_{j} \in \mathcal{PAC}(D_{i})$ }\\
0 \text{ if $D_{j} \notin \mathcal{PAC}(D_{i})$ }
\end{cases}
\end{equation}
$M(D_{i},D_{j})$ is the number of elements from the orbit of $D_{i}$ occur in any pseudo-accordiohedron of a quadrangulation in the orbit of $D_{j}$. That is,
\begin{equation} 
M(D_{i},D_{j})= \sum_{D_{k} = \sigma \cdot D_{i}} \delta(D_{j},D_{k}).
\end{equation}
Where the sum is over $\sigma$ in $\mathrm{Dih}_{n}$ and $\mathbb{Z}_{2}$. Notice the function $M(D_{i},D_{j})$ is same for all pseudo-quadrangulations from the same orbit. Thus it is a function on space of primitives. In terms of $M$ the equations which gives us weights is given by 
\begin{equation}\label{weighteq}
\sum_{D_{j} \in \mathcal{P}_{4,n} } \frac{|G_{D_{i}}|}{|G_{D_{j}}|} M(D_{i},D_{j} ) \alpha_{D_{j}} = 1 \hspace{1cm} \forall D_{i} \in \mathcal{P}_{4,n}.
\end{equation}
Where $G_{D}$ is the stabilizer of $D$ under the combined action of $\mathrm{Dih}_{n}$ and $\mathbb{Z}_{2}$ and  $\alpha_{D}$ is the weight of primitive $D$.

Let's work out the $n=4$ example. At $n=4$, there are two primitives, ${\cal Q}_{1}\, =\, \{14,\, 1_{L},\, \bar{1}    \bar{4},\, \overline{1}_{L}\, \}$  and ${\cal Q}_{3}\,=\{1_{L},\bar 1_{L}\},\{3_{L},\bar 3_{L}\}$. The sizes of stabilizers of $\mathcal{Q}_{1}$ and $\mathcal{Q}_{3}$ are $|G_{\mathcal{Q}_{1}}| = 1$ and $|G_{\mathcal{Q}_{3}}| = 4$. The values of function $M$ are as follows
\begin{align}
M(\mathcal{Q}_{1},\mathcal{Q}_{1}) &= 4  &    M(\mathcal{Q}_{1},\mathcal{Q}_{3}) &=  4 \endline 
M(\mathcal{Q}_{3},\mathcal{Q}_{1}) &= 1  &    M(\mathcal{Q}_{3},\mathcal{Q}_{3}) &=   2  .
\end{align}
Therefore, the equation for weights is,
\begin{align}
4 \alpha_{\mathcal{Q}_{1}} +  \alpha_{\mathcal{Q}_{3}} &= 1 \\
4 \alpha_{\mathcal{Q}_{1}} + 2 \alpha_{\mathcal{Q}_{3}} &= 1 
\end{align}
Which implies $\alpha_{\mathcal{Q}_{1}} =\frac{1}{4}$ and $\alpha_{\mathcal{Q}_{3}}=0$. Note that the four-point amplitude with quartic interaction is weighted sum of $20$ terms, however because of the function $M$ and equation \eqref{weighteq} we just have to analyse two pseudo-accordiohedra to deduce the weights. In general, we just have to analyse the pseudo-accordiohedra of primitives to deduce the weight.

In general, the full amplitude is given by the weighted sum 
\begin{equation}
    m_{n} = \frac{1}{2} \sum_{\mathcal{Q}_{i} \in \mathcal{P}_{4,n}} \sum_{\sigma} \alpha_{\mathcal{Q}_{i}} m_{n}^{\sigma \cdot \mathcal{Q}_{i} }.
\end{equation}
Where the sum is over the orbits of primitives $\mathcal{Q}_{i}$ and the overall half is to account for over counting due to $Y, \tilde{Y}$ doubling. The full four-point amplitude is given by 
\begin{equation}
    m_{4} = \frac{1}{2} \sum_{\sigma} \alpha_{1} m_{4}^{\sigma \cdot \mathcal{Q}_{1}} + \frac{1}{2}  \sum_{\sigma} \alpha_{3} m_{4}^{\sigma \cdot \mathcal{Q}_{3}}
\end{equation}
 As $\alpha_{\mathcal{Q}_{1}} =\frac{1}{4}$ and $\alpha_{\mathcal{Q}_{3}}=0$ from \eqref{q1restriction} we get 
\begin{align}
  m_{4} &=  \frac{1}{2} \bigg[\frac{1}{Y_{1}X_{1 \bar 2}}   +   \frac{1}{Y_{2}X_{1 \bar 2}}   +   \frac{1}{Y_{2}X_{2 \bar 3}}   +   \frac{1}{Y_{3}X_{2 \bar 3}}   +   \frac{1}{Y_{3}X_{3 \bar 4 }}   +   \frac{1}{Y_{4}X_{3 \bar 4 }}   +   \frac{1}{Y_{4}X_{14}}   +   \frac{1}{Y_{1} X_{14}}   \endline & \hspace{0.2cm} +  \frac{1}{Y_{1}Y_{3}}   +   \frac{1}{Y_{2}Y_{4}}   +  ( Y \rightarrow \tilde{Y} )   \bigg] 
\end{align}
In the abstract kinematic space ${\cal KL}_{n}$, $X_{ij},\, X_{i\bar{j}}\, i\, <\, j$ are independent and abstract variables.  But once we substitute the  physical values for $X_{i j}, X_{i\bar j}$ as
\begin{flalign}
\begin{array}{lll}
X_{ij}\, =\, (p_{i}\, +\, \ p_{j-1})^{2}\, \forall\, 1\, \leq\, i\, \leq\, n-1\\
X_{i\bar{j}}\, =\, (p_{j}\, +\, \dots\, +\, p_{n}\, +\, p_{1}\, +\, p_{i-1})^{2}\, \forall\, 1\,\leq\, i\, <\, j-1\, \leq\, n
\end{array}
\end{flalign}
we can then identify the abstract loop variables with ``Mandelstam variables" for loop integrands and write the (stripped) integrand as,
\begin{align}
 m_{4}  &=    \frac{1}{ \ell^{2} (\ell + p_{1} + p_{2})^{2}}   +   \frac{1}{(\ell + p_{1})^{2} (\ell + p_{1} + p_{2} + p_{3} )^{2}}  + \frac{1}{ \ell^{2}   (p_{2}+ p_{3} + p_{4})^{2}}  +   \frac{1}{ \ell^{2}   (p_{1}+ p_{2} + p_{3})^{2} }   \endline & \hspace{0.2cm}   +   \frac{1}{(\ell + p_{1})^{2}  (p_{2}+ p_{3} + p_{4})^{2}}   +   \frac{1}{(\ell + p_{1})^{2} (p_{3}+ p_{4} + p_{1})^{2}  } \endline & \hspace{0.2cm}  +   \frac{1}{(\ell + p_{1} + p_{2})^{2} (p_{3}+ p_{4} + p_{1})^{2}  }    +   \frac{1}{(\ell + p_{1} + p_{2})^{2}(p_{4}+ p_{1} + p_{2})^{2} }  \endline & \hspace{0.2cm}  +   \frac{1}{(\ell + p_{1} + p_{2} + p_{3} )^{2} (p_{4}+ p_{1} + p_{2})^{2} }   +   \frac{1}{(\ell + p_{1} + p_{2} + p_{3} )^{2}  (p_{1}+ p_{2} + p_{3})^{2} }   
\end{align}
We note that this is a stripped integrand with no momentum conservation imposed. The tadpole contributions need to be subtracted before impoing momentum conservation. 
\section{Outlook and Open questions}
The Amplituhedron program \cite{Arkani-Hamed:2013jha, Arkani-Hamed:2013kca, Franco:2014csa} to unearth the analytic structure of the S-matrix has seen many remarkable strides in recent years \cite{Arkani-Hamed:2018rsk, Arkani-Hamed:2017vfh}. In the space of non-super symmetric quantum field theories, with the work of Arkani-Hamed, Bai, He and Yan a clear synthesis of ideas appear to be emerging for scalar field theories. The synthesis is based on intertwining relationship between, various theories (classified by degree of the interaction), various polytopes and a very specific class of algebras , namely cluster algebras for $\phi^{3}$ theory and the so-called gentle algebras for generic scalar interactions. 
\begin{center}
\resizebox{\columnwidth}{!}{
\begin{tabular}{ |c | c | c | c | c |}
\hline
Order    &   Interaction  &  Polytope   &  Combinatorial Model   &  Algebra   \\
\hline
Tree & Cubic & Associahedra & Triangulations & Type-A cluster algebra \\
Tree & Quartic & Stokes Polytope & Quadrangulations & Gentle algebra \\
Tree & Polynomial & Accordiohedra & Dissections & Gentle algebra \\
1-Loop & Cubic & Type-D Associahedra & Pseudo-triangulations &  Type-D cluster algebra \\
1-Loop & Quartic & Pseudo-accordiohedra & Pseudo-quadrangulations & Colored gentle algebra.\\
\hline
\end{tabular}
}

\end{center}
Moreover, we now understand that in a precise sense (type-A or type-D) associahedron is a ``universal" polytope for all scalar theories whose amplitudes are generated by forms of various ranks on the associahedron. These forms are not arbitrary but are unique due to the principle of projectivity.\\ 
In this paper, we initiated a study of one-loop integrands in  massless $\phi^{4}$ theory as scattering forms. One of our main results is summarised in the last row in the table above. We also showed how the integrand can be thought of as a (weighted) sum of lower $\frac{n}{2}$ form on the $n$ dimensional  type-D associahedra in kinematic space \cite{Arkani-Hamed:2019vag}. The weights are determined by the combinatorics of pseudo-accordiohedra.\\ 
Many questions remain open and we list here just two of them. In addition to the singularities that we expect in the integrand of $\phi^{4}$ scattering amplitude, the canonical form on peudo-accordiohedra have singularities corresponding to radiative corrections to the external legs. As we know, these singularities make the (un-stripped, that is including momentum conserving delta function) amplitude infinite and in quantum field theories, we re-sum the radiative corrections to all loops and then perform mass renormalisation. The canonical forms on pseudo-accordiohedra should hence be thought of as integrands of the stripped amplitude and the relationship of integrand of the un-stripped amplitude with the scattering form remains to be investigated.\footnote{These issues  have been investigated in the context of CHY formula for one loop integrand in bi-adjoint $\phi^{3}$ theories recently \cite{Feng:2019xiq}}\\ 
Although we defined pseudo-accordiohedra using quadrangulation of the $2n$-gon with annulus, we suspect that these polytopes are already known in the mathematics literature as accordion complexes \cite{konk-palu}. In fact, these accordion complexes are quite general and may even  be relevant for higher loop integrands in $\phi^{4}$ theory. We leave investigation of these complexes for future work.

\section*{Acknowledgement}
We are deeply indebted to Vincent Pilaud for constantly guiding us through much of daunting mathematics literature on this subject, for his crucial insights at various stages of this work and constant encouragement. A special thanks to Frederic Chapoton for being patient with our questions over past few years and pointing us towards key references in the world of Accordiohedra and related subjects. We thank Pinaki Banerjee,  Renjan John and Prashanth Raman for many discussions on these subjects.

\appendix
\section{Geometric Realisations of $\mathcal{D}_{3}$ and $\mathcal{D}_{4}$}\label{appA}
We will start with $n\, =\, 3$ case and look at a dissection quiver associated to $\{\, 1 p_{L} \bar{1},\, 1 p_{R} \bar{1},\, 13,\, \bar{1}\, \bar{3}\, \}$. 
With these rules in place, let us write all the independent contstraints for aforementioned quiver.  
\begin{flalign}
\begin{array}{lll}
 X_{13}\, +\, X_{2\bar{1}}\, -\, X_{1\bar{1}}^{2}\, =\, c_{13}\\
 X_{2\bar{1}}\, +\, X_{2\bar{3}}\, -\, X_{2\bar{2}}^{0}\, =\, c_{13, 1p_{L}}\\
 X_{1\bar{1}}^{0}\, +\, X_{2\bar{2}}^{-2}\, -\, X_{1\bar{2}}\, -\, X_{2\bar{1}}\, =\, c_{(13),(\bar{1}\bar{3})}\\
 X_{3\bar{3}}^{0}\, +\, X_{2\bar{2}}^{0}\, -\, X_{2\bar{3}}\, -\, X_{3\bar{2}}\, =\, c_{v_{(\bar{1}p_{R}),(1p_{L})}}
\end{array}
\end{flalign}
In the final equation both the paths $X_{2\bar{3}}$ and $X_{\bar{2}3}$ are allowed by our criteria and so we consider both of them. There are six constraints in all as, the last two constraints split in four constraints as $X_{i\overline{i}}^{0}\, =\, Y_{i}\, +\, \tilde{Y}_{i}$. One can check that these are the only six independent constraints we obtain from this quiver.\\
We now look at  the wheel quiver which has no chords that are not central. In this case as well, there are precisely 6 independent constraints. These are,
\begin{flalign}
\begin{array}{lll}
 X_{i\overline{i+1}}\, +\, X_{i+1\overline{i+2}}\, -\, X_{i+1\overline{i+1}}\, =\, c_{1}\, 1\, \leq\, i\, \leq\, 3\\
 X^{-1}_{i+2 \overline{i+2}}\, +\, X^{+1}_{i+1\overline{i+1}}\, =\, c_{i,i+1}\, 1\, \leq\, i\, \leq\, 3
\end{array}
\end{flalign}
where $i+1\, =\, i - 3$.\\ 
We can now look at the wheel quiver where the reference dissection is $\{\, X_{i a_{L}}\, X_{\bar{i}, p_{R}}\, \}$. In this case, it can be shown that the constraints are 
\begin{flalign}
\begin{array}{lll}
 X_{i\overline{i+1}}\, +\, X_{i+1\overline{I+2}}\, -\, X_{i+1\overline{i+1}}\, =\, c_{1}\, 1\, \leq\, i\, \leq\, 3\\
 X^{+1}_{i+2 \overline{i+2}}\, +\, X^{-1}_{i+1\overline{i+1}}\, - X_{i+1\overline{i+2}}\, =\, c^{\prime}_{i,i+1}\, 1\, \leq\, i\, \leq\, 3
\end{array}
\end{flalign}
So with our dictionary $X_{i\bar{i}}^{-1}\, =\, \tilde{Y}_{i}$ and $X_{i\bar{i}}^{+1}\, =\, Y_{i}$,  We probably get expected constraints . 

We finally consider a quiver associated to a ``cyclic seed" where the reference dissection is spanned by $\{\, X_{1 p_{R}},\, X_{13},\, X_{3 p_{R}},\, \dots\, \}$. In this case we get the following set of independent constraints.
\begin{flalign}
\begin{array}{lll}
X_{2\bar{3}}\, +\, X_{1\bar{2}}\, -\, X_{2\bar{2}}^{0}\, =\, c_{2\bar{3}}\\[0.4em]
X_{\bar{1}\bar{3}}\, +\, X_{3\bar{2}}\, -\, X_{3\bar{3}}^{0}\, =\, c_{\bar{1} \bar{3}}\\
X_{2\bar{3}} + X_{\bar{3}\bar{1}} - X_{\bar{1}2}\, =\,C \\
X_{2\bar{1}}\, +\, X_{13}\, -\, X_{1\bar{1}}^{0}\, =\, C\\
X_{1\bar{1}}^{0}\, +\, X_{2\bar{2}}^{0}\, -\, 2\, X_{1\bar{2}}\, =\, C
\end{array}
\end{flalign}

We now consider the example of four dimensional polytope. In this case as well, we show for that pseudo triangulations which are degenerate or of wheel type, the gentle algebra generates the geometric realisation defined in \ref{typedrealisation}. 
\subsubsection*{$T_{2}$ quiver}
Constraints based on our rules on assigning degrees .
\begin{flalign}
\begin{array}{lll}
 X_{i+1\bar{i+2}}\, +\, X_{\bar{i+2}\bar{i+4}}\, -\, X_{i+2\bar{i+2}}\, =\, c_{i}\\[0.4em]
 X_{i+1\overline{i+2}}\, +\, X_{i+2\overline{i+3}}\, -\, X_{i+1\bar{i+3}}\, -\, X_{i+2\bar{i+2}}^{0}\, =\, c_{i,i+1} \hspace{1cm} 1\leq i \leq 4  \\[0.4em]
 Y_{i} + Y_{\overline{i+1}}\, -\, X_{i,\bar{i+1}}\, =\, c_{i\bar{i}}\,  \textrm{ three constraints}\\[0.4em]
 X_{1\overline{4}}\, +\, X_{\bar{3}4}\, -\, X_{1\bar{3}}\, -\, X_{4\bar{4}}^{0}\, =\, c_{1,4}\\[0.4em]
\end{array}
\end{flalign}
So we get 12 constraints.
\subsubsection*{Dissection quiver of $T_{1}$ : Two examples}
$\{\, X_{13},\, X_{14}\, X_{1 p_{L} \bar{1}}\, X_{1 p_{R} \bar{1}}\, \}$
The constraints we get are,
\begin{flalign}
\begin{array}{lll}
X_{13}\, +\, X_{24}\, -\, X_{14}\, =\, c_{13}\\
X_{24}\, +\, X_{3\bar{1}}\, -\, X_{\bar{1}1}\, =\, c_{24}\\
X_{2\bar{1}}\, +\, X_{14}\, -\, X_{24}\, -\, X_{1\bar{1}}\, =\, c_{13}\\
X_{2\bar{4}}\, +\, X_{1\bar{3}}\, -\, X_{2\bar{3}}\, =\, c_{2\bar{4}}\\
X_{2\bar{3}}\, +\, X_{3\bar{4}}\, -\, X_{2\bar{4}}\, - X_{3\bar{3}}\, =\, c_{2\bar{4}}\\
X_{1\bar{1}}\, +\, X_{2\bar{2}}\, -\, X_{1\bar{2}}\, -\, X_{2\bar{1}}\, =\, c_{1\bar{1}}\\
X_{2\bar{2}}\, +\, X_{3\bar{3}}\, -\, X_{2\bar{3}}\, -\, X_{3\bar{2}}\, =\, c_{2\bar{2}}\\
X_{4\bar{4}}\, +\, X_{3\bar{3}}\, -\, X_{3\bar{4}}\, -\, X_{4\bar{3}}\, =\, c_{3\bar{3}}\\
X_{1\bar{2}}\, +\, X_{2\bar{3}}\, -\, X_{1\bar{3}}\, -\, X_{2\bar{2}}\, =\, c_{1\bar{2}}
\end{array}
\end{flalign}
where the final constraint came from an L-shaped path which transverse lower left of the eye. \\
We consider reference dissection $\{\, X_{2\bar{1}},\, X_{3\bar{1}},\, Y_{1},\, \tilde{Y}_{1}\, \}$. We expect that the constraints we generate from this dissection 
\begin{flalign}
\begin{array}{lll}
 X_{13}\, +\, X_{24}\, -\, X_{14}\, =\, c_{1}\\
 X_{3\bar{4}}\, +\, X_{14}\, -\, X_{13}\, -\, X_{4\bar{4}}^{0} =\, c_{2}\\
 X_{3\bar{1}}\, +\, X_{4\bar{2}}\, -\, X_{3\bar{2}}\, =\, c_{3}\\
 X_{3\bar{2}}\, +\, X_{4\bar{3}}\, -\, X_{\bar{2}4}\, -\, X_{\bar{3}3}^{0}\, =\, c_{4}\\
 Y_{i}\, +\, \tilde{Y}_{i+1}\, -\, X_{i\bar{i+1}}\, =\, c_{i\bar{i}}\, 1\,\leq i \leq\, 3\\
\tilde{Y}_{i}\, +\, Y_{i+1}\, -\, X_{\bar{i}i+1}\, =\, c_{\bar{i}i}\, 1\, \leq i \leq\, 3\\
 X_{13}\, +\, X_{2\bar{4}}\, -\, X_{3\bar{4}}\, =\, c_{13}\\
 X_{2\bar{4}}\, +\, X_{13}\, -\, X_{3\bar{4}}\, =\, c_{2\bar{4}}
\end{array}
\end{flalign}
{\bf Final example} \\
We consider reference dissection $\{\, X_{2\bar{1}},\, X_{3\bar{1}},\, Y_{1},\, \tilde{Y}_{1}\, Y_{2}, \tilde{Y}_{2} \}$. The set of independent constraints that we get from the gentle algebra is
\begin{flalign}
\begin{array}{lll}
 X_{13}\, +\, X_{24}\, -\, X_{14}\, =\, c_{1}\\
 X_{14} + X_{3\bar{4}}\, -\, X_{13}\, -\, X_{4\bar{4}}^{0} = c\\
 X_{3\bar{1}}\, +\, X_{4\bar{2}}\, -\, X_{3\bar{2}}\, =\, c\\
 X_{3\bar{2}} +\, X_{4\bar{3}}\, -\, X_{2\bar{4}}\, -\, X_{3\bar{3}}^{0}\, =\, c\\
 X_{3\bar{3}}^{0}\, +\, X_{4\bar{4}}^{0}\, -\, X_{3\bar{4}}\, -\, X_{\bar{3}4}\, =\, c\\
 X_{2\bar{4}}\, +\, X_{13}\, -\, X_{3\bar{4}}\, =\, c\\
 X_{3\bar{2}}\, +\, X_{2\bar{1}}\, -\, X_{3\bar{1}}\, -\, X_{2\bar{2}}^{0}\, =\, c\\
 X_{2\bar{2}}\, +\, X_{3\bar{3}}\, -\, X_{2\bar{3}}\, -\, X_{\bar{2}3}\, =\, c\\
 \tilde{Y}_{1} + Y_{2} - X_{\bar{1}2}\, =\, c\\
 \tilde{Y}_{4}\, +\, Y_{1}\, -\, X_{14}\, =\, c
\end{array}
\end{flalign}

\end{document}